\documentclass[aps,prb,twocolumn,superscriptaddress]{revtex4}
\usepackage{amsmath,amssymb,bm}
\usepackage{graphicx}
\usepackage{epstopdf}
\usepackage{latexsym}
\usepackage{subfigure}
\usepackage{color}

\newcommand{\ccc}{{Cs$_2$CuCl$_4$}}
\newcommand{\ccb}{{Cs$_2$CuBr$_4$}}

\newcommand{\vs}[1]{\bm #1}
\newcommand{\calg}[1]{\cal #1}

\newcommand{\vectorize}[2]{{\begin{array}{c} #1 \\ #2 \end{array}}}

\begin{document}
\title{Extreme sensitivity of a frustrated quantum magnet: \ccc}
\author{Oleg A. Starykh}
\affiliation{Department of Physics and Astronomy, University of Utah, Salt Lake City, UT 84112}
\author{Hosho Katsura}
\affiliation{Kavli Institute for Theoretical Physics, University of
  California, Santa Barbara, CA 93106-9530}
\author{Leon Balents}
\affiliation{Kavli Institute for Theoretical Physics, University of
  California, Santa Barbara, CA 93106-9530}

\date{\today}
\begin{abstract}
  We report a thorough theoretical study of the low temperature phase
  diagram of \ccc, a spatially anisotropic spin $S=1/2$ triangular
  lattice antiferromagnet, in a magnetic field.  Our results, obtained
  in a quasi-one-dimensional limit in which the system is regarded as a set of
  weakly coupled Heisenberg chains, are in excellent agreement with
  experiment.  The analysis reveals some surprising physics.  First, we
  find that, when the magnetic field is oriented within the triangular layer, 
  spins are actually most strongly correlated within planes 
  {\sl  perpendicular} to the triangular layers. 
    This is  despite the fact that the inter-layer exchange coupling in \ccc\
    is about an order of magnitude smaller than the weakest (diagonal) exchange 
    in the triangular planes themselves.  Second, the phase diagram in such orientations is
  exquisitely sensitive to tiny interactions, heretofore neglected, of
  order a few percent or less of the largest exchange couplings.  These
  interactions, which we describe in detail, induce entirely new phases,
  and a novel commensurate-incommensurate transition, the signatures of
  which are identified in NMR experiments.  We discuss the differences
  between the behavior of \ccc\ and an ideal two-dimensional triangular
  model, and in particular the occurrence of magnetization plateaux in
  the latter.  These and other related results are presented here along
  with a thorough exposition of the theoretical methods, and a
  discussion of broader experimental consequences to \ccc\ and other
  materials.
\end{abstract}

\maketitle

\section{Introduction}
\label{sec:introduction}

The spin-$1/2$ nearest-neighbor Heisenberg antiferromagnet on the two
dimensional triangular (hexagonal) lattice is the simplest theoretical
model for frustrated quantum magnetism.\cite{Anderson_RVB}   Without
additional perturbations, the model is believed to order at zero
temperature into a 3-sublattice coplanar ground 
state.\cite{Huse_Elser_1988_PRL, Bernu1992PRL,Sorella1999PRL}  
However, one may expect a strong sensitivity to
additional perturbations to the isotropic triangular lattice Hamiltonian.  The material \ccc\ provides
an interesting example of a spatially {\sl anisotropic} spin-$1/2$
triangular antiferromagnet.\cite{Coldea2001PRL}~ For several years, neutron scattering,\cite{Coldea2003PRB}
magnetization, and specific heat measurements\cite{RaduPRL2005, TokiwaPRB2006} on \ccc\ have intrigued the
community with unexpected behavior.  These experimental properties have
been suggested by a variety of authors, including the experimentalists
themselves, to indicate exotic physics such as a spin-liquid ground
state, unconventional ``spinon'' excitations at higher energies, and
quantum criticality.  Theoretical work on this material has been
intense.~\cite{Chung2003PRB,IsakovPRB2005PRB,AliceaPRL2005,
kohno2009dps,kohno07:_spinon_and_tripl_in_spatial,Starykh2007,PhysRevLett.103.197203}

\begin{figure}[htb]
\begin{center}
\vspace{.2cm}
\includegraphics[width=0.7\columnwidth]{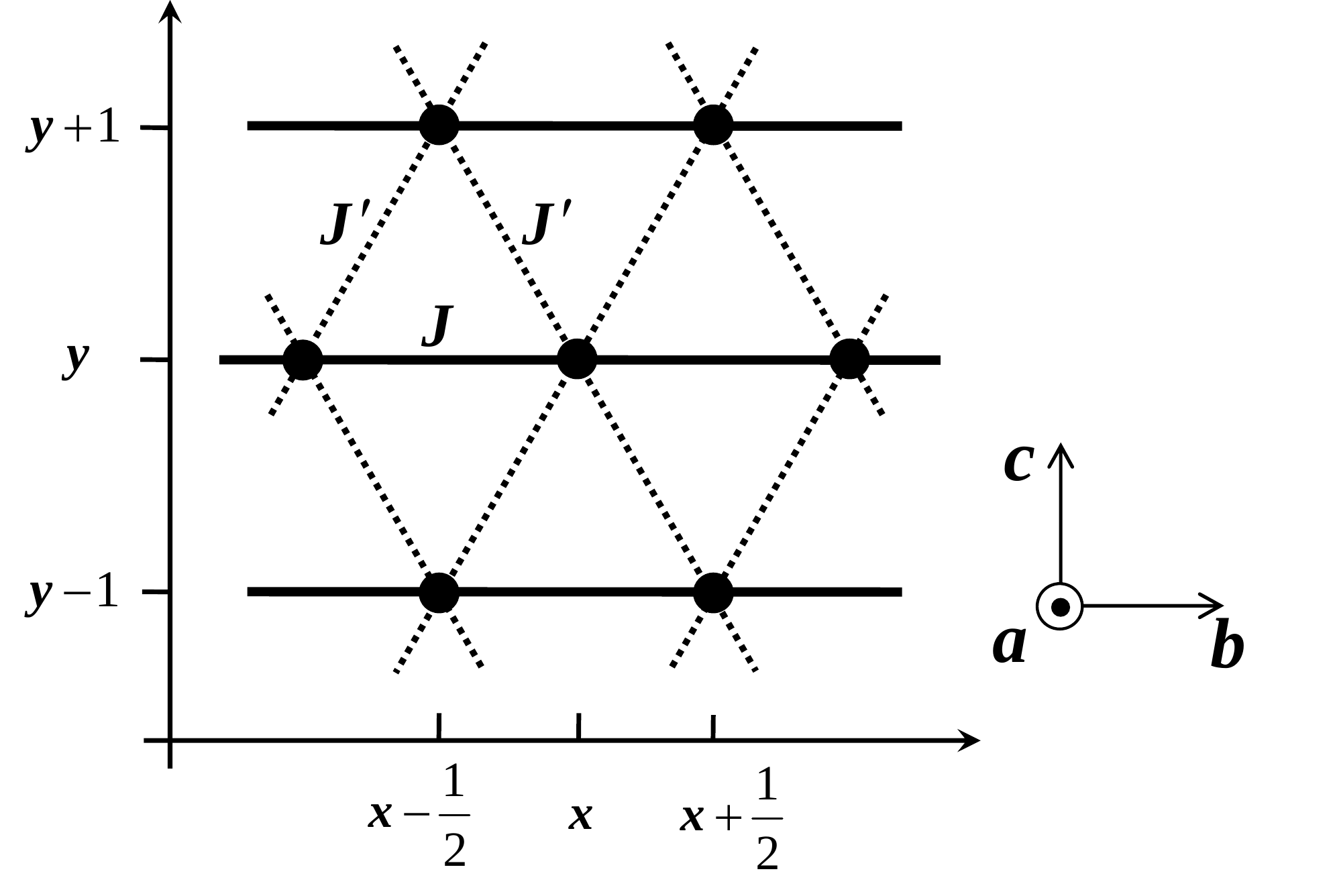}
\caption{Magnetic sites and exchange couplings in a 2D triangular layer in \ccc: 
on-chain bonds $J$ (thick lines $\parallel b$), frustrating diagonal bonds $J'$ (dotted lines). 
Stacked layers are coupled by nearest-neighbor exchange $J''$. Crystallographic $a$,$b$,$c$ axes are indicated.}
\label{fig: planar_lattice}
\end{center}
\end{figure}

An advantage of \ccc\ is that the small exchange constants and high
degree of magnetic isotropy allow for a fairly accurate determination of
several of the largest Hamiltonian parameters, by comparison with the
measured single magnon dispersion relation above an approximately
fully-polarized state.  The approximate Hamiltonian, including a
magnetic field, determined by the experimentalists in this way is
\begin{eqnarray}
  \label{eq:Hexpt}
  H & = & \frac{1}{2} \sum_{ij} \left[ J_{ij} {\bm S}_i \cdot
    {\bm S}_j - {\bm D}_{ij} \cdot {\bm S}_i \times {\bm S}_j \right]
  - {\bm h}\cdot \sum_i {\bm S}_i.~~~
\end{eqnarray}
Here $i,j$ are sites of a stack of triangular lattices (see Figs. \ref{fig: planar_lattice} and \ref{fig: DM_distribution}).
The principle exchange interactions determined in Ref. \onlinecite{Coldea2002PRL} are 
$J=0.374 ~{\rm meV}$ on nearest-neighbor bonds parallel to the $b$ axis,
significantly smaller $J'=0.128~{\rm meV}\approx 0.34 J$ on diagonal
bonds in the $b$-$c$ plane, and quite small $J''=0.045 J$ 
along vertical bonds between adjacent triangular layers.  A
Dzyaloshinskii-Moriya (DM) coupling was also measured along the diagonal
bonds, with ${\bm D}_{ij}=-{\bm D}_{ji}=\pm D \hat{a}$ where the sign in
specific directions are as indicated in Fig. \ref{fig: DM_distribution}, and
$D=0.020$meV$=0.05J$.

The relative smallness of the couplings other than $J$ suggests that one
may perhaps fruitfully regard this Hamiltonian as one of Heisenberg spin
chains along the $b$ axis defined by $J$, which are weakly coupled
together by the remaining interactions.\cite{nersesyan1998,bocquet2001}  This point
of view was validated in
Refs.~\onlinecite{kohno2009dps,kohno07:_spinon_and_tripl_in_spatial,Starykh2007,PhysRevLett.103.197203},
where it was shown that much 
of the observed low energy ordering {\sl and} the high energy inelastic
neutron scattering data on \ccc\ can be calculated starting from known
exact results for 1d Heisenberg chains.  Indeed, numerical approaches in
Refs.\onlinecite{DN_Sheng_PRB2006,Yunoki_Sorella_PRB2006,Hayashi_Ogata_JPSJ2007,
Pardini_Singh_PRB2008,Jiang_Sheng_Balents_PRB2009,Heidarian_Sorella_PRB2009,Tay_Motrunich} 
showed that, due to frustration, the diagonal interaction may be
as large as $J'/J < 0.7$ while still retaining approximate
quasi-one-dimensionality.  

While this approach has been quite successful for \ccc, most notably in
directly confronting data {\sl without any adjustable parameters}, there
remain some puzzling features in the experiments.  One of the most
striking ones is the drastic difference in the low
temperature phase diagrams of the material in magnetic fields aligned
along the three different principle axes of the crystal.  Though some
aspects of these differences were explained in Ref.~\onlinecite{Starykh2007},
based upon the ``standard'' model in Eq.~(\ref{eq:Hexpt}), other glaring
discrepancies remain.  In this paper, we resolve these outstanding
differences between theory and experiment by correcting the standard
model of \ccc.

It is important to emphasize that the corrections to
Eq.~(\ref{eq:Hexpt}) {\sl must} be small, because the standard model
does an excellent job in explaining a large volume of experimental
data.  The parameters in Eq.~(\ref{eq:Hexpt}) were determined by high
field measurements of single-magnon spectra,\cite{Coldea2002PRL} which leave little room for
doubt of their correctness with relatively small error bars.  Moreover,
the same model, used at zero and intermediate fields, is quite
successful in reproducing the full inelastic neutron spectrum,
containing both continuum and magnon/triplon (sharp) contributions.\cite{kohno07:_spinon_and_tripl_in_spatial,PhysRevLett.103.197203}
Nevertheless, in some field orientations, entirely different low temperature
phases are observed in experiment than are predicted by the standard
model. Thus, we must somehow explain major {\sl qualitative}
differences in the ground states of \ccc\ in a field by very small
corrections to $H$, of no more than a few percent!

A key message of this paper is that, indeed, the frustration and
quasi-one-dimensionality of this problem can and do amplify tiny terms
in the Hamiltonian to the point where they actually control the ground
state.  Sensitivity to small perturbations is of course an often-cited
characteristic of frustrated systems.  However, the extent to which this
sensitivity can be fully characterized in the problem under
consideration here is, to our mind, unprecedented.  Using the methods of
bosonization, the renormalization group, and chain mean-field theory
(CMFT), we are able to distinctly identify the hierarchy of emergent low
energy scales that control the very complex ordering behavior of the
anisotropic triangular antiferromagnet, in a magnetic field and with a
variety of very weak symmetry-breaking terms.

\begin{figure}[htb]
\begin{center}
\vspace{.2cm}
\includegraphics[width=\columnwidth]{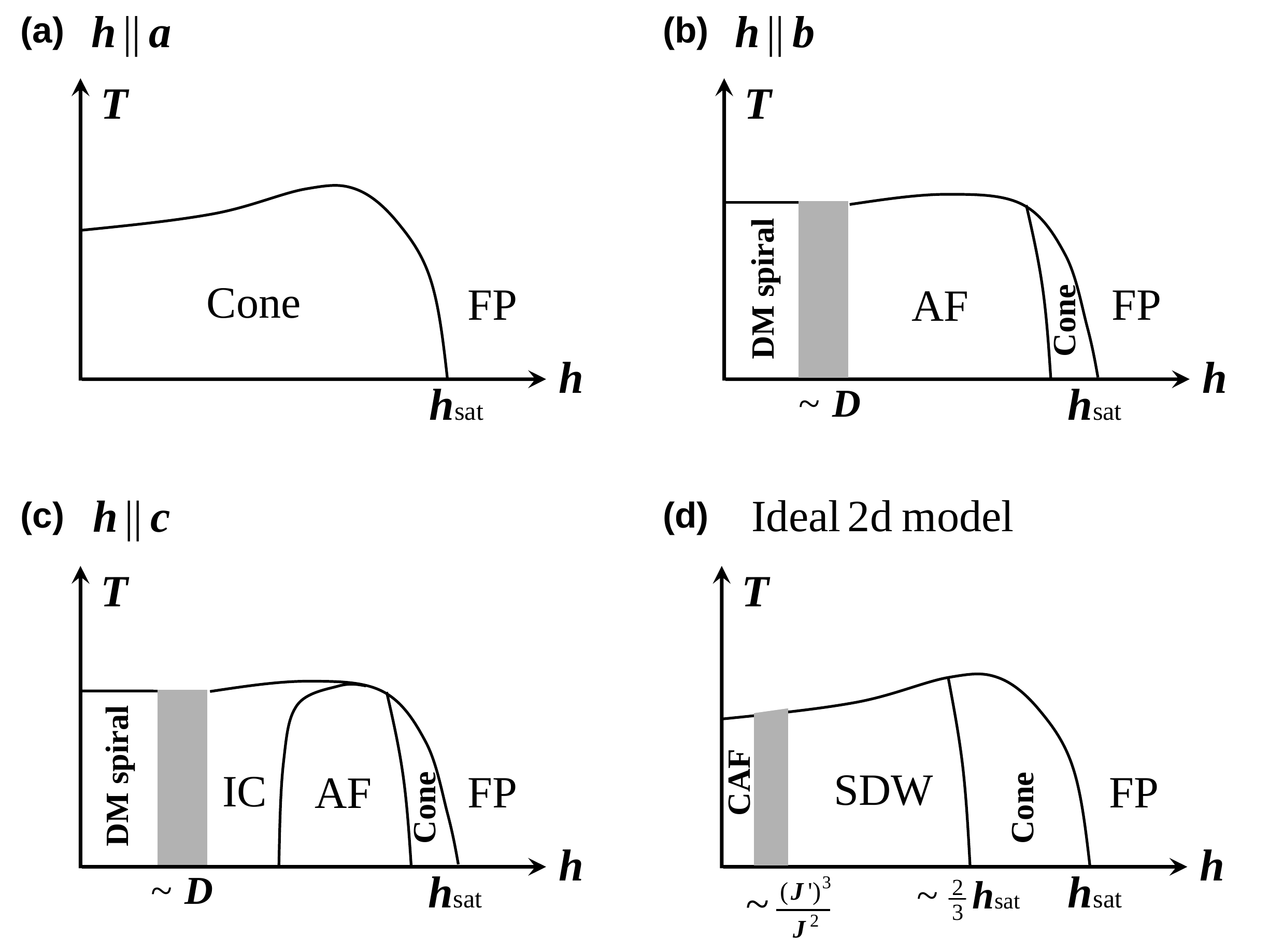}
\caption{Schematic phase diagrams in the temperature-magnetic field plane 
for fields along (a) the $a$ axis, (b) the $b$ axis, and (c) the $c$ axis. 
(d) Schematic phase diagram for the ideal 2d ($J$-$J'$) model.  Here we
use the abbreviations: FP = fully polarized state; AF = commensurate
antiferromagnetic state; IC = incommensurate state; 
CAF = collinear antiferromagnetic state; 
and SDW = spin density wave state.  The shaded areas denote the regions
in which Dzyaloshinskii-Moriya and exchange non-trivially compete.  For
these regions, we do not have reliable theoretical predictions at this
point.}
\label{fig: magnetic_phases}
\end{center}
\end{figure}

This paper contains many results, and a thorough presentation of the
methods required to obtain them.  To briefly summarize, we have
determined the ground state phase diagrams for the ideal two-dimensional
anisotropic triangular antiferromagnetic Heisenberg model, and for the
model appropriate to \ccc, in all three distinct field
orientations, over most of the range of applied magnetic fields.  In the
former, we find spin density wave (SDW) and cone states, and in the SDW,
a family of quantized magnetization plateaux.  In the latter, we
find several phases, including an incommensurate cone state, an
commensurate coplanar antiferromagnetic state, and a second
incommensurate phase, descended from the antiferromagnetic one.  The
occurence of these phases depends crucially on the field orientation,
and matches well with experiments on \ccc.  The associated phase
diagrams in the temperature-magnetic field plane are shown schematically
in Fig.~\ref{fig: magnetic_phases}.  Details of each phase and its properties
can be found in the appropriate section of the main text.

\begin{figure}[htb]
  \centering
  \includegraphics[width=0.95\columnwidth]{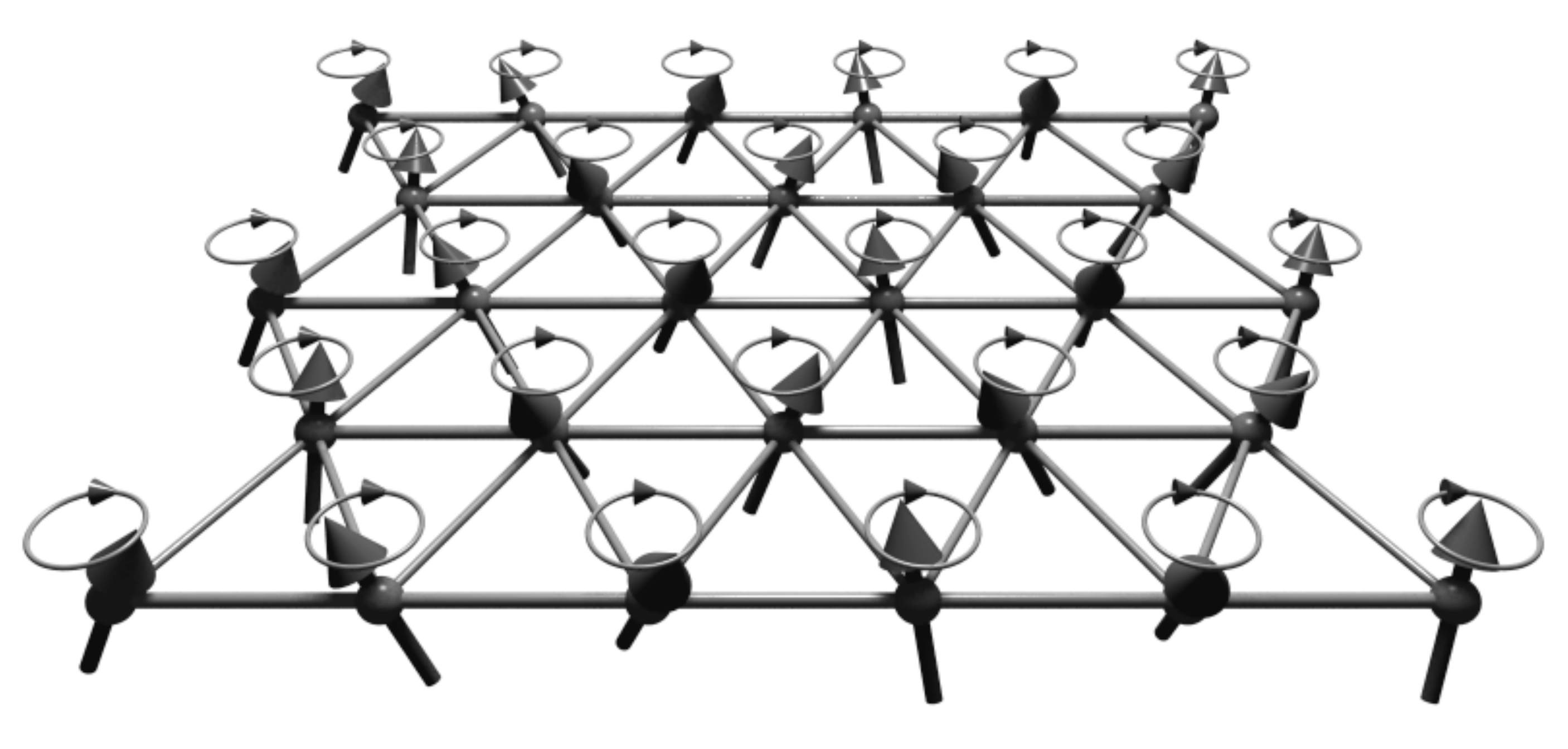}
  \caption{A single triangular layer of the cone state, illustrated for a field
    along the $a$ axis.  Circles with arrows indicate the sense of
    precession of the spins, as one moves along the $x$ axis.  This is most easily seen intuitively by
    comparing every other spin, which compensates for the natural
    staggering due to the underlying N\'eel correlations of the 1d
    chains.  Note that in the cone state, all spins precess in the same
    sense on all chains within a plane.  For fields along $a$, however,
    this sense alternates between successive vertical layers, owing to
    the staggering of $D$.  Within a single layer, cone states for
    fields along other axes are identical to this one after a global
    spin rotation.}
  \label{fig:cone-state}
\end{figure}

\begin{figure}[htb]
  \centering
  \includegraphics[width=0.95\columnwidth]{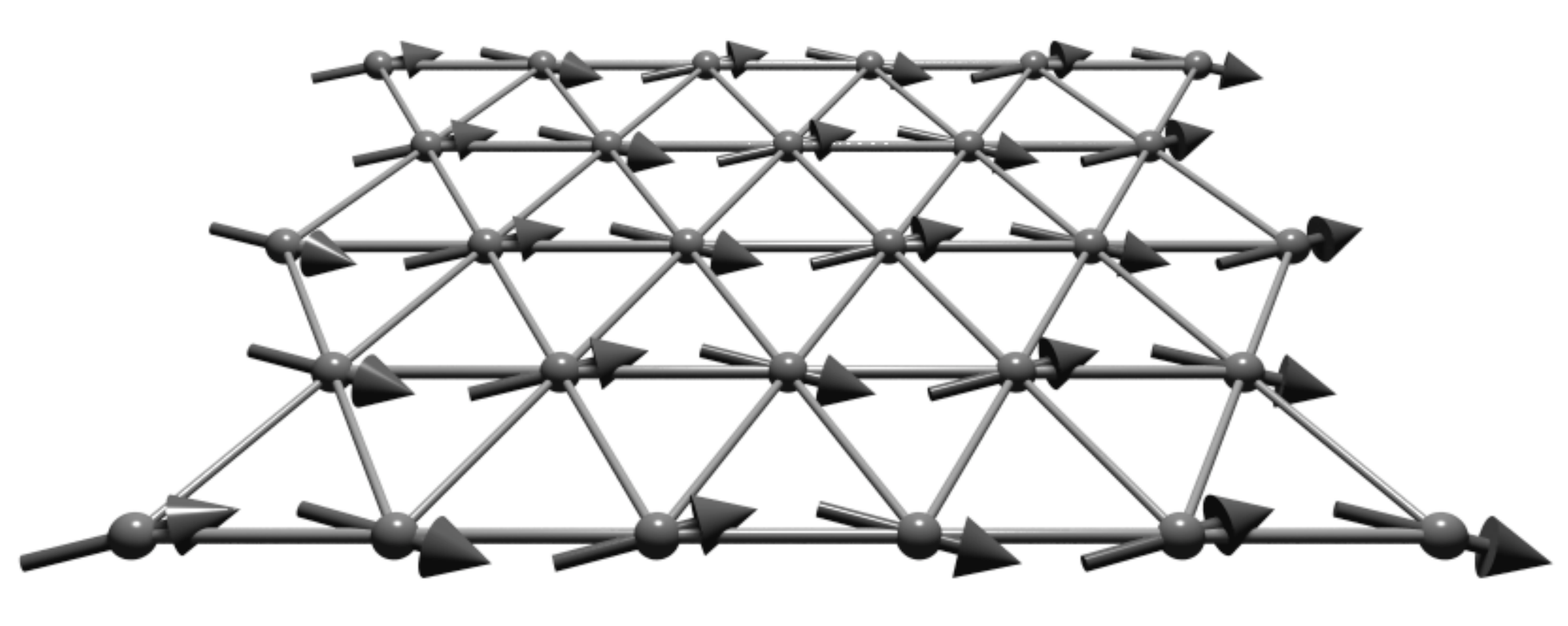}
  \caption{A single layer of the AF state, illustrated for a field along
    the $b$ axis.  The spins lie in a plane spanned by the $b$ axis and
    a second axis within the $a-c$ plane but at a non-zero angle to both
    the $a$ and $c$ axes.  The component of the spin normal to $b$ is
    antiparallel on successive even (or odd) chains, so that the patten
    has period $\Delta y=4$ along the $c$ axis.  This corresponds to a
    doubled cystallographic unit cell in this direction.}
  \label{fig:af-state}
\end{figure}

\begin{figure}[htb]
  \centering
  \includegraphics[width=0.95\columnwidth]{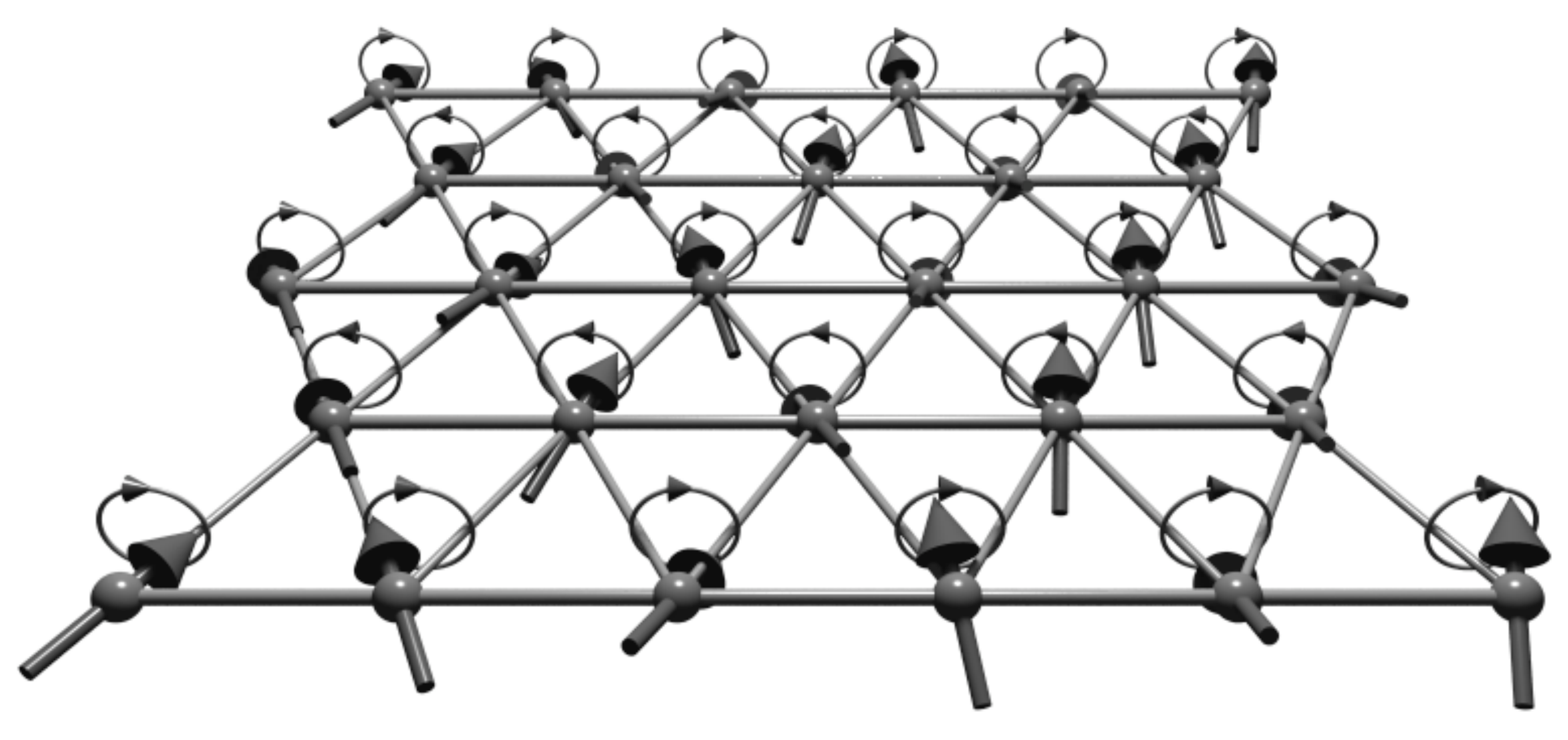}
  \caption{A single layer of the IC state, illustrated for a field along
    the $c$ axis.  Circles with arrows indicate the sense of precession
    of the spins, as one moves along the $x$ axis.  This is most easily
    seen intuitively by comparing every other spin, which compensates for
    the natural staggering due to the underlying N\'eel correlations of
    the 1d chains.  Note that in the IC state, by contrast to the cone
    state, spins on alternate chains in the plane precess in the
    opposite sense. }
  \label{fig:ic-state}
\end{figure}
One noteworthy highlight is that, remarkably,
when the magnetic field is in the $b$-$c$ plane, the spin correlations
impugn the popular interpretation of \ccc\ as a two-dimensional
``anisotropic triangular lattice'' antiferromagnet.  In fact, in this
very wide regime, in the ground state, the spins are more correlated in
the $a$-$b$ planes, {\sl perpendicular} to the triangular layers, than
they are within those layers!  Taking into account these correlations is
crucial to obtaining the proper low temperature phase diagram.  They
lead to an enhanced sensitivity to some very weak second neighbor and
effective ``biquadratic'' interactions, which are needed to stabilize
the antiferromagnetic and incommensurate states mentioned above.

The remainder of the paper is organized as follows.  In
Sec.~\ref{sec:background}, we present some necessary background,
including the standard model Hamiltonian for \ccc, the low energy
properties of a single Heisenberg chain, the results of a space group
analysis of allowed DM interactions in \ccc, and a summary of the
general scaling and Chain Mean Field Theory (CMFT) approach to studying
competing interactions.  In Sec.~\ref{sec:ideal-2d-model}, we determine
the ground state behavior of the ``ideal'' model of an isolated
spatially anisotropic triangular Heisenberg antiferromagnet, often
presumed to apply to \ccc.  It clearly disagrees with experimental
results for all three field orientations, as already pointed out in
Ref. \onlinecite{Starykh2007}.  In Sec.~\ref{sec:field-along-axis}, we
successfully apply the standard model to the case with a magnetic field
applied perpendicular to the triangular plane, the DM interaction $D$
playing a crucial role in reconciling the behavior with experiment.
Next, in Sec.~\ref{sec:field-along-b}, we study the case of magnetic
field along the $b$ (chain) axis.  The $D$ term becomes rapidly
negligible in this orientation, and we argue instead that the
inter-plane coupling $J''$ becomes dominant.  A subtle hierarchy of
energy scales (see Fig.~\ref{fig:energyscales}) is exposed, which leads
to the establishment of competing antiferromagnetic and cone phases in
this case.  Then, in Sec.~\ref{sec:field-along-c}, we consider the field
along the final principal axis, $c$, where an additional
symmetry-allowed DM coupling plays a key role.  It leads to the new
incommensurate phase and an interesting commensurate-incommensurate
phase transition.  Having established all the ground state phases, we
discuss some experimental consequences in Sec.~\ref{sec:exper-cons}.  We
give the explicit spin structures, describe the NMR lineshapes, which
provide a telling confirmation of the theoretical results, and establish
the nature of the $T>0$ phase diagram.  We conclude with some brief
discussion in Sec.~\ref{sec:conclusions}.  Several appendices present
details of calculations underlying some of the results in the main text.

\section{Background}
\label{sec:background}

\subsection{Explicit hamiltonian and coordinates}
\label{sec:expl-hamilt-coord}

For the bulk of this paper, we will adopt a simple orthogonal coordinate
system, with $x$ along the crystallographic $b$ direction, parallel to
the chains, $y$ along the $c$ axis, perpendicular to the chains within
the triangular plane, and $z$ along the $a$ axis, perpendicular to the
triangular planes.  Moreover, we will adopt a simplified geometry, which
respects the topology and interactions between spins, but does not
precisely reproduce the actual locations of Cu$^{2+}$ ions.  A
description of the actual ion locations is given in Appendix \ref{sec:dmv}.  In our
simplified geometry, the spins form a set of regular triangular
lattices stacked uniformly along the $z$ axis, with spacing $1$ between
spins in the $x$ and $z$ directions, and spacing $\Delta y=1$ between chains in
the triangular plane.

In this representation, the lattice Hamiltonian for the standard model
is
\begin{equation}
  \label{eq:117}
  H_{\rm sm} = \sum_{xyz}\Big\{ J {\bm S}_{x,y,z}\cdot {\bm S}_{x+1,y,z}
  - {\bm h}\cdot {\bm S}_{x,y,z} \Big\} + H_1 + H_2,
\end{equation}
with
\begin{eqnarray}
  \label{eq:1}
  H_1 & = & \sum_{xyz}
J' {\bm S}_{x,y,z}\cdot\left({\bm S}_{x-\frac{1}{2},y+1,z}+{\bm S}_{x+\frac{1}{2},y+1,z}\right) 
\label{standard_hamiltonian}
\end{eqnarray}
and
\begin{eqnarray}
  \label{eq:118}
  H_2 & = &  \sum_{xyz}\Big\{ J'' {\bm S}_{x,y,z}\cdot {\bm S}_{x,y,z+1} \\
& & + D (-1)^z {\hat z} \cdot {\bm S}_{x,y,z}\times \big(
  {\bm S}_{x-\frac{1}{2},y+1,z}-{\bm S}_{x+\frac{1}{2},y+1,z}\big) \Big\}.
\nonumber 
\end{eqnarray}

\subsection{Review of low energy properties of Heisenberg chains}
\label{sec:review-low-energy}

We give a brief synopsis of known results on the low energy
theory of the single Heisenberg chain in a field,
\begin{eqnarray}
  \label{eq:singlechain}
  H_{1d} & = & J \sum_x {\bm S}(x)\cdot {\bm S}(x+1) - h \sum_x S^z(x).
\end{eqnarray}
Here $x$ are taken as integers, and we have taken the $z$ axis in spin
space along the field.  In Eq.~(\ref{eq:singlechain}) the
magnetization 
$M = \sum_x \frac{1}{N}S^z(x) = \langle S^z(x) \rangle$
is conserved ($N$ is the number of spins), and so it is convenient to
work at fixed $M$.  The equilibrium relation between magnetization and
field $M(h)$ is known from the Bethe ansatz solution, see
Ref.~\onlinecite{PhysRev.133.A768}. 

For any $M$ less than full saturation, $|M|<1/2$, the low energy
theory can be described in Abelian bosonization by a single massless
free scalar field $\theta$ and its ``dual'' $\phi$ (related to the
canonical momentum conjugate to $\theta$)
\begin{equation}
  \label{eq:duals}
[\theta(x), \phi(x')] = - i \Theta(x-x'),  
\end{equation}
with $\Theta(x)$ the Heaviside step function.  The low energy
Hamiltonian is then
\begin{equation}
  \label{eq:H-bos-0}
  H_0 = \int dx ~\frac{v}{2} \Big((\partial_x \phi)^2 + (\partial_x
  \theta)^2\Big)   ,
\end{equation}
where the velocity $v$ depends on $M$, see Fig.~9 of Ref.~\onlinecite{affleck1999field}.  At a
given $M$, the fluctuations of the ``longitudinal'' spin component
along the field axis are gapless at wavevectors $k_x=0,\pi\pm
2\delta$, with $\delta=\pi M$.   Similarly, the
``transverse'' spin fluctuations perpendicular to the field axis are
gapless at $k_x=\pm 2\delta, \pi$.  The lattice
spin operator is decomposed thereby according to 
\begin{eqnarray}
  \label{eq:1dspindecomp}
  S^z(x) & \sim & M + \mathcal{S}^z_0(x) \\ &&+ e^{i(\pi-2\delta)x}
    \mathcal{S}^z_{\pi-2\delta}(x) 
     +e^{-i(\pi-2\delta)x}
    \mathcal{S}^z_{\pi+2\delta}(x) , \nonumber \\
  S^+(x) & \sim &  e^{-i2\delta x} \mathcal{S}^+_{2\delta}(x)
    + e^{i 2\delta x} \mathcal{S}^+_{-2\delta}(x) + (-1)^x
  \mathcal{S}^+_\pi(x). \nonumber
\end{eqnarray}
Here the scaling operators $\mathcal{S}^z_0, \mathcal{S}^z_{\pi\pm
  2\delta}$ describe longitudinal spin fluctuations, and
$\mathcal{S}^\pm_{\pm 2\delta}, \mathcal{S}^\pm_\pi$ transverse ones.
These should be assumed to vary slowly with $x$ (and time). 
Note that the operators $\mathcal{S}^\mu_k$ ($\mu=z, \pm$) do not mean 
the Fourier components of $S^\mu(x)$. 
They can be expressed in terms of bosonic fields as follows
\begin{eqnarray}
  \label{eq:bosonizespinops}
  \mathcal{S}^z_0(x) & = & \beta^{-1}\partial_x \phi, \\
  \mathcal{S}^z_{\pi-2\delta}(x) & = & -\frac{i}{2} A_1 e^{-2\pi i \phi/\beta},
  \\
  \mathcal{S}^+_{\pm 2\delta}(x) & = & \pm \frac{i}{2} A_2 e^{i\beta \theta}
  e^{\pm i 2\pi \phi/\beta}, \\
  \label{eq:bosonizespinops+3}
  \mathcal{S}^+_\pi(x) & = & A_3 e^{i\beta\theta}.
\end{eqnarray}
The parameter $\beta$ 
is obtained by solving the integral equations,\cite{Bogoliubov-Izergin-Korepin, Qin-Fabrizio, Cabra-Honecker-Pujol}
and the amplitudes $A_1,A_2,A_3$ have been determined
numerically in Ref.~\onlinecite{hikihara2004correlation}.  We note that the above
effective field theory describes the long-distance correlations of the
spin chain, and is a good approximation beyond some cut-off length
$a_0$.  For ``generic'' values of the magnetization, this is of the order
of lattice spacing; however, it diverges near saturation
($|M|\rightarrow 1/2$), where it scales roughly as the distance
between flipped spins (antiparallel to the field), $a_0 \sim
(1/2-|M|)^{-1}$.

The parameter $\beta = 2\pi R$ is related to the ``compactification
radius'' $R$. At zero magnetization $M=h=0$, the $SU(2)$ invariant
Heisenberg chain has $2\pi R^2 = 1$. In the field $\beta$ and $R$
decrease (see Fig. \ref{fig:Beta_curve} of Appendix \ref{sec table}) 
toward the limit $2\pi R^2=1/2$ as
$|M|\rightarrow 1/2$.  The scaling dimensions for general $M$ are
given in terms of $R$, and in the limits of zero and full
polarization, are listed in Table~\ref{tab:dims}.  
\begin{table}
\vspace{0.1in}
\begin{tabular}{||c||c|c|c||} \hline\hline
Operator & $\Delta$ & $M=0$ & $M\rightarrow 1/2$ \\
\hline\hline 
$\mathcal{S}^z_0$ & $1$ & $1$ & $1$ \\
\hline
$\mathcal{S}^z_{\pi \pm 2\delta}$ & $1/4\pi R^2$ & $1/2$ & $1$ \\
\hline
$\mathcal{S}^\pm_{\pm 2\delta}$ & $\pi R^2+1/4\pi R^2$ & $1$ & $5/4$ \\
\hline
$\mathcal{S}^\pm_{\pi}$ & $\pi R^2$ & $1/2$ & $1/4$ \\
\hline \hline
\end{tabular}
\caption{Scaling dimensions of scaling fields associated with spin
  fluctuations in the one dimensional Heisenberg chain at
  magnetization $M$.  The third and fourth columns give the scaling
  dimensions in the limit of zero and full polarization,
  respectively.} \label{tab:dims} 
\end{table}

In addition to the scaling fields above, which appear in the expansion
of spin operators, we will also make use of the spin current, for the
component of spin along the field axis.  This has the form
\begin{eqnarray}
  \label{eq:19}
  J^z(x) & = & \frac{-i}{2} \left[ S^+(x) S^-(x+1) -
    S^-(x)S^+(x+1)\right]\nonumber \\
&&  \sim \mathcal{J}^z(x).
\end{eqnarray}
Like the spin density $\mathcal{S}_0^z$, this has a simple bosonization
formula,
\begin{equation}
  \label{eq:20}
  \mathcal{J}^z = \frac{v}{\beta J} \partial_x \theta.
\end{equation}
We note that at $M=0$, $v/J=\pi/2$ takes a simple value, but at $M>0$
the coefficient decreases continuously.

It may also be useful to connect with the limit of zero field
$M=\delta=0$, which is probably more familiar.  Here the Hamiltonian has
SU(2) symmetry.  In this limit, the three operators ${\rm
  Re}[\mathcal{S}_\pi^+]$, ${\rm Im}[\mathcal{S}_\pi^+]$, and ${\rm
  Re}[\mathcal{S}^z_{\pi-2\delta}]$ become unified into the three
components of the N\'eel field ${\bm N}$ (scaling dimension 1/2), while the other three operators, ${\rm
  Re}[\mathcal{S}_{2\delta}^+]$, ${\rm Im}[\mathcal{S}_{2\delta}^+]$,
and $\mathcal{S}_0^z$, become the uniform magnetization operator ${\bm
  L} = {\bm J}_R + {\bm J}_L$, where ${\bm J}_{R/L}$ are the chiral spin
currents (scaling dimension 1).  The remaining operator, ${\rm
  Im}[\mathcal{S}^z_{\pi-2\delta}]$ becomes the staggered dimerization
field $\varepsilon$ (scaling dimension 1/2).

\subsection{DM-ology}
\label{sec:dm-ology}

Here we present the DM terms which correct the standard model, as
allowed by the space group symmetry of the lattice.  They are derived in
Appendix~\ref{sec:dmv}.  Since \ccc\ is an $S=1/2$ system with a single unpaired
electron in a non-degenerate orbital, we expect that spin-orbit effects
to be perturbative.  In this limit, the leading effect is to generate DM
terms on the same bonds on which exchange interactions are present, and
which are proportional to both the exchange coupling on that bond, and
to the strength of spin orbit interactions.  As a consequence, we need to
consider DM terms only on bonds with reasonably strong exchange,
that is, the intra-layer triangular lattice bonds.   These come in two
types: the on-chain bonds and the diagonals.  We denote the DM vectors
on the former bonds by ${\bm D}$ and on the latter by ${\bm D}'$, in
analogy to $J$ and $J'$ exchange couplings on the same bonds.

\begin{figure}[htb]
\begin{center}
\vspace{.5cm}
\includegraphics[width=0.8\columnwidth]{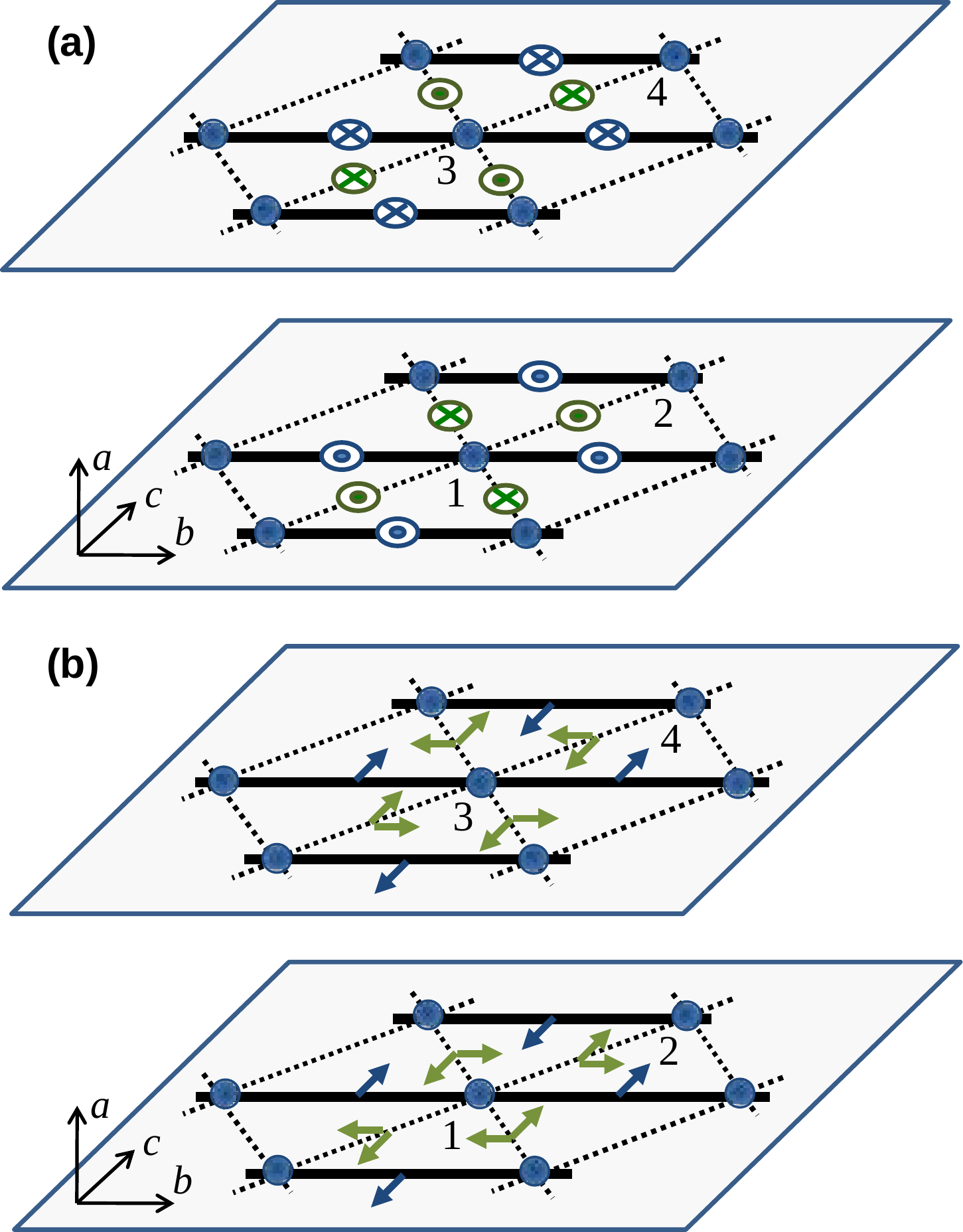}
\caption{Distribution of the DM vectors. Sites 1,2,3, and 4 correspond
  to $(x,y,z)$, $(x+\frac{1}{2},y+1,z)$, $(x,y,z+1)$, and
  $(x+\frac{1}{2},y+1,z+1)$, respectively.  (a) The signs $\otimes$,
  $\odot$ refer to the direction of the DM vectors along $a$ axis.  (b)
  The arrows indicate the direction of the DM vectors along $b$ or $c$
  axis.  We use the convention (see Eq.~\eqref{eq:11}) in which the
  first spin in the cross product ${\bm D}_{ij}\cdot {\bm S}_i \times
  {\bm S}_j$ is the one with smaller $x$ for the on-chain bonds (i.e. it
  is on the left side of the bond), and it is the one with smaller $y$
  on the diagonal bonds (i.e. it is the lower of the two spins).  Note
  that a different convention is used in
  Refs.~\onlinecite{Coldea2002PRL,veillette2005incomm}.}
\label{fig: DM_distribution}
\end{center}
\end{figure}

The space group symmetry of the lattice determines the pattern of
relative signs of the DM vectors on each of these bonds (see Appendix \ref{sec:dmv}).  We find the
following form
\begin{eqnarray}
  \label{eq:11}
  H_D & = & \sum_{xyz} \Big\{ {\bm D}_{y,z}\cdot {\bm S}_{x,y,z}\times
  {\bm S}_{x+1,y,z} \nonumber
\\ && + {\bm D}^+_{y,z}\cdot {\bm S}_{x,y,z}\times
  {\bm S}_{x+1/2,y+1,z} \nonumber \\ && + {\bm D}^-_{y,z}\cdot {\bm S}_{x,y,z}\times
  {\bm S}_{x-1/2,y+1,z}\Big\},
\end{eqnarray}
where
\begin{eqnarray}
  \label{eq:18}
  {\bm D}_{y,z} & = & D_a (-1)^z \hat{a} + D_c (-1)^y \hat{c}, \label{DM_chain}\\
  {\bm D}^\pm_{y,z} & = & \pm D'_a (-1)^z \hat{a} + D'_b (-1)^{y+z}
  \hat{b} \pm D'_c (-1)^{y+z}\hat{c},\nonumber \label{DM_diagonal}\\ &&
\end{eqnarray}
and $D'_a\equiv D$ is the DM term in the standard model. 
The relative signs of the DM vectors are graphically shown in
Fig. \ref{fig: DM_distribution}. 

One sees that symmetry allows five distinct DM couplings: $D=D'_a$,
$D'_b$, $D'_c$, $D_a$ and $D_c$.  Of these, only $D'_a$ and $D_c$ will be
invoked in the body of this paper.  The remaining three can be safely
neglected, as explained in Appendix~\ref{sec:negl-dm-coupl}.

\subsection{Competing Interactions, Scaling, and Chain Mean Field Theory}
\label{sec:renorm-group-appr}

We assume that {\sl all} inter-chain couplings are weak.  In this case,
a scaling analysis based on a perturbative RG treatment is appropriate.
This proceeds in a standard way.  One integrates out short-distance
modes (i.e. small $x$ or large $k_x$), progressively reducing the large
momentum cutoff from its ``bare'' value $\Lambda$ (of order the inverse
lattice spacing, which we take in turn to be $O(1)$) to $\Lambda
e^{-\ell}$, where $\ell \in [0,\infty]$ is the logarithmic RG scaling
variable.  Equivalently, the corresponding real space cutoff is
$L=ae^\ell$, where $a$ is a microscopic length (which we take to be
$O(1)$).  As we integrate out modes, the couplings themselves are
renormalized. For each given coupling constant $\gamma_{i}$, which
appears in the Hamiltonian as $H=H_0+\cdots + \int\!  dx \, \gamma_{i}
\mathcal{O}_i$, we can define a {\sl dimensionless} coupling
$\breve{\gamma}_i = \gamma_i/(v\Lambda_\ell^2)$, which is measured
relative to the typical magnitude of the terms in the free bosonic field
theory.  Equivalently, division by both a factor of $v\Lambda_\ell$ (the
typical energy at this scale) and an additional factor of $\Lambda_\ell$
(a typical inverse length) are needed to render $\gamma_i$, which is an
energy density, dimensionless.  To linear order, each dimensionless
coupling ``flows'' according to the RG equation
\begin{equation}
  \label{eq:RGgeneral}
  \partial_\ell \breve{\gamma}_{i} = (2-\Delta_{i})\breve{\gamma}_{i}.
\end{equation}
Note that the factor of $2$ in this equation, which comes from the
normalization by $1/\Lambda_\ell^2$, is equivalent to the space-time
dimensionality of the (1+1)-dimensional field theory of the spin
chains.  In RG schemes in which space-time is rescaled to keep the
cutoff fixed, this factor arises directly from that rescaling.  We
prefer to formulate the RG without rescaling in this paper, so that all
lengths, times, energies, etc. are explicit.

Many of the above operators may be {\sl relevant} in the RG sense.  This
means that, with increasing $\ell$, the dimensionless coupling constants
{\sl increase}, which implies $\Delta<2$ for that coupling.  The RG is
valid only so long as the {\sl largest} of these dimensionless couplings
remains small.  Crudely, then, we may determine the length scale $\xi$
at which inter-chain coupling becomes significant by the point
$\ell=\ell^*$ {\sl at which the first operator renormalizes to become of
  $O(1)$}, where $\xi=a e^{\ell^*}$.  The length $\xi$ defines a
correlation length, below which the dynamics is approximately
one-dimensional.

If this first ``diverging'' operator is unique, one can often identify
the nature of the associated instability of the decoupled chains.  In
many cases, this can be done by dropping the other operators, and
treating the remaining one by a type of mean field theory.  Sometimes it
can be treated in a semiclassical fashion.  None of these approaches are
rigorous, but they are eminently reasonable, and are 
likely to correctly predict the nature of the resulting state.

$ $From this reasoning, we see that both the bare magnitude and the scaling
dimension (relevance) of the different interactions are important in
determining the low energy state of the system.  To be more concrete,
consider two candidate operators, $\mathcal{O}_1,\mathcal{O}_2$.  Their
renormalized coupling constants obey
\begin{equation}
  \label{eq:32}
  \breve{\gamma}_i(\xi) = \frac{\gamma_i}{v} \xi^{2-\Delta_i}.
\end{equation}
Here, since $\Lambda=\Lambda_{\ell=0}$ is $O(1)$, we replace
$\breve{\gamma}_i(\ell=0) = \gamma_i(\ell=0)/v=\gamma_i/v$.  Setting
$\breve{\gamma}_i(\xi_i)= C$, an $O(1)$ constant, we obtain
\begin{equation}
  \label{eq:33}
  \xi_i = \left(\frac{C v}{\gamma_i}\right)^{\frac{1}{2-\Delta_i}}.
\end{equation}
We expect that operator 1 (2) is dominant if $\xi_1$ ($\xi_2$) is the
shorter length.  Hence the boundary between the two regimes, in which
one or the other operator dominates, occurs when $\xi_1=\xi_2$, or
\begin{equation}
  \label{eq:34}
  \gamma_1^{2-\Delta_2}= (C v)^{\Delta_1-\Delta_2} \gamma_2^{2-\Delta_1}.
\end{equation}
Note that, although the scaling of this boundary is determined by this
argument, the precise location is not, due to the ambiguity of $C$.  We
can understand this conclusion also from the scaling of the ground state
energy density, $\mathcal{E}_0$, which obeys
\begin{equation}
  \label{eq:35}
  \mathcal{E}_0(\gamma_1,\gamma_2) = b^{-2} \mathcal{E}_0\left(\frac{\gamma_1}{v}
  b^{2-\Delta_1},\frac{\gamma_2}{v} b^{2-\Delta_2}\right).
\end{equation}
Choosing $b = (\frac{\gamma_1}{v})^{-1/(2-\Delta_1)}$, we obtain
\begin{equation}
  \label{eq:36}
    \mathcal{E}_0(\gamma_1,\gamma_2) = \left(\frac{\gamma_1}{v}\right)^{\frac{2}{2-\Delta_1}} f( \gamma_2/ (\gamma_1/v)^{(2-\Delta_2)/(2-\Delta_1)}),
\end{equation}
where $f({\calg X})$ is a universal scaling function.  If there is a phase
transition as $\gamma_1$ and $\gamma_2$ are varied, there must therefore be a
non-analyticity at ${\calg X}={\calg X}^*$ in $f({\calg X})$, for some ${\calg X}^*$.  This gives an
equivalent condition to Eq.~(\ref{eq:34}).  To precisely determine the
phase boundary, we need to know ${\calg X}^*$ (or $C$), which means we need
knowledge of $f({\calg X})$.  Such scaling functions are generally determined by
the full RG flows out of the scale-invariant theory, and not just
perturbative data.

To resolve this ambiguity, we turn to an approach which is equivalent to
the former one at the {\sl scaling} level, but which is more
quantitative.  A natural choice is the Chain Mean Field Theory (CMFT),
in which inter-chain couplings are treated by a self-consistent
Weiss-type decoupling, using the exact solutions of perturbed but
decoupled individual chain problems \cite{schulz-cmft}. 
Our CMFT approach is described in detail in Appendix~\ref{sec:cmft}. 
In principle, it can be employed to determine a
full mean-field phase diagram.  Here, we will mostly use it in more
limited ways, as convenient.  To address the ambiguity discussed above,
we use the CMFT to compute a putative ordering temperature, $T_i$, for
each channel driven by an operator $\mathcal{O}_i$.  The instability
which sets in first upon lowering the temperature, i.e. with maximal
$T_i$, is assumed to be dominant.  Another application of CMFT will be
to compute the magnitude of the ordering induced by a coupling $\gamma_i$, at
zero temperature.  This will be useful in making quantitative estimates
of more subtle smaller energy scales, as we will see below.

\section{Ideal 2d model}
\label{sec:ideal-2d-model}

In this section, we consider the behavior of the ideal 2d model,
described by the standard model in Eqs.~(\ref{eq:Hexpt},\ref{eq:1}) with
$J''=D=0$.  

\subsection{Continuum limit}
\label{sec:continuum-limit}

For this case, the technology of the previous section can be
straightforwardly applied.  We begin with the na\"ive procedure of
simply inserting the decompositions in Eq.~(\ref{eq:1dspindecomp}) into
the microscopic inter-chain lattice Hamiltonian in Eq.~(\ref{eq:1}) (We
will assess the need to go beyond this approximation later).
Specifically, we have
\begin{eqnarray}
  \label{eq:2}
   S_{x,y,z}^z & \sim &  M + \mathcal{S}^z_{y,z;0}(x) \\
&& + e^{i(\pi-2\delta)x} \mathcal{S}^z_{y,z;\pi-2\delta}(x) + e^{-i(\pi-2\delta)x} \mathcal{S}^z_{y,z;\pi+2\delta}(x) , \nonumber \\
S_{x,y,z}^+& \sim &  e^{-i2\delta x} \mathcal{S}^+_{y,z;2\delta}(x)\nonumber \\
&& 
    + e^{i 2\delta x} \mathcal{S}^+_{y,z;-2\delta}(x) + e^{i\pi x}
  \mathcal{S}^+_{y,z;\pi}(x). \nonumber
\end{eqnarray}
It might appear that a slightly different formula should be applied for
odd chains, since in those cases in our convention the $x$ coordinates
are half-integer rather than integer.  However, the differences can be
removed by constant shifts of $\theta$ and $\phi$ for the odd chains,
without any further changes.  Hence we can uniformly apply
Eq.~(\ref{eq:2}) to all chains. 

Having dropped the $D$ term, the behavior of the model at zero
temperature is independent of the direction of the field, and is a
function only of the magnetization $M$ and the magnitude of $J'$.
Inserting the decomposition of Eq.~(\ref{eq:1dspindecomp}) into
$H_1$ in Eq.~(\ref{eq:1}), one need keep only terms which do not
oscillate, the condition corresponding to momentum conservation.  Using
the slowly-varying nature of the scaling operators in $x$ (but {\sl not}
in $y$ and/or $z$), one may take a continuum limit in $x$ by gradient expansion to
obtain the lowest non-vanishing terms of each type.  One finds:
\begin{widetext}
\begin{eqnarray}
  \label{eq:H1dec}
  H_1 & \approx & J' \sum_{y,z} \int\! dx\, \Big\{ 2 M^2 + 2
  \mathcal{S}_{y,z;0}^z\mathcal{S}_{y+1,z;0}^z +
  2\sin \delta\left[ \mathcal{S}_{y,z;\pi-2\delta}^z 
  \mathcal{S}_{y+1,z;\pi+2\delta}^z+ {\rm h.c.}\right] \nonumber \\
  & & +\frac{1}{2} \left[-i\mathcal{S}_{y,z;\pi}^+ \partial_x
  \mathcal{S}_{y+1,z;\pi}^- + {\rm h.c.}\right] + \cos\delta \left[
   \mathcal{S}_{y,z;2\delta}^+
    \mathcal{S}_{y+1,z;2\delta}^- + \mathcal{S}_{y,z;-2\delta}^+
    \mathcal{S}_{y+1,z;-2\delta}^- + {\rm h.c.}\right] \Big\}
\end{eqnarray}

Let us now assess the importance of each of the terms in
Eq.~(\ref{eq:H1dec}).  This is accomplished by ranking each of the terms in order of
increasing scaling dimension, or equivalently, decreasing relevance in
the RG sense.  Formally, the most relevant term is the first ($M^2$)
one, which is a $c$-number constant and hence of dimension zero.  Though
it is a ``trivial'' constant (at fixed magnetization) and hence does
not affect the dynamics of the system, it is indeed the dominant
correction to the ground state energy of the weakly coupled chains.
Being positive, it implies an increase of this energy with increasing
$M$, and hence a suppression of the $M(h)$ curve at fixed external
field $h$.  This is calculated in Ref.\onlinecite{Starykh2007}, and
reproduced in Appendix~\ref{sec table}.  The
result agrees very well with experimental data on \ccc.

Consulting Table~\ref{tab:dims}, one sees that of the remaining terms,
the third and fourth terms are presumably most important.  The third
term involves $\mathcal{S}^z_{\pi\pm 2\delta}$ operators, whose
scaling dimensions approach the minimal value of $1/2$ at small
magnetization.  Because this term lacks any derivatives, it achieves
nearly the smallest total scaling dimension ($\approx 2 \times 1/2 =
1$) for small $M$.  The fourth term contains a derivative (which adds
$1$ to its scaling dimension), but contains $\mathcal{S}^\pm_\pi$
operators, whose scaling dimensions decrease from $1/2$ towards $1/4$
near saturation.  Thus the total scaling dimension of the fourth term
decreases from $2\times 1/2 + 1=2$ at small $M$ towards $2\times
1/4+1=1.5$ near saturation.  This makes it less relevant than the
third term at small $M$, but more relevant than it near saturation.
The remaining (second and fifth) terms have larger scaling dimensions
for all values of the magnetization.  

We therefore drop these less relevant terms, as well as the constant
contribution to the energy to obtain $H_1 \rightarrow H'_1$, with
\begin{eqnarray}
  \label{eq:H1p}
  H'_1 & = & \sum_{y,z} \int\! dx\, \Big\{ 
  \gamma_{\rm sdw} \mathcal{S}_{y,z;\pi-2\delta}^z 
  \mathcal{S}_{y+1,z;\pi+2\delta}^z 
  -i \gamma_{\rm cone}\mathcal{S}_{y,z;\pi}^+ \partial_x
  \mathcal{S}_{y+1,z;\pi}^- + {\rm h.c.} \Big\},
\end{eqnarray}
\end{widetext}
with $\gamma_{\rm sdw} = 2 J'\sin\delta$, $\gamma_{\rm cone} = J'/2$.
Equivalently, using the bosonization formulae in
Eq.~(\ref{eq:bosonizespinops}), one can rewrite $H'_1$ in sine-Gordon
form,
\begin{eqnarray}
  \label{eq:Hsg1}
  H'_1 & = &  \sum_{y,z} \int\! dx\, \Big\{ {\tilde \gamma}_{\rm sdw} \cos [2\pi
  (\phi_{y,z}-\phi_{y+1,z})/\beta ] \\ &&  - {\tilde \gamma}_{\rm cone}
  (\partial_x \theta_{y,z}+ \partial_x \theta_{y+1,z}) 
  \cos [\beta(\theta_{y,z} - \theta_{y+1,z})] \Big\},\nonumber
\end{eqnarray}
with ${\tilde \gamma}_{\rm sdw} = J' A_1^2 \sin \delta$ and ${\tilde \gamma}_{\rm cone} = J' A_3^2
\beta/2$.

\subsection{Phases of \eqref{eq:Hsg1}}
\label{sec:phases}

The names of these coupling constants have been chosen to reflect
their probable consequences.  For small magnetization, where
$\gamma_{\rm sdw}$ is most strongly relevant, one expects collinear ``spin
density wave'' (SDW) ordering of spins along the $z$ (field) axis, with
\begin{equation}
  \label{eq:44}
  \langle \mathcal{S}^z_{y,z;\pi\pm 2\delta}\rangle = |\psi| (-1)^y e^{i\alpha_z},  
\end{equation}
which minimizes the SDW interaction term.  Here $\alpha_z \in [0,2\pi]$
can be arbitrary for each $z$, since the layers are decoupled. Near saturation, where
$\gamma_{\rm cone}$ becomes more relevant, one expects a ``spiraling'' order
of the components of the spins transverse to the field, 
\begin{equation}
  \label{eq:45}
  \langle  \mathcal{S}^\pm_{y,z;\pi}(x) \rangle =   |\psi| (\sigma_z)^y
   e^{i \sigma_z q_0 x}e^{i\Theta_z},
\end{equation}
with some $q_0>0$, and where $\sigma_z=\pm 1$ and $\Theta_z$ are independent for each $z$.
Some
incommensurate pitch $q_0$ is preferred by the derivative in
Eq.~(\ref{eq:H1p}), but is expected to be small as it is disfavored by
the single chain Hamiltonian.  Because of the non-zero magnetization
along the $z$ axis, the spins in this phase sweep out a ``cone'' as one
moves along the $b$ (chain) axis in real space.

\begin{figure}[h]
  \centering
  \includegraphics[width=3.4in]{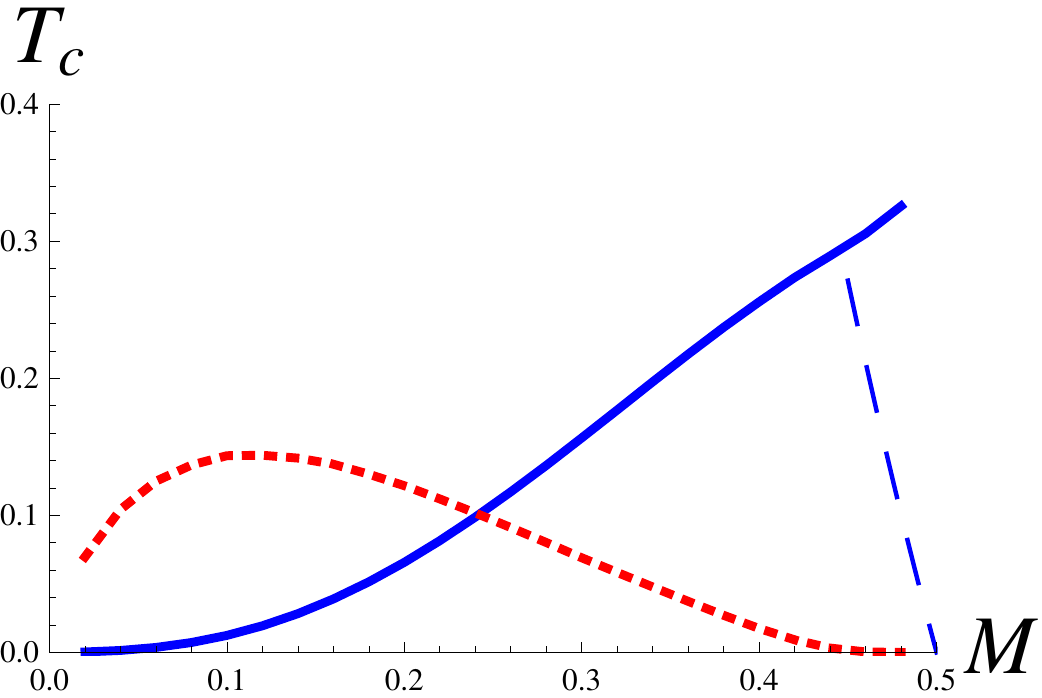}
  \caption{(Color online) Putative ordering temperatures within chain mean field theory for SDW
    interaction, ($T_{\rm sdw}$, dotted
     (red) line) and  cone ($T_{\rm cone}$, solid (blue) line)
    interactions, within the ideal two-dimensional model. 
Dashed blue line, emanating from $M=1/2$ point, represents $T_{\rm cone} \propto (1/2-M)$ (Eq.~\eqref{eq:38}) which describes the cross-over between 1d behavior obtained by bosonization with CMFT, and 2d one, 
which takes over for $M \geq 0.45$. The larger
    ordering temperature is expected to be physical, while the putative
    lower temperature transition will be suppressed by the dominant competing order.} 
  \label{fig:RG1}
\end{figure}

Following the logic in Sec.~\ref{sec:renorm-group-appr}, the dominant
interaction, at each magnetization, is the one whose putative ordering
temperature is largest.  The estimated ordering temperatures for the SDW
and cone states, calculated from CMFT, are shown in Fig.~\ref{fig:RG1}.
One can see that the two curves cross at $M\approx 0.24$, which
therefore separates a region of SDW state at lower magnetization from a
cone state at higher magnetization.

This change from an SDW to a cone state is primarily due to the
variation of scaling dimensions with magnetization.  As $M$ increases,
the spin correlations are increasingly XY-like, i.e. enhanced transverse
to the field and suppressed parallel to it.  While the SDW term is
obviously more relevant than the cone term near $M=0$, due to the
derivative in the latter, the change in scaling dimensions eventually
compensates.  Equating the two scaling dimensions, one finds $R=R_c$,
with $2\pi R_c^2 = (\sqrt{5}-1)/2\approx 0.62$ (the golden ratio!).
This occurs at magnetization $M_c \approx 0.32$ (i.e about $65$ percent
of the saturation value).  This approximately recovers the more accurate
estimate $M_c \approx 0.24$, obtained above.  Similar estimate holds for
the SDW-to-cone phase transition in a single zig-zag ladder.\cite{kolezhuk2005}

\subsection{Critical temperature}
\label{sec:critical-temperature}

Further details of the behavior of the critical temperatures in
Fig.~\ref{fig:RG1} can be understood physically.  The SDW critical
temperature increases from small values at small $M$, to a maximum
around $M=0.1$, above which it again decreases.  The initial rise is due
to the partial relieve of frustration of the inter-chain $J'$ coupling by
increasing incommensurability of the longitudinal spin correlations.
The ultimate decay of $T_{\rm sdw}$ is due simply to the suppression of
longitudinal spin correlations as the chain becomes more XY-like.  The
same increasing XY-like behavior leads to the growth of $T_{\rm cone}$
with $M$.   

The two endpoints, $M\rightarrow 0$ and $M\rightarrow 1/2$, require 
special consideration.  Approaching zero field,  the
dominant SDW interaction vanishes.  This case requires a subtle
analysis of fluctuation-generated interactions, which was undertaken in
Ref.~\onlinecite{Starykh2007}.  There it was observed that corrections to the
na\"ive continuum limit are crucial to obtain the correct behavior,
which is {\sl neither} an SDW state nor a spiral, but rather a
commensurate, collinear, antiferromagnet (CAF). This CAF state
replaces SDW as the ground state near the $M=0$ limit, see Fig.~\ref{fig: magnetic_phases}d, which is the
reason for $T_{\rm sdw/cone}$ curves in Figure~\ref{fig:RG1} (as well
as in most other $T_c$ vs. $M$ figures in the paper) start not at $M=0$
but at a finite $M=0.02$ value. We will not go into
further detail on this point here, but simply mention that another 
instance of fluctuation-generated couplings will be encountered later in
Secs.~\ref{sec:field-along-b}-\ref{sec:field-along-c}.

The other limit, $M\rightarrow 1/2$, can be attacked differently.  At
$M=1/2$, one has full spin saturation, and the state is unique and
trivial.  Single spin-flip magnon excitations can be found exactly
including the effects of $J'$. One may obtain in this limit a cone
state as a magnon condensate, as in Refs. \onlinecite{nikuni-shiba,veillette2006commensurate}. In this
formulation, it is clear that the critical temperature for the ordering
must vanish as $M\rightarrow 1/2$.  However, this is not observed in
Fig.~\ref{fig:RG1}.  This is due to a non-commuting order of limits.  In
the vicinity of saturation, scaling (see Sec.~\ref{sec:cmft-limits}) in
fact predicts that {\sl all} physical quantities are functions of the
combination
\begin{equation}
  \label{eq:37}
  \Xi = \frac{J'/J}{\tfrac{1}{2}-M},
\end{equation}
when $\tfrac{1}{2}-M \ll 1$.  The bosonization analysis carried out
above is valid for $\Xi\ll 1$.  However, for $\Xi\gg 1$ a different
behavior obtains.  Specifically, the critical temperature is expected to
scale (up to logarithmic corrections) according to
\begin{equation}
  \label{eq:38}
  T_{\rm cone} =  (\tfrac{1}{2}-M)^2\mathcal{F}\left[\frac{J'/J}{\tfrac{1}{2}-M}\right]
\end{equation}
where $\mathcal{F}[\Xi]\sim \Xi^2$ for $\Xi\ll1$ and $\mathcal{F}[\Xi]
\sim \Xi$ for $\Xi \gg 1$.  One can see that this form indeed vanishes
on approaching saturation.  The maximum of $T_{\rm cone}$ should be
obtained by differentiating, occurs when $\mathcal{F}(\Xi)= \Xi
\mathcal{F}'(\Xi)$, which implies $\Xi$ of $O(1)$.  Hence the maximum
$T_{\rm cone}$ occurs very close to saturation, where $\tfrac{1}{2}-M
\sim J'/J$, and its presence is not captured in the bosonization result
plotted in Fig.~\ref{fig:RG1}.  Hence the temperature $T_{\rm cone}$ is
overestimated, leading to an underestimate of the magnetization of the
crossing point from Fig.~\ref{fig:RG1}.  This effect is probably small,
however, since this occurs relatively far from saturation.

\subsection{SDW and magnetization plateaus}
\label{sec:plateau}

The above considerations treat only the most relevant terms in the
effective Hamiltonian.  This, however, neglects some important physics
in the SDW phase.  In particular, it misses {\em commensurability}
effects, when the SDW period can become ``locked'' (i.e. fixed over a
region of field and temperature) to a multiple of the lattice constant.

Microscopically, this effect arises from ``umklapp" processes, which
distinguish quasimomentum from true momentum, violating absolute
conservation of the former.  The $x$ component of the quasimomentum is
important here, and hence umklapp events carry momentum $2\pi$.  Since
the SDW carries momentum $\pi \pm 2\delta$, a umklapp event occurs 
when a number $k$ of SDW quanta are absorbed or emitted, adding to $\pm
2\pi$.  This condition is rigorously derived below, where we outline symmetry
considerations which fix the form of the allowed microscopic umklapp Hamiltonian
completely.

\subsubsection{Symmetry constraints}
\label{sec:symmetry-constraints}

We start by analyzing how $\phi_y(x)$ transforms under discrete lattice
symmetries.\cite{miles2008}  For that, we re-write
\eqref{eq:1dspindecomp} as \begin{eqnarray}
S_y^z(x) &\sim& M + \beta^{-1}\partial_x \phi_y(x) + \nonumber\\
&& - A_1 \sin[\frac{2\pi \phi_y(x)}{\beta} - (\pi - 2\delta) x] .
\label{eq:umk1}
\end{eqnarray}
It then follows that translation along the chain transforms $\phi_y(x)$ as
\begin{equation}
\phi_y(x) \to \phi_y(x+1) - \frac{\beta}{2\pi} (\pi -2\delta) ,
\label{eq:translation-x}
\end{equation}
while translation along the north-east diagonal ($y\to y+1, x\to x+1/2$)
changes it to
\begin{equation}
\phi_y(x) \to \phi_{y+1}(x+1/2) - \frac{\beta}{4\pi} (\pi -2\delta) .
\label{eq:translation-y}
\end{equation}
Spatial inversion ($x\to -x$) changes it as well:
\begin{equation}
\phi_y(x) \to \frac{\beta}{2} - \phi_y(-x) .
\label{eq:inversion}
\end{equation}
In addition, $\phi_y(x)$ is defined modulo $\beta$ so that
\begin{equation}
\phi_y(x) \to \phi_y(x) + \beta
\label{eq:period}
\end{equation}
must be respected also.

We now specify the general form for the $k$-th order umklapp term:
\begin{equation}
H_{\rm umk}^{(k)} = \sum_y \int dx ~t_k(y) \cos[\frac{2\pi k}{\beta} \phi_y(x) + \omega_k].
\label{eq:umk2}
\end{equation}
where $t_k \sim O(J)$ is the bare amplitude and $\omega_k$ is yet undetermined phase. 
The periodicity requirement, Eq. \eqref{eq:period}, implies that $k$ must be an {\em integer}.
The translation in Eq. \eqref{eq:translation-x} changes the argument of cosine in Eq. \eqref{eq:umk2}
into $2\pi k \phi_y/\beta + \omega_k - k(\pi-2\delta)$ which implies that
\begin{equation}
k (\pi-2\delta) = 2\pi \nu .
\label{eq:umk3}
\end{equation}
Since $\delta = \pi M$, the above equation implies that allowed values of the
magnetization are given by
\begin{equation}
M^{(k,\nu)} = \frac{1}{2} \Big(1 - \frac{2 \nu}{k}\Big),
\end{equation}
where $\nu$ and $k$ are positive integers.  
This condition is equivalent to the 
magnetization quantization condition for a single spin chain.\cite{oshikawa1997} 
However, we will see that
$\nu$ and $k$ are {\sl not} arbitrary in the two-dimensional triangular lattice.

The remaining symmetries, translation along the diagonal of the
triangular lattice, Eq. \eqref{eq:translation-y} , 
and spatial inversion, Eq. \eqref{eq:inversion}, require that $t_k(y) = (-1)^{y \nu} t_k$ and 
$\omega_k = -\pi/2$ for {\em odd} $k$ and $\omega_k = 0$ for {\em even} $k$. 
As a result, the most general form of the umklapp term, consistent with lattice symmetries
and involving single chains and no spatial derivatives, reads
\begin{eqnarray}
\label{eq:umk6}
H_{\rm umk}^{(k={\rm odd})} &=& \sum_y \int dx ~t_k (-1)^{y \nu} \sin[\frac{2\pi k}{\beta} \phi_y(x)], \\
H_{\rm umk}^{(k={\rm even})} &=& \sum_y \int dx ~t_k (-1)^{y \nu} \cos[\frac{2\pi k}{\beta} \phi_y(x)].
\label{eq:umk6b}
\end{eqnarray}

\subsubsection{Energetic constraints}
\label{sec:energ-constr}

To  proceed, we bring out the energetics associated with the underlying
SDW order by making the shift 
\begin{equation}
\phi_y(x) \to \phi_y(x) + (-1)^y \frac{\beta}{4}
\label{eq:umk7}
\end{equation} so as to minimize the leading SDW term in
Eq. \eqref{eq:Hsg1} (see Appendix~\ref{sec:cmft-sdw}).  In terms of the
transformed fields, the 
SDW ground state corresponds to an $x$- and $y$-independent
configuration of $\phi_y(x) = \phi_0$.  This shift modifies
Eq. \eqref{eq:umk6} as 
\begin{equation} 
H_{\rm umk}^{(k={\rm odd})} = \sum_y \int dx ~t_k
(-1)^{y (\nu +1)} \sin[\frac{\pi k}{2}] \cos[\frac{2\pi k}{\beta}
\phi_y(x)] .  
\end{equation} 
We immediately conclude that {\em odd}-$k$ umklapp
processes must have {\em odd} $\nu$ in order to be able to gain some
energy -- otherwise the sum oscillates with $y$ and does not give an
extensive contribution.  Similarly, Eq.\eqref{eq:umk6b} becomes 
\begin{equation} H_{\rm umk}^{(k={\rm
    even})} = \sum_y \int dx ~t_k (-1)^{y \nu} \cos[\frac{\pi k}{2}]
\cos [\frac{2\pi k}{\beta} \phi_y(x)] , 
\end{equation}
implying that now $\nu$ must be {\em even}.

This interesting result allows us to finally represent {\em all} allowed umklapp terms in a
single compact equation
\begin{eqnarray}
H_{\rm umk}^{(k)} &=& \sum_y \int dx ~\tilde{t}_k\cos[\frac{2\pi k}{\beta} \phi_y(x)] ,
\label{eq:umk8}
\end{eqnarray}
where $\tilde{t}_k=t_k  (\cos[\frac{\pi k}{2}] + \sin[\frac{\pi
  k}{2}])$, and Eq.~\eqref{eq:umk8} must be supplemented by the
important constraint 
\begin{equation}
  \label{eq:87}
  \nu = k \qquad \textrm{(mod 2)},
\end{equation}
i.e. $\nu$ {\em must have the same parity as} $k$.

\subsubsection{Allowed plateaux}
\label{sec:allowed-plateaux}

It is natural to consider the plateaux in order of increasing $k$ --
this will be directly related to the robustness of the plateau (see below).
One finds that the first non-trivial possibility
(different from the non-polarized $M=0$ or the fully polarized $M=1/2$ limits), 
corresponds to $\nu=1, k =3$ when $M=1/6 = \frac{1}{3} (\frac{1}{2})$. 
Assuming for concreteness $\tilde{t}_{k=3} > 0$ we find that $\phi_0 =
\beta (2 n + 1)/6$ (with $n=0,1,2$)
minimizes Eq.~\eqref{eq:umk8} (and, by construction, the SDW interaction in Eq.~\eqref{eq:Hsg1}).
Working backwards through the chain of the transformations we find that
Eq.~\eqref{eq:umk1} predicts (recall that $(\pi - 2\delta) = 2\pi/3$ here)
\begin{equation}
\langle S_y^z(x) \rangle_{M=1/6} = M + A_1 (-1)^y \cos[\frac{2\pi x}{3}
- \frac{\pi (2 n +1)}{3}].
\end{equation}
This equation describes the famous \cite{chub1991} {\em up-up-down}
(uud) spin configuration of the $1/3$-magnetization plateau, i.e. with
two-thirds of the sites having a larger spontaneous magnetic moment than
the remaining third. It also correctly predicts relative arrangement of 
down-spins on neighboring chains: the system gains energy by 
coupling every down-spin with a pair of up-spins on adjacent chains.
The resulting pattern has down-spins in the centers of hexagons 
formed by up-spins.  For the other sign, $\tilde{t}_{k=3}<0$, one finds
instead of the uud configuration one in which two-thirds of the sites
have a {\sl smaller} spontaneous moment than the remaining third.  This
corresponds to the ``quantum'' magnetization plateau suggested in
Ref.~\onlinecite{hida2005quantum}, where the magnetic unit cell is
composed of a spin singlet on a pair of sites accompanied by an up-pointing spin.  

Other possible plateaux include $M = 3/10 = \frac{3}{5} (\frac{1}{2})$ ($k=5, \nu=1$)
and $M = 5/14 = \frac{5}{7} (\frac{1}{2})$ ($k=7, \nu=1$).
Importantly, several of the smaller-$k$ plateaux are excluded due to 
the `mismatch' between the parities of $k$ and $\nu$ numbers. These include
$k=4, \nu=1$ which leads to $M = 1/4$ (one-half plateau) and
$k=6, \nu=1$ which would result in $M = 1/3$ (two-thirds plateau).

\subsubsection{Effective two-dimensional sine-Gordon model}
\label{sec:plateaux-width}

We now use the RG to derive an effective two-dimensional sine-Gordon model.
Our starting point is given by the following Hamiltonian:
\begin{eqnarray}
H_{\rm plateau}^{(k)} &=& \sum_y \int dx \{ \frac{v}{2} (\partial_x\phi_y)^2 - 
\tilde{\gamma}_{\rm sdw} \cos[\frac{2\pi}{\beta}(\phi_y - \phi_{y+1})] \nonumber\\
&&  - \frac{v q(h)}{\beta} \partial_x \phi_y(x) + \tilde{t}_k \cos[\frac{2\pi k}{\beta}\phi] \},
\label{eq:umk9}
\end{eqnarray}
which incorporates the shift Eq. \eqref{eq:umk7}.
Observe the appearance of the new, linear in spatial derivative term, 
which is added here \cite{miles2008} to describe variation of the magnetic field $h$
near the optimal plateau value $h^{(k,\nu)}$. The optimal field is defined
by the condition that the magnetization in the absence of the umklapp
term, $M_0(h)$, is given by the plateau's value,
$M_0(h^{(k,\nu)}) = M^{(k,\nu)}$. Then $q(h) = 2\pi k (M_0(h) - M^{(k,\nu)})$.

We then iteratively integrate out high-energy modes, reducing the momentum cutoff 
from initial $\Lambda_0 \sim 1$ to $\Lambda_{\rm sdw}= \Lambda_0 e^{-\ell_{\rm sdw}}$, 
as described in the Appendix~\ref{sec:biquadratic}.
The new, reduced cutoff $\Lambda_{\rm sdw}$ is determined by the
condition that the renormalized SDW coupling, 
$\tilde\gamma_{\rm sdw} (\Lambda_{\rm sdw}/\Lambda_0)^{\Delta_{\rm sdw}}$
becomes comparable to the contribution of the gradient term to the
energy density at the same scale, $v \Lambda_{\rm sdw}^2$.
Here $\Delta_{\rm sdw} = 2/(4\pi R^2)$ is the scaling dimension of the
SDW cosine in Eq.~\eqref{eq:umk9}. This leads to the estimate
$\Lambda_{\rm sdw}/\Lambda_0 \sim (\frac{\tilde\gamma_{\rm sdw}}{v\Lambda_0^2})^{1/(2 - \Delta_{\rm sdw})}$.
At this scale, SDW coupling is of the order $v\Lambda^2 \sim
(\tilde{\gamma}_{\rm sdw}^2/v^{\Delta_{\rm sdw}})^{\frac{1}{2-\Delta_{\rm sdw}}}$ and the SDW term should be minimized. 
Therefore, the argument of cosine is small
and we can approximate 
\begin{equation}
\cos[\frac{2\pi}{\beta}(\phi_y - \phi_{y+1})] \to 1 -
\frac{1}{2}\left( \frac{2\pi}{\beta} \right)^2 (\partial_y \phi(x,y))^2 . 
\end{equation}
The umklapp term has a scaling dimension, 
$\Delta_k = (2\pi k/\beta)^2/(4\pi) = k^2/(4\pi R^2)$, which
grows quadratically with the $k$, and is thereby strongly suppressed by 
high-energy fluctuations:
\begin{eqnarray}
\tilde{t}_k(\ell_{\rm sdw}) &=&\tilde{t}_k(0) \Big(\frac{\Lambda_{\rm
    sdw}}{\Lambda_0}\Big)^{\frac{k^2}{4\pi R^2}} \nonumber\\ 
&&\approx v \Big(\frac{J'}{v}\Big)^{\frac{k^2} {8\pi R^2 -2}} .
\label{eq:plateau2}
\end{eqnarray}
At this stage it is convenient to define a rescaled field, $\varphi = 2\pi k
\phi/\beta + \pi \Theta(\tilde{t}_k)$, which includes a shift to achieve
a definite sign of the umklapp term ($\Theta(x)$ is the Heaviside step
function).  This gives the {\em two-dimensional} sine-Gordon Hamiltonian
of the $k$-th plateau,
\begin{eqnarray}
H_{\rm plateau}^{(k)} &=& \int dx dy \Big\{ \frac{u}{2} (\partial_x\varphi)^2 + \frac{c_y}{2} (\partial_y \varphi)^2 +\nonumber\\
&&  - \frac{v q(h)}{2\pi k} \partial_x \varphi_y(x) - |\tilde{t}_k| \cos[\varphi] \Big\}.
\label{eq:umk10}
\end{eqnarray}
Here $u = v (\beta/2\pi k)^2$, $c_y \sim k^{-2}(\tilde{\gamma}^2_{\rm sdw}/v^{\Delta_{\rm sdw}})^{\frac{1}{2-\Delta_{\rm sdw}}}$.

\subsubsection{Plateaux width}
\label{sec:plateaux-width-1}

At $T=0$, the two-dimensional sine-Gordon Hamiltonian Eq. \eqref{eq:umk10}
can be analyzed classically.\cite{chaikin_book}  We review this
standard analysis as it is important both here and in
Sec.~\ref{sec:comm-incomm-trans}.  The classical sine-Gordon model exhibits two phases:
commensurate, describing the plateau, with $\langle\partial_x\varphi
\rangle =0$; and incommensurate, with $\langle\partial_x \varphi
\rangle\neq 0$. The incommensurate state, which describes the SDW phase
with field-dependent ordering momentum, is achieved for sufficiently
strong $|q|\geq q_c$. The critical value $q_c$ is determined by the
condensation of {\em kinks}, when $E_{\rm kink} =0$. Here the kink
represents the solution of Eq. \eqref{eq:umk10} interpolating between two
degenerate minima of cosine potential: $\varphi(x=-\infty,y) = 0$ and
$\varphi(x=+\infty,y)=2\pi$, for all $y$. One immediately observes that
the linear derivative term in Eq. \eqref{eq:umk10} contributes $-v q(h)/k$
to the kink's energy (per unit length in the $y$ direction). The rest
follows from standard steps,\cite{chaikin_book} which show that energy
of the kink, relative to the energy of the uniform plateau state with,
for example, $\varphi=0$, is given by
\begin{equation}
E_{\rm kink} = 8 \sqrt{u |\tilde{t}_k|} - \frac{v |q|}{k} .
\end{equation}
Thus
\begin{equation}
q_c = \frac{4 \beta}{\pi} \sqrt{\frac{|\tilde{t}_k|}{v}} \sim  \Big(\frac{J'}{v}\Big)^{\frac{k^2} {4(4\pi R^2 -1)}}  .
\label{eq:umk11}
\end{equation}
Since in the relevant range of the magnetic field $d M_0/d h$ is constant,
the plateau width in field units, $\delta h_{(k,\nu)}$ is directly proportional to $q_c$.  

Focusing on the $1/3$-magnetization plateau we can estimate, with the
help of Figure \ref{fig:Beta_curve}, 
that $2\pi R^2 \approx 3/4$ at $M=1/6$.  This leads to $\delta h_{(3,1)} \sim
(J'/v)^{9/2}$.  The next most robust plateau is at $3/5^{\rm th}$ of the
saturation magnetization ($k=5, \nu=1$), for which $\delta h_{(5,1)}
\sim (J'/v)^{25/2}$.  The existence of this plateau is unclear, since
this magnetization is close to the boundary of the SDW state, and
indeed, the calculation in Figure~\ref{fig:RG1} predicts that it falls
outside the stability range of the SDW state.  It is, however, {\sl
  inside} the SDW phase as estimated from the pure scaling dimension
criterion $2\pi R^2> (\sqrt{5}-1)/2$.  Thus this plateau still seems a
reasonable candidate for observation in some anisotropic triangular
materials.  

Other plateaux, such as, for example, the one-half magnetization one
($M=1/4$), are much narrower due to the {\em equal parity requirement},
Eq.~(\ref{eq:87}), which implies $k=8, \nu=2$.  Then $\delta h_{(8,2)}
\sim (J'/v)^{32}$, making it very hard to observe indeed.  These
arguments make it clear that the $1/3$ plateau ($M=1/6$) is drastically
more robust than others, and thereby it is expected to be much more
commonly observed.  We also reiterate that it should persist all the way
down to the decoupled chains limit, $J'=0$.

It is interesting to compare our findings with those in
Refs. \onlinecite{okunishi2003,hikihara2010} which studied magnetization
plateaux in a single zig-zag ladder, made of two spin chains (with
exchange $J$) coupled in a triangular (zig-zag) fashion by exchange
$J'$. This geometry can also be viewed as a single chain with first and
second-neighbor interactions $J_1=J'$ and $J_2=J$. In this
one-dimensional system a robust $1/3$-plateau is found to exist in the
intermediate exchange region $0.487 \leq J/J' \leq 1.25$. In particular,
it does not seem to extend far into the $J' \to 0$ limit, although a very
narrow sleeve of the plateau phase cannot be reliably excluded by the
current numerical studies. We note however that the much reduced extent of the long-range
ordered plateau region, in
comparison with the quasi-2d predictions above, simply reflects the reduced stability of the
crystalline (more specifically SDW in this case) order in 1d systems (at
$T=0$).  In more technical terms, the magnetization plateau requires pinning
of both the `center-of-mass' and relative combinations of $\phi_{1,2}$
fields in a two-chain system which, in turn, requires significant
modification of the chain Luttinger parameter $K$ from its bare value of
$1/2$, in notations of Ref.~\onlinecite{hikihara2010}, by various marginal
(density-density type) inter-chain terms.  Such modifications
generically require $J' \sim O(1)$, which is the reason for the absence
of the plateau in the $J' \to 0$ limit in this case.

\subsubsection{Critical behavior of the wavevector}
\label{sec:crit-behav-wavev}

Our description can be extended to the neighborhood of the plateau-SDW
transition, where the ordering momentum $|q(h)| > q_c$  shows abrupt variation with magnetic field.
Near the commensurate state, the incommensurate SDW phase can be
understood as a {\em soliton lattice}, with a finite linear density 
$n_s$ of solitons.\cite{chaikin_book,lee-2009} In the dilute limit $n_s w \ll 1$,
where $w = \sqrt{u/|t_k|}$ is the width of the soliton, the solitons {\em repel} each other,
with an exponentially decaying potential $U e^{-x/w}$.
The pre-factor $U = 32\sqrt{u |t_k|}$ can be obtained by calculating the energy
of two solitons separated by a distance $x$ with the help of \eqref{eq:umk10}.
As a result, the energy density of the dilute soliton lattice is given by
\begin{equation}
E_{\rm sol.lat.} = 2\pi v (q_c - q) n_s + n_s U e^{-1/(n_s w)} .
\label{eq:umk12}
\end{equation}
Here the last term represents the repulsion between the nearest solitons
of the lattice. The optimal concentration $n_s^*$, for $q > q_c$,
follows by minimizing Eq.~(\ref{eq:umk12}) (with logarithmic accuracy):
\begin{equation}
n_s^* = \frac{1}{w \ln[\frac{U}{2\pi v (q-q_c)}]} .
\label{eq:umk13}
\end{equation}
This implies that the shift of the ordering momentum $\delta Q = 2\pi n_s^*$ from its
commensurate value $Q_{k,\nu} = 2\pi \nu/k$ inside the plateau vanishes with 
an {\em infinite slope}, according to
\begin{equation}
\delta Q = \frac{-2\pi}{w \ln | (h - h^{(k,\nu)})/\delta h_{(k,\nu)}|},
\label{eq:umk14}
\end{equation}
where the last expression is written with logarithmic accuracy.

\section{Standard model: field (including zero) along $a$ axis}
\label{sec:field-along-axis}

The predictions of the last section for the idealized model with
$J''=D=0$, unfortunately do not agree with experiments on \ccc.  At zero
magnetic field, the ground state is actually an incommensurate spiral,
similar to that predicted at high fields in the previous section, rather
than the collinear state produced by the fluctuation-generated
interactions.  Moreover, the zero field spiral ground state appears to
continuously evolve on increasing fields along the $a$ axis, with no
intervening phase transition before reaching a fully polarized
ferromagnetic state at the saturation field.  The SDW state predicted in
the previous section is entirely absent.

This behavior, however, is readily explained by the standard model {\sl
  including} the $D$ and $J''$ terms, as discussed in
Ref. \onlinecite{Starykh2007}.  To proceed, we again apply the
decompositions, Eq.~(\ref{eq:2}), to the these terms.  Following the
logic of the previous section, we keep only the most relevant
contributions.  This gives
\begin{widetext}
  \begin{eqnarray}
    \label{eq:3}
    && H'_2  =  \sum_{y,z} \int\! dx\, \Big\{ -D (-1)^z \left(
    \mathcal{S}^+_{y,z;\pi} \mathcal{S}^-_{y+1,z;\pi} +
    \mathcal{S}^-_{y,z;\pi} \mathcal{S}^+_{y+1,z;\pi} \right)  +\gamma''_z \left( \mathcal{S}^z_{y,z;\pi-2\delta}
    \mathcal{S}^z_{y,z+1;\pi+2\delta} + \mathcal{S}^z_{y,z;\pi+2\delta}
    \mathcal{S}^z_{y,z+1;\pi-2\delta}\right) \nonumber \\
  & & + \gamma''_{\delta-z}  \left(  \mathcal{S}^z_{y,z;\pi-2\delta}
    \mathcal{S}^z_{y,z+1;\pi-2\delta}e^{-4i\delta x} + \mathcal{S}^z_{y,z;\pi+2\delta}
    \mathcal{S}^z_{y,z+1;\pi+2\delta}e^{4i\delta x} \right)  + \gamma''_{\pm}\left(\mathcal{S}^+_{y,z;\pi} \mathcal{S}^-_{y,z+1;\pi}+\mathcal{S}^-_{y,z;\pi} \mathcal{S}^+_{y,z+1;\pi}\right)\Big\},
\end{eqnarray}
with $\gamma''_z=\gamma''_{\delta-z}=2\gamma''_\pm = J''$.
\end{widetext}

At zero field, we can simplify, using SU$(2)$ symmetry and $\delta=0$:
\begin{eqnarray}
  \label{eq:4}
  H'_2 & = & \sum_{y,z} \int\! dx\, \Big\{ -D (-1)^z \left( N^+_{y,z}
    N^-_{y+1,z} + N^-_{y,z}  N^+_{y+1,z} \right) \nonumber \\
  & & + J'' {\bm N}_{y,z}\cdot {\bm N}_{y,z+1} \Big\}.
\end{eqnarray}

\subsection{Competition between $D$ and $J''$}
\label{sec:comp-betw-d}

In Eq.~\eqref{eq:4} it is evident that both $D$ and $J''$ induce strongly relevant
perturbations, with scaling dimension $1$.  These are more relevant than
any terms na\"ively present in zero field (i.e. the two terms studied in
the previous section), and much larger than the fluctuation-induced
correction ($\sim (J')^4/J^3$) with the same scaling dimension, which
drives the formation of the collinear antiferromagnetic state \cite{Starykh2007} in their absence.  They also
become more relevant with increasing field.  Hence we expect that these
terms should control the actual ordering in \ccc\ for this field
orientation.
\begin{figure}[h]
  \centering
  \includegraphics[width=3.4in]{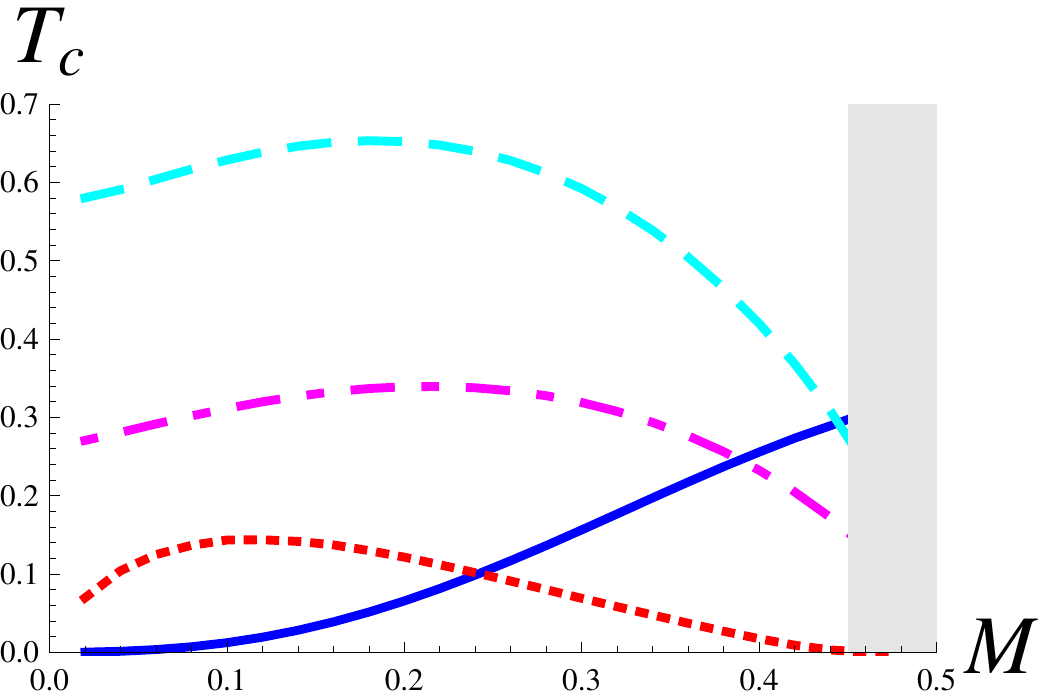}
  \caption{(Color online) Same as Fig.~\ref{fig:RG1}, but with the chain mean field
    ordering temperature due to interlayer interactions ($T_{\rm il}$,
    dot-dashed (magenta) line) and DM interaction ($T_{\rm D}$, dashed (cyan) line) included.  Note that the
    highest two temperatures are $T_{\rm D}$ and $T_{\rm il}$, with the former
    substantially larger.  Thus for a magnetic field along the $\hat{a}$
    axis, where the $D$ term is present, it is expected to dominate the
    ordering except in the region close to saturation, $M\gtrsim 0.45$
    or so.  That later region, which is described by a 2d scaling, is represented by the shaded rectangle.}
  \label{fig:RG2}
\end{figure}

However, it is not so obvious which of the two is dominant.  Indeed,
they actually compete.  This can be seen as follows.  The $D$ term is
minimized (for $D>0$) by configurations in which 
\begin{equation}
  \label{eq:39}
  \langle N_{y,z}^+ \rangle_{D} = (-1)^{yz} N e^{i\beta\vartheta_z}, 
\end{equation}
where the classical phase $\vartheta_z$ may
depend upon $z$.  For such configurations, however, the $J''$ term
oscillates in sign with $y$, and hence averages to zero.  To instead
minimize the $J''$ term, one requires configurations in which 
\begin{equation}
  \label{eq:40}
  \langle  {\bm N}_{y,z}\rangle_{J''} = (-1)^z N {\hat {\bm n}}_y, 
\end{equation}
where the unit vector
${\hat{\bm n}}_y=(n^1_y, n^2_y, n^3_y)$ may depend upon $y$.  For all such configurations, the
$D$ term vanishes when averaged over $z$.

The balance of this competition is determined by the
relative magnitudes of $D$ and $J''$.
We rely again on the chain mean field method, which indicates that the
DM term dominates for the parameters of \ccc\ -- as shown by the fact
that the associated mean-field ordering temperature in
Fig.~\ref{fig:RG2} is largest.  This is essentially due
to the fact that the DM interaction acts on twice as many bonds as does
the interlayer exchange.   

A physical distinction between the two candidate states, which may be
compared to experiment, is in their vector chirality, which is
concentrated on the diagonal bonds of the triangular lattice.  Define 
\begin{equation}
  \label{eq:30}
  \chi^z_{y,z,\pm}(x) = {\hat z}\cdot\langle {\bm S}_{y,z}(x) \times {\bm S}_{y\pm 1,z}(x+\tfrac{1}{2})\rangle  .
\end{equation}
This quantity is non-zero in both phases.  In the continuum limit, one
obtains
\begin{equation}
  \label{eq:31}
  \chi^z_{y,z,\pm} \sim \frac{1}{2}\left( N^+_{y,z} N^-_{y\pm 1,z}+ {\rm h.c.}\right).
\end{equation}
Let us compare this chirality for the states favored by $D$ and $J''$.
For the $D$ term, one obtains $\chi^z_{y,z,\pm} \sim (-1)^z N^2$, which
is constant in the triangular planes but alternates between layers.
This staggering of chirality along the crystallographic $a$ axis is
observed experimentally in zero field.  For the $J''$ term, one obtains
instead $\chi^z_\pm \sim N^2 (n_y^1 n_{y+1}^1 +
n_y^2 n_{y+1}^2)$, which can vary within the triangular
planes but is the same in every such layer.  Thus experiment supports
the $D$-induced order but not the $J''$ one, in agreement with the
calculation described above.

Now consider non-zero field.  In this case, returning to
Eq.~(\ref{eq:3}) with $\delta\neq 0$, we see that the third term is
oscillatory, and can be dropped.  The second term is less relevant than
the first and fourth, and thus is also subdominant.  One is still left
with a competition between the first and fourth terms, again controlled
by the balance of DM and inter-layer exchange.  Since several terms that
had formed part of the latter coupling for $h=0$ are now removed, we
should expect that the DM will be relatively enhanced and continue to
win the competition for all values of the field.  This gives a natural
explanation for the continuity of the ordered phase across the range of
fields observed in experiments (with this field orientation).     It is
also interesting to note that the scaling dimension of the
$\mathcal{S}^+_{y,z;\pi}$ operator decreases with increasing field,
actually making the DM coupling more relevant.  This is reflected in an
increasing critical temperature for the ordered phase with increasing
field, up to a maximum which occurs relatively close to saturation.  Again,
consulting Fig.~\ref{fig:RG2}, we see that indeed the $D$ term dominates
the ordering except very close to saturation ($M>0.45$ or so).  In that
narrow field window, the larger bare value of the cone interaction
($O(J')$) is sufficient to overcome its larger scaling dimension.

\subsection{Incommensuration of the ordered state}
\label{sec:incomm-order-state}

Na\"ively, it would appear from the above analysis that, in the region
$M<0.45$, a commensurate ordered state is induced by the $D$ term.
However, while the $D$ term is indeed dominant in this regime, we still
need to take into account the subsidiary effects of the $J'$
interaction.  It turns out that the cone coupling $\gamma_{\rm cone}$ does
not actually compete with $D$, so that it introduces a weak
incommensuration in the ordered state.  

To see this, we apply the expectation value in Eq.~\eqref{eq:39} to the
cone interaction in Eq.~\eqref{eq:H1p} (using $\mathcal{S}^\pm_{y,z;\pi}
= N^\pm_{y,z}$ in zero field), assuming $\vartheta_z$ is a slowly-varying
function of $x$:
\begin{equation}
  \label{eq:41}
  H'_1 \rightarrow -2 \gamma_{\rm cone} \beta N^2  \sum_{y,z} \int\! dx \,
  (-1)^z \partial_x \vartheta_z .
\end{equation}
The (average) phase $\vartheta_z$ is obviously the classical analog of $\theta_z$, so
that there is a gradient cost obtained from Eq.~\eqref{eq:H-bos-0},
which should be added to the above term to obtain
\begin{equation}
  \label{eq:42}
  H_{\rm eff} = \sum_{y,z} \int\! dx \, \Big\{ \frac{v}{2}
  (\partial_x \vartheta_z)^2 -2 \gamma_{\rm cone} \beta N^2  (-1)^z \partial_x \vartheta_z \Big\}.
\end{equation}
This is easily minimized with respect to $\vartheta_z$:
\begin{equation}
  \label{eq:43}
 \partial_x \vartheta_z = 2\frac{\gamma_{\rm cone}}{v} \beta N^2 (-1)^z
 \equiv \frac{q_0}{\beta}(-1)^z.
\end{equation}
Here $q_0$ is the induced incommensurability.  For zero
magnetization, we have $q_0 = (4J'/J) N^2$.  However,
Eq.~\eqref{eq:43} also applied to $M>0$, if $N$ is replaced by the
magnitude of the expectation value of $\mathcal{S}^\pm_{y,z;\pi}$.  In
both cases, the result is that
\begin{equation}
  \label{eq:46}
  \langle \mathcal{S}^+_{y,z;\pi}\rangle = (-1)^{yz} N e^{i (-1)^z q_0
    x+ i \Theta_z},
\end{equation}
where $\Theta_z$ is arbitary for each layer, since we have up to now
neglected inter-layer coupling.  Let us compare to what is expected in
the cone-dominated regime, $M>0.45$.  Here we should apply the ansatz in
Eq.~\eqref{eq:45}, which minimizes the cone interaction, to the $D$ term
in Eq.~\eqref{eq:3}.  One obtains
\begin{equation}
  \label{eq:47}
  H'_2 \rightarrow \sum_{y,z}\int\! dx\, (-D) |\psi|^2 (-1)^z \sigma_z,
\end{equation}
which is minimized by taking $\sigma_z={\rm sgn}(D)(-1)^z$.  
Suppose that $D$ is positive, which is always possible if we redefine $z$. 
Then Eq.~\eqref{eq:45}
becomes
\begin{equation}
  \label{eq:48}
  \langle \mathcal{S}^+_{y,z;\pi}\rangle = (-1)^{yz} |\psi| e^{i (-1)^z
    q_0 x+i\Theta_z} .
\end{equation}
Comparing Eq.~\eqref{eq:46} and Eq.~\eqref{eq:48}, we see the forms are
identical.  Thus, the regimes $M<0.45$ and $M>0.45$ are actually
smoothly connected, and distinguished only by which interaction controls
the largest part of the ordering energy.  This means that $D$ and $J'$
do not really compete.  Indeed, in Appendix~\ref{sec:cmft-dm}, we show
that in  CMFT both interactions together increase the critical
temperature of the cone state.

\subsection{Interlayer correlations}
\label{sec:interl-corr}

The expression in Eq.~(\ref{eq:48}) contains an undetermined phase,
$\Theta_z$, for each layer.  One may look to $J''$, which has been
neglected in obtaining the form in Eq.~(\ref{eq:48}), to fix these
phases.  However, at the na\"ive level of first order perturbation
theory, this is not the case.  In particular, taking the expectation
value of the $J''$ term in Eq.~(\ref{eq:4}) or the corresponding
$\gamma''_\pm$ term in Eq.~(\ref{eq:3}), one finds a oscillating result,
which vanishes upon summation over $y$.  This indicates that the effects
of $J''$ on the undetermined phases is second order in $J''$.

Such second order effects can be considered as a fluctuation-induced
interaction, which can be derived in a similar way as in the calculation of
Appendix~\ref{sec:biquadratic}.  One obtains
\begin{equation}
  \label{eq:66}
  \Delta H = - J''_2 \sum_{y,z} \int\! dx\,\cos (\theta_{y,z}-\theta_{y,z+2}),
\end{equation}
with {\sl ferromagnetic} $J''_2 \sim (J'')^2/v >0$.  Taking its expectation value, this
terms splits the large phase degeneracy, leaving only two undetermined
values, 
\begin{equation}
  \label{eq:67}
  \Theta_z = \left\{ \begin{array}{cc} \Theta_0 & \textrm{for $z$ even}
      \\ \Theta_1 & \textrm{for $z$ odd} \end{array}\right.
\end{equation}
Some bare microscopic second neighbor exchange might contribute to
$J''_2$, but experiments indicate that the net result remains
ferromagnetic, as there is no enlargement of the unit cell in the $a$
($z$) direction. 

For the standard model, the two remaining phase degeneracies are
protected by symmetry.  The {\sl overall} $U(1)$ phase,
$\Theta_0+\Theta_1$ is of course expected to be arbitrary, owing to
rotational symmetry of the Hamiltonian about the $a$ ($z$) axis.  The
{\sl relative} phase, $\Theta_0-\Theta_1$, is protected by translation
symmetry, $x\rightarrow x+1$, under which $\Theta_z \rightarrow
\Theta_z+ (-1)^z q_0$.

\section{Field along $b$ axis}
\label{sec:field-along-b}

\begin{figure}[h]
  \centering
 \includegraphics[width=3.4in]{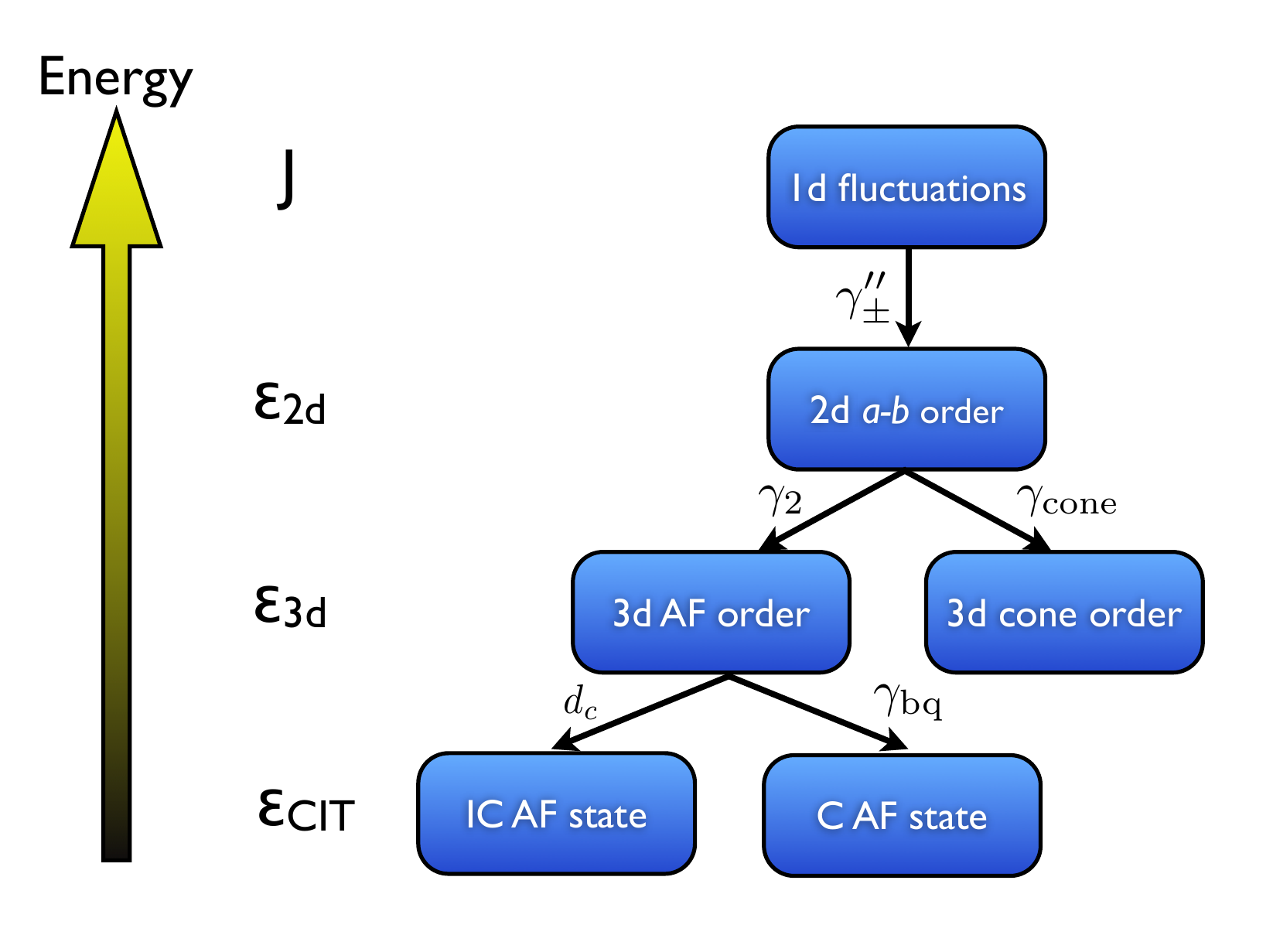}
\caption{Cascade of energy scales operative for fields in the $b$-$c$
  plane.  Symbols adjacent to the arrows indicate the interactions
  responsible for the associated (partial) ordering.  The cascade can be
  quantified by different ``condensation'' energy densities, which give
  the lowering of the energy density due to the establishment of the
  associated partial order.  At the highest
  energies, between $J$ and $\varepsilon_{2d} \sim J'' |\psi|^2$, the system exhibits
  one-dimensional fluctuations.  Here $|\psi|$ is the amplitude of the
  order parameter, Eq.~(\ref{eq:8}).  Between $\varepsilon_{2d}$ and
  $\varepsilon_{3d}$, the spins order within $a$-$b$ planes, but the planes are not
  registered.  Below $\varepsilon_{3d} \sim {\rm Max}\{ J_2|\psi|^2,
  \frac{(J')^2}{J} |\psi|^4\}$, full three-dimensional
  order develops, which may be of antiferromagnetic (AF) or cone type.
  In the former case, there may be yet another lower energy scale,
  $\varepsilon_{\rm CIT }$ (which is of the same order as $\varepsilon_{3d}$), below which the
  magnetic structure is completely determined.  This may be commensurate
  or incommensurate, the latter occuring only for fields along the $c$
  axis, and is driven by DM interactions.}
  \label{fig:energyscales}
\end{figure}

In this section and the next, we will discuss the 
physics determining the ordered ground states when the magnetic field is
normal to the crystallographic $a$ axis.  These cases are much more
complex than above, because, as we will see, the ordering is determined
by several distinct interactions which are important at different energy
scales.  The ``cascade'' of energy scales, which must be considered in
turn, from largest to smallest, is indicated graphically in
Fig.~\ref{fig:energyscales}.  

\subsection{Irrelevance of $D$ term}
\label{sec:irrelevant-d-term}

Having understood that the DM interaction $D$ dominated the physics for
fields along the $a$ axis, we first consider its role in this field
orientation.  Remarkably, the change in orientation has a drastic
effect.  With a field in the $b$-$c$ plane, the $D$ term always involves
one spin component parallel to the field and one perpendicular to it.
Consulting the decomposition of spin operators in Eq.~(\ref{eq:2}), we
immediately see that the dominant fluctuations of these two spin
components are always incommensurate.  As a consequence, in the
continuum limit all terms directly arising from $D$ {\sl oscillate} with
a $e^{\pm 2 i \delta x}$ factor.  This makes them formally strongly
irrelevant.  More physically, upon coarse-graining over length scales
shorter than $\pi/\delta$, these terms average to zero.  Thus, provided
that $D$ does not scale to strong coupling under the RG before this
scale is reached, they become negligible.  This will be true everywhere
except the low field limit.  Specifically, since it has
scaling dimension $1$, the renormalized $D$ term at this scale is of
order $D (\pi/\delta)$.  Demanding this be small compared to $v$, we
obtain the criterion $\delta=\pi M \gtrsim D/J$ for it to be negligible.  

Thus for most of the magnetic field range, we can drop the $D$ term.
This gives a simple reason why the experimental behavior in \ccc\ for
this field orientation is completely different from that with field
along the $a$ axis.  It is tempting to expect instead that the ideal 2d
model discussed in Sec.~\ref{sec:ideal-2d-model} should apply.  A
comparison to experiments strongly suggests, however, that this is not the
case, except perhaps at high fields.   Unfortunately, no published low temperature NMR
or neutron data are available in this field orientation.
However, magnetization and specific heat measurements \cite{TokiwaPRB2006} show a
single phase occupying the region below the high field cone state and
above the low field DM-dominated spiral.    By comparison to the same
measurements along the $c$ axis, which are strikingly similar, it
appears most likely that this intermediate phase represents a
commensurate state.  In the ideal 2d model, we would be forced to
interpret it instead as an incommensurate SDW.  This seems untenable, as
such an SDW phase should also show a $1/3$ magnetization plateau (see
Sec.~\ref{sec:plateau}), of which there is no sign.  Thus we conclude
that the SDW phase has been superseded by another competing state.  In
the following, we find an explanation for this competing state as a
commensurate ``antiferromagnetic'' (AF) phase.

\subsection{Role of interlayer interactions}
\label{sec:interl-inter}

To explain this, we must take into account the one remaining interaction
present in the standard but not ideal model -- the exchange $J''$
between triangular planes.  Like the (now absent) $D$ term, it is
strongly relevant, and unlike the $D$ term, it is not averaged out in
any field orientation.  The appropriate continuum limit has already been
given, the dominant piece being the last term of Eq.~(\ref{eq:3}),
reproduced here:
\begin{equation}
  \label{eq:5}
  H'_2 = \gamma''_{\pm} \sum_{y,z} \int\! dx\,  \left(\mathcal{S}^+_{y,z;\pi} \mathcal{S}^-_{y,z+1;\pi}+\mathcal{S}^-_{y,z;\pi} \mathcal{S}^+_{y,z+1;\pi}\right),
\end{equation}
with $\gamma''_\pm=J''/2$.  Written explicitly using bosonization, it
becomes
\begin{equation}
  \label{eq:27}
  H'_2 = {\tilde \gamma}''_\pm \sum_{y,z} \int\! dx\, \cos [\beta (\theta_{y,z}-\theta_{y,z+1})],
\end{equation}
with ${\tilde \gamma}''_\pm = 2 A_3^2 \gamma''_\pm$.  This
coupling is more relevant than either the SDW or cone interaction, in
the entire range of magnetization.  We may therefore expect that it
scales to strong coupling unambiguously before any competing
interactions.  To check this, we again consult the comparison of
critical temperatures in Fig.~\ref{fig:RG2}.  Neglecting the effects of
the $D$ term, as we have just discussed, we see that the interlayer
interaction is clearly dominant for all magnetizations below about 80
percent of the saturation value.  Above this magnetization, the larger
bare value of the cone interaction, which is of order $J'$ rather than
$J''$, overcomes the difference in scaling dimensions and controls the
physics.  In this high magnetization regime, the physics is therefore
very similar to that described in the previous sections, and an
incommensurate cone state is expected.

In the remainder of this section, we focus on the main field regime,
where $\gamma''_\pm$ is dominant.  The latter obeys the RG equation
(using the dimensionless coupling $\breve{\gamma}''_\pm =
\gamma''_\pm/(v\Lambda_\ell^2)$ as discussed in Sec.~\ref{sec:renorm-group-appr})
\begin{equation}
  \label{eq:6}
  \partial_\ell \breve{\gamma}''_\pm = (2-2\Delta_\pm) \breve{\gamma}''_\pm,
\end{equation}
where $\Delta_\pm = \pi R^2$ is the scaling dimension of the
$\mathcal{S}^\pm_{\pi}$ fields.
Integrating this to the scale $\ell=\ln (\xi'')$ such that $\breve{\gamma}''_\pm
(\ell) \sim v$ defines the length scale 
\begin{equation}
  \label{eq:7}
  \xi'' \sim (v/J'')^{1/(2-2\Delta_\pm)}.
\end{equation}
For lengths shorter than $\xi''$, one-dimensional fluctuations are
significant and approximately those of free chains.  On longer length
scales, we expect that $\gamma''_\pm$ drives ordering of the
$\mathcal{S}^\pm_{y,z;\pi}$ fields.  $H'_2$ in Eq.~(\ref{eq:5}) is
minimized by configurations of the form
\begin{equation}
  \label{eq:8}
  \left\langle \mathcal{S}^\pm_{y,z;\pi}\right\rangle = |\psi| (-1)^z e^{\pm i \beta \vartheta_y},
\end{equation}
where $|\psi|$ is a real number giving the magnitude of the spontaneous
moment, and $\vartheta_y$ is a {\sl classical} phase that can be chosen
{\sl independently} for each vertical $a$-$b$ plane specified by $y$.
Note that longitudinal order is strongly suppressed at this scale,
$\langle \mathcal{S}^z_{y,z}\rangle = 0$, by the uncertainty principle
(in bosonization this follows from the duality of the $\theta$ and
$\varphi$ fields).    We expect by scaling that $|\psi| \sim
(\xi'')^{-\Delta_\pm} \ll 1$,
reflecting the suppressed magnitude of magnetic order by 1d
fluctuations.  Hence 
\begin{equation}
  \label{eq:28}
  |\psi| = \sigma(M) \left( \frac{J''}{v}\right)^{\frac{\Delta_\pm}{2-2\Delta_\pm}},
\end{equation}
where the prefactor $\sigma(M)$ is computed by CMFT in Appendix~\ref{sec:cmft-T=0}.  We
estimate $|\psi|\approx 0.25-0.3$ over most of the field range.  From
this, we can estimate the lowering of the energy density due to the
establishment of such two-dimensional order, simply by taking the
expectation value of Eq.~(\ref{eq:5}):
\begin{equation}
  \label{eq:12}
  \varepsilon_{2d} \sim J'' |\psi|^2.
\end{equation}
An experimental measure of this energy density is the XY spin
stiffness along the $a$ axis, which is of the same order,
$\rho_{s;a} \sim \varepsilon_{2d}$.  Note that the spin stiffness along
the $b$ axis is much larger, of order $\rho_{s;b} \sim v$.

\subsection{Exchange coupling of $a$-$b$ planes}
\label{sec:coupling-a-b}

The arbitrary choice of $\vartheta_y$ for every $y$ is a consequence of the
fact that the dominant interaction, $\gamma''_\pm$, does not couple
different $a$-$b$ planes.  Less relevant interactions can and do remove
this arbitrariness, ultimately determining the precise nature of the
ordered state.

To study this, we first include exchange interactions between chains
within the $b$-$c$ planes.  In the standard model, this is only the $J'$
coupling along the nearest-neighbor diagonals.  However, it was argued
in Ref.~\onlinecite{Starykh2007} that it is important to also take
into account weak exchange $J_2$ between spins on {\sl second-neighbor}
chains separated by distance $\Delta y=2$.  While clearly $J_2 \ll J'$,
it is important because it is unfrustrated, unlike the $J'$
interaction.  

In the continuum limit, these couplings lead to the Hamiltonian
\begin{eqnarray}
  \label{eq:9}
  H'_3 & = & \sum_{y,z} \int\! dx \, \Big\{ -i \gamma_{\rm cone}\mathcal{S}_{y,z;\pi}^+ \partial_x
  \mathcal{S}_{y+1,z;\pi}^-  + {\rm h.c.}  \nonumber\\
  && + \gamma_2 \mathcal{S}_{y,z;\pi}^+
  \mathcal{S}_{y+2,z;\pi}^- + {\rm h.c.} \Big\} ,
\end{eqnarray}
with $\gamma_2=J_2/2$ (and $\gamma_{\rm cone} = J'/2$ as given earlier).  

Taking the expectation values using Eq.~(\ref{eq:8}), the resulting
renormalized Hamiltonian can then be treated classically, and minimized
to find the ground state.  It is evident that the ``twist'' interaction
$\gamma_{\rm cone}$ favors an incommensurate state with $k_x \neq \pi$.
To describe this requires allowing for non-zero gradients
$\partial_x\vartheta_y$.  While such configurations are not ground states
in the absence of $\gamma_{\rm cone}$, they are low in energy, because a
small gradient comprises a soft (Goldstone) mode.  The magnitude of the
associated incommensurability is determined by a balance of $\gamma_{\rm
  cone}$ with the gradient terms in $H_0$, which of course favor
commensurate order at $k_x=\pi$.  We therefore include the latter, and
write the entire effective Hamiltonian explicitly in the $\vartheta_y$
variables, which we will allow to be $x$-dependent but {\sl independent}
of $z$ according to Eq.~(\ref{eq:8}).  The total energy becomes
\begin{eqnarray}
  \label{eq:10}
  E_b & = & L_z \sum_y \int \! dx\, \Big\{ \frac{v}{2} (\partial_x
  \vartheta_y)^2 + g_2 \cos [\beta (\vartheta_y - \vartheta_{y+2})] \nonumber \\
  & & -  g_{\rm cone}  (\partial_x \vartheta_{y}+ \partial_x \vartheta_{y+1}) 
  \cos [\beta(\vartheta_{y} - \vartheta_{y+1})] \Big\},
\end{eqnarray}
with $g_2 = 2 \gamma_2 |\psi|^2$ and $g_{\rm cone} = \gamma_{\rm cone}
|\psi|^2 \beta$.

Now we can see that, for $g_2>0$, which is expected from
antiferromagnetic superexchange, the two interactions strongly compete.
In this case, the minima of the $g_2$ term are states with 
\begin{equation}
  \label{eq:25}
  \vartheta_y = \frac{\pi y}{2\beta} + \frac{\Theta_{{\rm mod}(y,2)}}{\beta} .
\end{equation}
Here $\Theta_0,\Theta_1$ define the overall phase on the even and odd
chains, respectively.  
Inserting this into the twist term, one finds a vanishing
result due to cancellations when the sum over $y$ is carried out, even
if $\Theta_0$ and $\Theta_1$ are allowed to have gradients. Hence this
solution has energy density equal to $-g_2$.  This is a commensurate
``antiferromagnetic'' (AF) state.  Conversely, the solutions which
minimize the twist term have $\vartheta_y = \kappa x$, for which
the $g_2$ term is {\sl maximized} rather than minimized.  Here
$\kappa=2 g_{\rm cone}/v$ is determined by minimizing the full energy,
leading to the energy density $-2g_{\rm cone}^2/v + g_2$. This is the
incommensurate cone state.  Comparing the energies of the two states,
one finds that the AF state obtains for $g_2 > g_{\rm cone}^2/v$.
This requires a minimum value of second neighbor exchange for the
commensurate state, $J_2 > J_{2}^*$, where
\begin{equation}
  \label{eq:29}
  J_{2}^* = \frac{\beta^2|\psi|^2}{4} \frac{(J')^2}{v}.
\end{equation}
For \ccc, $J_2^*$ is very small, and is in fact only a few percent ($\leq 5\%$) of
$J$ for the relevant field range.  Moreover, we argue below that the
above value of $J_2^*$ is actually an overestimate, as it neglects a
fluctuation-generated interaction which is of the same order.  Thus, an
exceedingly tiny second neighbor coupling $J_2$, likely undetectable
directly, qualitatively changes the ground state.  In general,
re-expressing the minimum energy density in terms of bare variables, we
have
\begin{equation}
  \label{eq:13}
  \varepsilon_{3d} \sim -{\rm Max}\left\{ J_2 |\psi|^2,
    \frac{\beta^2|\psi|^4}{4} \frac{(J')^2}{v}\right\}.
\end{equation}
This energy scale determines the spin stiffness along the $c$ axis,
$\rho_{s;c} \sim \varepsilon_{3d}$.  

\subsection{Locking of even and odd $a$-$b$ layers}
\label{sec:locking-even-odd}

When $J_2$ is dominant in establishing three-dimensional AF order, it is
ineffective in coupling the even and odd layers.  As a consequence,
there remains an artificial degeneracy of solutions, specifically, one
may make opposite rotations of the phases $\Theta_1$ and $\Theta_2$.
This rotation is not a true symmetry of the microscopic theory.
However, the simplest possible coupling of phases in neighboring chains,
of the form $\cos \beta(\vartheta_y - \vartheta_{y+1})$, {\sl is}
prohibited by reflection symmetry, see \eqref{eq:57} and \eqref{eq:59}.  Instead, the leading possible
coupling between neighboring chains is of the form
\begin{equation}
  \label{eq:14}
  H_{\rm bq} = {\bf +} g_{\rm bq} \sum_{y,z} \int \! dx\, \cos [2\beta (\vartheta_{y,z}
  - \vartheta_{y+1,z})].
\end{equation}
Here we have already assumed Eq.~(\ref{eq:8}), and taken the average of
the fluctuation-generated interaction.  See
Appendix~\ref{sec:biquadratic} for details.  
For classical XY spins with phase $\beta \vartheta$, this interaction would
correspond to a biquadratic coupling $({\bm S}_i \cdot {\bm S}_j)^2$,
between spins on neighboring chains.  Such fluctuation-generated
biquadratic interactions are indeed familiar from the theory of
frustrated magnets, and are a manifestation of ``order by
disorder''.\cite{shender,Henley_JAP}  In that context, it is well-known that
fluctuations generally favor collinear states, which requires $g_{\rm
  bq}>0$.  This is indeed confirmed by the microscopic calculation in
Appendix~\ref{sec:biquadratic}, which leads to the estimate
\begin{equation}
  \label{eq:15}
  g_{\rm bq} \sim \frac{(J')^2}{v} |\psi|^{4}.
\end{equation}

In the AF phase, we may use the solutions for $\theta_{y,z}$ determined
above, and hence rewrite Eq.~(\ref{eq:14}) as
\begin{equation}
  \label{eq:16}
  E_{\rm bq} =  -g_{\rm bq} L_x L_y L_z \cos[2(\Theta_0-\Theta_1)].
\end{equation}
Clearly states with $\Theta_0=\Theta_1+ n \pi$ are preferred, which implies
commensurate, collinear, AF order.  The condensation energy density
associated with the selection of the collinear order is thus
$\varepsilon_{\rm CIT} \sim g_{\rm bq}$ (the reason for the choice of this subscript will
become clear in the next section).  Physically, this energy scale
determines the gap of the antisymmetric pseudo-Goldstone mode corresponding to
$\Theta_0-\Theta_1$, which is of order 
\begin{equation}
  \label{eq:17}
  \Delta_{\rm as} \sim \sqrt{v\varepsilon_{\rm CIT}} \sim J' |\psi|^{2}.
\end{equation}
This can potentially be measured as an emergent low energy (but gapped)
mode in neutron scattering.  

Comparing Eq.~\eqref{eq:13} with Eq.~\eqref{eq:15} and
Eq.~\eqref{eq:16}, we observe that the energy gain due to $g_{\rm bq}$
term is of the same order as the energy gain of the incommensurate cone
state.  This is not a coincidence as the both effects have their common
origin in the inter-chain exchange $J'$.  This suggests that, even
in the absence of any micrscopic $J_2$ exchange, a  collinear state could be energetically preferred to
the cone state.   However, the RG approach used to obtain
Eq.~(\ref{eq:15}) is not accurate in determining the $O(1)$ numerical
prefactor, which is essential for making such a comparison
quantitatively.  Thus we at present can only speculate that this might
be the case.  Even if not, these considerations imply that the 
interaction $J_2$ needed to induce the AF state is even lower than
the estimate in Eq.~(\ref{eq:29}).

\section{Field along $c$ axis}
\label{sec:field-along-c}

Experimentally, this field orientation shows the most complex phase
diagram.  In addition to the commensurate ``AF'' phase seen for fields
along the $b$ axis, a broad region of incommensurate phase is also
clearly observed in NMR measurements \cite{takigawa-poster} (and defined by earlier
magnetization measurements \cite{TokiwaPRB2006}).  Within the model used up to now, the
difference in phase diagrams for fields along the $b$ and
$c$ axes is inexplicable: the Hamiltonian has a symmetry under
spin rotations within the $b$-$c$ plane.

\subsection{DM interaction on chain bonds}
\label{sec:dm-interaction-chain}

Therefore additional spin-rotational symmetry breaking interactions {\sl must} be
included to explain this discrepancy.  We therefore turn to the general
set of allowed DM interactions in Sec.~\ref{sec:dm-ology} for our
consideration.  As we saw in the previous section, DM terms whose
D-vector is orthogonal to the applied field average out rapidly in the
presence of an applied field.  Hence we need consider only components of
the D-vectors along the $c$ axis.  There are two independent such terms:
$D_c$ and $D'_c$.  Given that DM terms are generally proportional to the
corresponding exchange, we expect $D_c$ to be the largest of the two,
and we focus on its effects (in Appendix~\ref{sec:d_c} we explain in
detail why $D'_c$ can be neglected).  It introduces the perturbation
\begin{eqnarray}
  \label{eq:21}
  H_c  & \sim & D_c \sum_{y,z} (-1)^y \int \! dx\,
  \mathcal{J}_{y,z}^z(x) \nonumber \\
  & \sim & d_c \sum_{y,z} (-1)^y \int \! dx\, \partial_x \theta_{y,z},
\end{eqnarray}
where $d_c = v D_c/(\beta J)$.

Notably, $H_c$ is linear in the boson fields, and hence, in the absence
of any other interactions, the term $d_c$ can be taken into account
exactly.  Moreover, it is actually a pure boundary term, whose effect on
the energy depends solely on the {\sl winding numbers},
$[\theta_{y,z}(\infty) - \theta_{y,z}(-\infty)]/(2\pi \beta)$, and
vanishes in the zero winding number sector.  However, $H_c$ favors
sectors with non-vanishing winding numbers (proportional to $L_x$, in
fact).  

\subsection{$D_c$ does not compete with $J$,$J'$, and $J''$}
\label{sec:d_c-does-not}

To understand the degree of competition of $H_c$ with the other
interactions, it is instructive to consider the shifted variables
\begin{equation}
  \label{eq:22}
  \tilde\theta_{y,z}(x) = \theta_{y,z}(x) + (-1)^y \frac{d_c}{v} x.
\end{equation}
With this shift, $d_c$ is ``eliminated'' from the free Hamiltonian, up
to a constant: $H_0[\theta]+H_c[\theta] = H_0[\tilde\theta]+ {\rm
  const.}$.  Physically, this change of variables corresponds to a shift
of the dominant wavevector of correlations from $k_x=\pi$ to $k_x = \pi \pm
\beta d_c/v=\pi \pm D_c/J$.  Significantly, the dominant
$\gamma''_\pm$ coupling is invariant under the shift:
$H'_2[\theta]=H'_2[\tilde\theta]$.  Thus $J''$ and $D_c$ do not compete.
The same is true for the $\gamma_2$ ($J_2$) interaction.  Thus, in the region
where the AF phase appears for the case of field along the $b$
axis, since the $J''$ and $J_2$ couplings dominate, we expect the
energetics is unchanged at the three highest energy scales in Fig.~\ref{fig:energyscales}.  

\subsection{Commensurate-Incommensurate Transition}
\label{sec:comm-incomm-trans}

Differences do appear, however, once the cone and biquadratic
interactions are considered, as these are {\sl not} invariant under the
shift in Eq.~(\ref{eq:22}).  We focus on the putative AF region, for
which we may assume the decomposition in Eq.~(\ref{eq:25}).  Allowing
for small gradients in $\Theta_\pm(x,y,z) = \Theta_0(x,y,z)\pm
\Theta_1(x,y,z)$, we obtain the continuum hamiltonian $H=H_++H_-$, with
\begin{eqnarray}
  \label{eq:23}
  H_+ & =  \int \! d^3{\bm r}\, \Big\{ &\sum_\mu\frac{c_\mu}{2}(\partial_\mu
  \Theta_+)^2 \Big\}, \\
  H_- & =  \int \! d^3{\bm r}\, \Big\{ &\sum_\mu\frac{c_\mu}{2}(\partial_\mu
  \Theta_-)^2 + \frac{d_c}{2\beta}\partial_x \Theta_- \nonumber \\
&& - g_{\rm bq} \cos (2 \Theta_-)\Big\} ,
\end{eqnarray}
where $c_x = v/4\beta^2$, $c_y = g_2 $, and $c_z = {\tilde \gamma}''_\pm/4$. 

Here the cone interaction has dropped out, and the low energy
Hamiltonian has decomposed into two decoupled parts.  The first, $H_+$,
is simply the Hamiltonian of a free massless boson.  It describes the
Goldstone mode $\Theta_+$ associated with spin rotations about the field
axis.  The second part, $H_-$, is the familiar sine-Gordon model,
discussed earlier in Sec.~\ref{sec:plateaux-width}.  In this case it is
in three dimensions, but this has no significant consequences.  As in
Sec.~\ref{sec:plateaux-width}, the sine-Gordon model describes a
commensurate phase (here, the AF state) and an incommensurate (IC) one,
separated by a {\sl Commensurate-Incommensurate Transition}, or CIT.

The results for the CIT can be taken over directly from
Sec.~\ref{sec:plateau}, with the mapping $\Theta_- \rightarrow
\varphi/2$.  Here, it is $d_c$ which plays the role of the tuning
parameter, favoring the IC phase, for $|d_c|>d_c^*$, where
\begin{equation}
  \label{eq:26}
  d_c^* = \frac{4\sqrt{v g_{\rm bq}}}{\pi}.
\end{equation}
On entering the IC phase, the system forms a soliton lattice, with a
corresponding incommensurate wavevector $q_0$ (measured relative to the
AF state).  Note that the IC phase found here is thus a smooth
deformation of the AF state, which makes it quite distinct from the cone
state, which is also incommensurate.  The incommensurate wavevector
grows rapidly after the CIT, which can be seen by translating
Eq.~(\ref{eq:umk14}) to the current case:
\begin{equation}
  \label{eq:88}
 q_0 =  \frac{\langle \partial_x \Theta_-\rangle}{2} \sim
 \frac{\pi}{4\beta} \sqrt{\frac{g_{bq}}{v}} \frac{1}{\ln [(|d_c|-d_c^*)/d_c^*]},
\end{equation}
where the brackets $\langle \partial_x \Theta_-\rangle$ indicates the
spatial average.  Once the above logarithm is not large, the solitons are strongly
overlapping, and Eq.~(\ref{eq:88}) is no longer valid.  Instead, one may
simply minimize the energy neglecting $g_{\rm bq}$, which gives 
\begin{equation}
  \label{eq:89}
  q_0 = \frac{\beta d_c}{v} = \frac{D_c}{J}.
\end{equation}
To summarize, $q_0$ varies from its maximal value given in
Eq.~(\ref{eq:89}) at the low field end of the IC phase, and {\sl
  decreases} with increasing field, vanishing asymptotically according
to Eq.~(\ref{eq:88}) at the CIT to the AF phase.  Because the variation
in Eq.~(\ref{eq:88}) is so rapid, very likely $q_0$ appears
approximately constant in most of the IC phase, dropping precipitously
to zero in a narrow region near the CIT.

\section{Experimental Consequences}
\label{sec:exper-cons}

In this section, we consider a few key experimental consequences of the
analysis of the previous sections.  First, we give explicit expressions
for the spin structures in the various phases predicted there, which
should be useful for comparison to neutron scattering measurements.
Next, we derive the nuclear magnetic resonance (NMR) lineshapes in each
of these phases, using these expressions, and compare to experiments by
Takigawa and collaborators.  Finally, we describe the phase diagrams in
the magnetic field--temperature plane, for the different field
orientations.  

\subsection{Explicit spin structures}
\label{sec:expl-spin-struct}

Here we reconstruct explicit formulae and plots of the spin ordering
patterns in the various phases discussed earlier.  

\subsubsection{Cone state}
\label{sec:cone-state}

First consider the
incommensurate ordered ``cone'' state, described in
Sec.~\ref{sec:incomm-order-state}, which occurs for any field along the
$a$ axis.  This is described by Eq.~(\ref{eq:46}).  Using
Eq.~(\ref{eq:2}), we can express the microscopic spin operator's
expectation value
\begin{eqnarray}
  \label{eq:49}
&&   \left\langle S_{xyz}^x \right\rangle_{\rm cone} =  (\left\langle
      S_{xyz}^+\right\rangle e^{i\pi x} + {\rm c.c.})/2  \\
    & &=  (-1)^{yz} N \cos[(\pi + (-1)^z q_0 )x + \Theta_z] \nonumber \\
    & &=  (-1)^{yz} N \cos[(\pi + q_0) x + ((-1)^z-1)\pi x + (-1)^z
    \Theta_z]. \nonumber
\end{eqnarray}
To further simplify, we note that the $x$ coordinate takes integer
values for even $y$ and half-integer values for odd $y$.  As a
consequence, $((-1)^z-1)\pi x$ is an integer multiple of $2\pi$ {\sl
  unless} $y$ is odd {\sl and} $z$ is odd.  This allows this factor
inside the cosine to be removed in favor of an overall $(-1)^{yz}$
factor in front of it, which cancels the one already present.  Therefore
one finds, finally
\begin{equation}
  \label{eq:49a}
  \left\langle S_{xyz}^x \right\rangle_{\rm cone} = N \cos[(\pi + q_0)x + \tilde\Theta_z],
\end{equation}
where $\tilde\Theta_z= (-1)^z\Theta_z$.  Similar manipulations for the y
component of the spins give
\begin{equation}
  \label{eq:49b}
  \left\langle S_{xyz}^y \right\rangle_{\rm cone} = (-1)^z  N \sin[(\pi + q_0)x + \tilde\Theta_z].
\end{equation}
and of course, one has
\begin{equation}
  \label{eq:50}
  \left\langle S_{xyz}^z \right\rangle_{\rm cone} = M.
\end{equation}

\subsubsection{Antiferromagnetic phase}
\label{sec:commensurate-phase}

Here we simply apply Eq.~(\ref{eq:25}), and use $\Theta_0=\Theta$,
$\Theta_1=\Theta + \pi n$ (with $n=0,1$) as preferred in the
commensurate ``antiferromagnetic'' (AF) phase by Eq.~(\ref{eq:16}).  This gives
\begin{eqnarray}
  \label{eq:51}
  \left\langle S_{xyz}^x \right\rangle_{\rm AF} & = & (-1)^z|\psi| \cos[\pi
  x + \sigma \pi y/2 + \Theta], \\
  \left\langle S_{xyz}^y \right\rangle_{\rm AF} & = & (-1)^z|\psi|
   \sin[\pi  x + \sigma \pi y/2 + \Theta], 
\end{eqnarray}
where the $\sigma=\pm 1 = (-1)^n$.  These equations describe a state in
which the $x-y$ components of the spins are {\sl collinear}.  This may
not be obvious, but is true because the combination $\pi x + \sigma \pi y/2$ is
always an integer multiple of $\pi$, owing to the fact that $x$ is
integer (half-integer) for even (odd) $y$.  Combined with the constant
uniform magnetization, Eq.~(\ref{eq:50}), these equations describe a
{\sl co-planar} spin state, distinct from the three-dimensional cone
configuration.  We note, however, that small perturbations due to the
various DM interactions will probably disrupt this ideal coplanarity.
The commensurate nature of the ordering is, however, robust.

\subsubsection{Incommensurate phase for fields along $c$ axis}
\label{sec:incomm-phase-fields}

Here we consider the incommensurate phase which is discussed in
Sec.~\ref{sec:field-along-c}.  For simplicity, we will ignore the narrow
but subtle region in the vicinity of the CIT, where a non-trivial
soliton lattice should be taken into account.  The basic symmetry of
this phase is well described by the ``smooth'' regime (corresponding to
strongly overlapping solitons), where we simply treat the
incommensuration as linear shift of the phase fields, i.e. we take
$\tilde\theta_{y,z}$ in Eq.~(\ref{eq:22}) as constant.   The preceding
formula now are modified to 
\begin{eqnarray}
  \label{eq:52}
   \left\langle S_{xyz}^x \right\rangle_{\rm IC} = (-1)^z|\psi|
   \cos[(\pi+(-1)^y q_0) x + \pi y/2 + \Theta_y],~~~~ \\
  \left\langle S_{xyz}^y \right\rangle_{\rm IC} = (-1)^z|\psi|
  \sin[(\pi+(-1)^y q_0) x + \pi y/2 + \Theta_y],~~~~
\end{eqnarray}
where $q_0= \beta d_c/v$ -- see Eq.~\eqref{eq:89} --
and $\Theta_y$ is a phase taking two distinct arbitrary values for even and odd $y$.  

\subsection{NMR lineshape}
\label{sec:nmr-lineshape}

Recent NMR experiments by Takigawa and collaborators \cite{takigawa-poster} have revealed
numerous phases and transitions in \ccc\ in magnetic fields.  Here we
wish to address the signatures of the phases predicted in this paper in
the NMR lineshape.  The basic approach is to consider the Hamiltonian of
a given nuclear spin ${\bf I}_i$ to be the sum of two effective fields 
\begin{equation}
  \label{eq:53}
  H_a = ({\bf h}^{\rm ext}_i + {\bf h}^{\rm hf}_a)\cdot {\bf I}_i,
\end{equation}
where ${\bf h}^{\rm ext}_i$ is the effective field on the nucleus $i$
due to the external field ${\bf H}$, factoring in any anisotropies of
the nuclear g-tensor (which are believed to be small \cite{private}).
The remaining ``hyperfine field'' ${\bf h}^{\rm hf}_i$ represents
transferred hyperfine interactions with nearby electronic spins.  The
NMR resonance frequency of this particular nucleus is simply
proportional to the magnitude of the total effective field.  The
simplest approximation, which we take here, is to assume in addition
that $|{\bf h}^{\rm hf}_i| \ll |{\bf h}^{\rm ext}_i|$. This is certainly
so in intermediate and high-field regions, which we focus on. For lower
fields, of the order of $1 - 2$ T, this may not be such a good
approximation.\cite{private} But even in this case the off-diagonal
contribution (see \eqref{eq:54} below), which is central to our
consideration, should be smaller than the diagonal one and a modified
expansion in off-diagonal components of ${\bf h}^{\rm hf}_i$ should be
possible. With these assumptions in mind, and disregarding the g-factor
anisotropy (so that ${\bf h}^{\rm ext}_i \propto {\bf H}$), we can
approximate the shift due to the hyperfine interaction by
\begin{equation}
  \label{eq:70}
  \Delta\nu_i \propto {\bf h}^{\rm hf}_i \cdot {\bf h}^{\rm ext}_i /|{\bf h}^{\rm ext}_i |  = {\bf h}^{\rm hf}_i \cdot {\bf\hat H},
\end{equation}
where ${\bf\hat H}= {\bf H}/|{\bf H}|$.
%
\begin{figure}[htb]
\begin{center}
\vspace{.2cm}
\includegraphics[width=0.7\columnwidth]{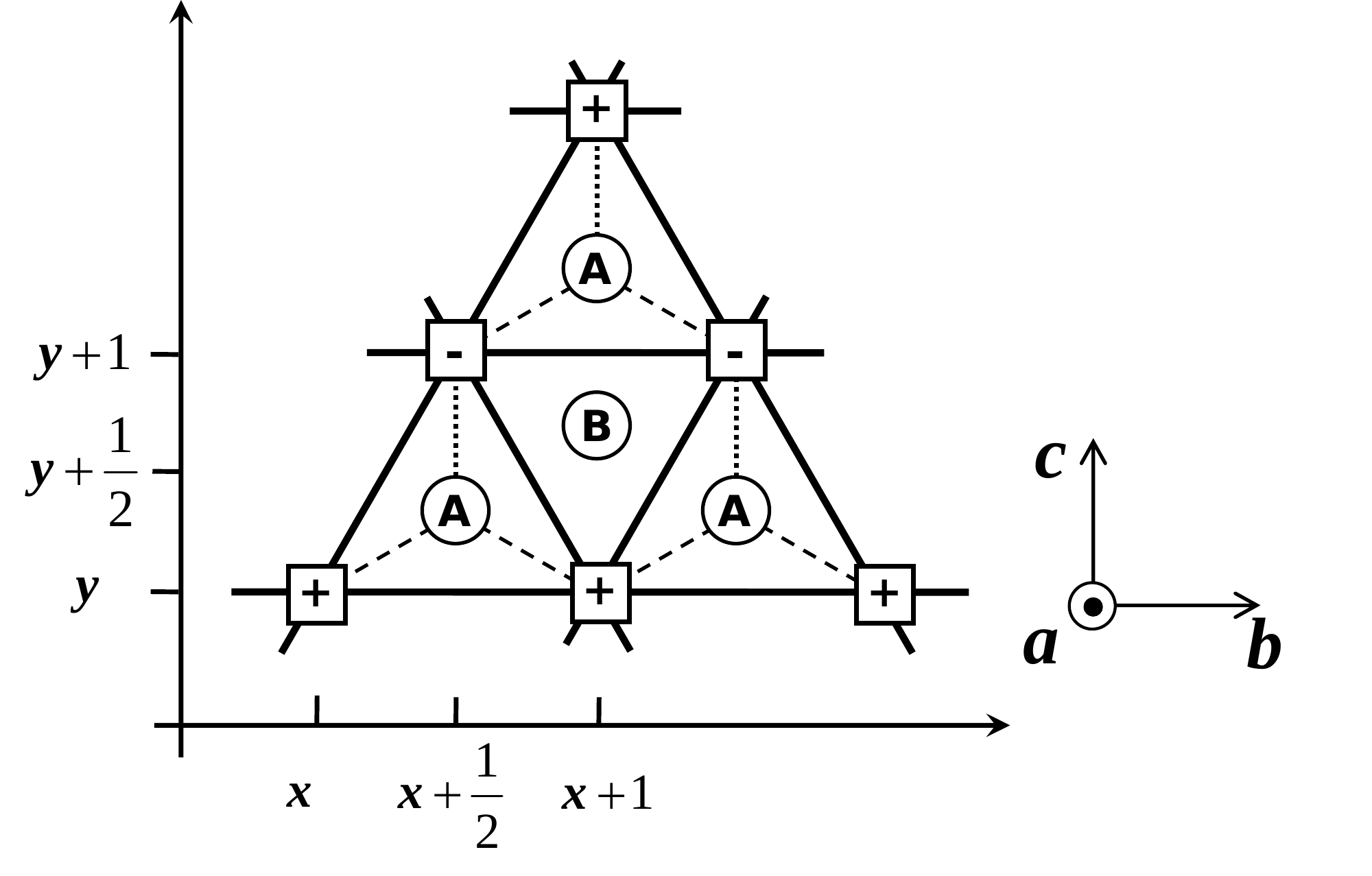}
\caption{Scheme of the trasferred hyperfine interaction for Cs(A). 
The signs $+$/$-$ refer to the relative signs of the off-diagonal entry 
$\kappa_{ij}$ in Eq. (\ref{eq:54}). $\kappa_{ij}$ takes equal value for 
two sites of a triangle as shown by the dashed lines while that for the other site can be different as shown by the dotted line.}
\label{fig: NMR_Cs_A_ions}
\end{center}
\end{figure}  
In \ccc, there are two inequivalent Cs sites measured in the Cs NMR
measurements.  We focus on the Cs(A) site, which is located slightly
above or below the center of a triangle of Cu spins.\cite{NMRpaper}\
The NMR response of the\cite{private} Cs(B) site, which is coupled to
Cu spins in the two planes adjacent to it, is more difficult to analyze
at present, but we do expect the qualitative conclusions derived below
to remain valid for this situation as well. In the Cs(A) case, the
hyperfine field of nucleus $i$ located between spins $j$ at the sites of
this triangle should be given by the sum of three transferred
contributions:
\begin{equation}
  \label{eq:72}
  {\bf h}^{\rm hf}_i = \sum_{j \; {\rm nn}\; i} {\bf K}_{ij} \langle {\bf S}_j \rangle,
\end{equation}
where ${\bf K}_{ij}$ is a tensor describing the anisotropic transferred
hyperfine exchange from the Cu spin at site $j$ to the nucleus $i$
(see Fig. \ref{fig: NMR_Cs_A_ions}).
According to recent measurements,\cite{NMRpaper} in these tensors, the only significant
off-diagonal entry is $[{\bf K}_{ij}]^{ac}=[{\bf K}_{ij}]^{ca} =\kappa_{ij}$, and,
moreover, $\kappa_{ij}$ takes equal values for the two sites $j$ of the
triangle which are on the same chain.

For magnetic fields along $a$ and $c$, this off-diagonal transferred
exchange is crucial in determining the NMR lineshape.  Let us see how
this occurs.  In either of these cases, we define, as usual the $z$ axis
of spin along the field axis.  Let us then take the $x$ axis of spin
along the other of the two, i.e. for ${\bf\hat H}={\hat a}$, take
$S^x=S^c$, and conversely, if ${\bf\hat H}={\hat c}$, then
$S^x=S^a$.   From Eq.~(\ref{eq:70}), the NMR shift is entirely
determined by the $z$ component of the hyperfine field.  This, in turn,
is given by
\begin{equation}
  \label{eq:54}
  \left[{\bf h}^{\rm hf}_i \right]^z = \sum_{j \; {\rm nn}\; i} \left( \left[{\bf
      K}_{ij}\right]^{zz} \langle S^z_j \rangle + \kappa_{ij} \langle S^x_j \rangle \right).
\end{equation}
In all of the phases predicted for \ccc, the component of the spins
parallel to the field is constant, and equal to the average
magnetization $M$ (this is {\sl not} true in the SDW phase, which is
expected in the ideal 2d case of Section~\ref{sec:ideal-2d-model}).  Therefore the first term in
Eq.~(\ref{eq:54}) gives a constant contribution to the shift, which is
the same for all Cs(A) nuclei.  Thus
\begin{equation}
  \label{eq:73}
  \Delta\nu_i  \propto {\rm const.}+ \sum_{j \; {\rm nn}\; i}
   \kappa_{ij} \langle S^x_j \rangle .
\end{equation}
Using the experimentally determined form of the
hyperfine couplings, and dropping the constant, one has
\begin{eqnarray}
  \label{eq:74}
  && \Delta\nu_{x+\frac{1}{2},y+\frac{1}{2},z}  \\
&& \nonumber (-1)^y \left(\kappa_1 \left[ \langle
    S^x_{x,y,z} \rangle + \langle
    S^x_{x+1,y,z} \rangle \right] + \kappa_2 \langle
    S^x_{x+\frac{1}{2},y+1,z} \rangle \right).
\end{eqnarray}
Here we have absorbed the proportionality constant in the shift into the
definitions of $\kappa_1$ and $\kappa_2$. 
We are now in a position to evaluate the NMR lineshape for the different
magnetic phases.
\begin{figure}[htb]
\begin{center}
\vspace{.2cm}
\includegraphics[width=0.98\columnwidth]{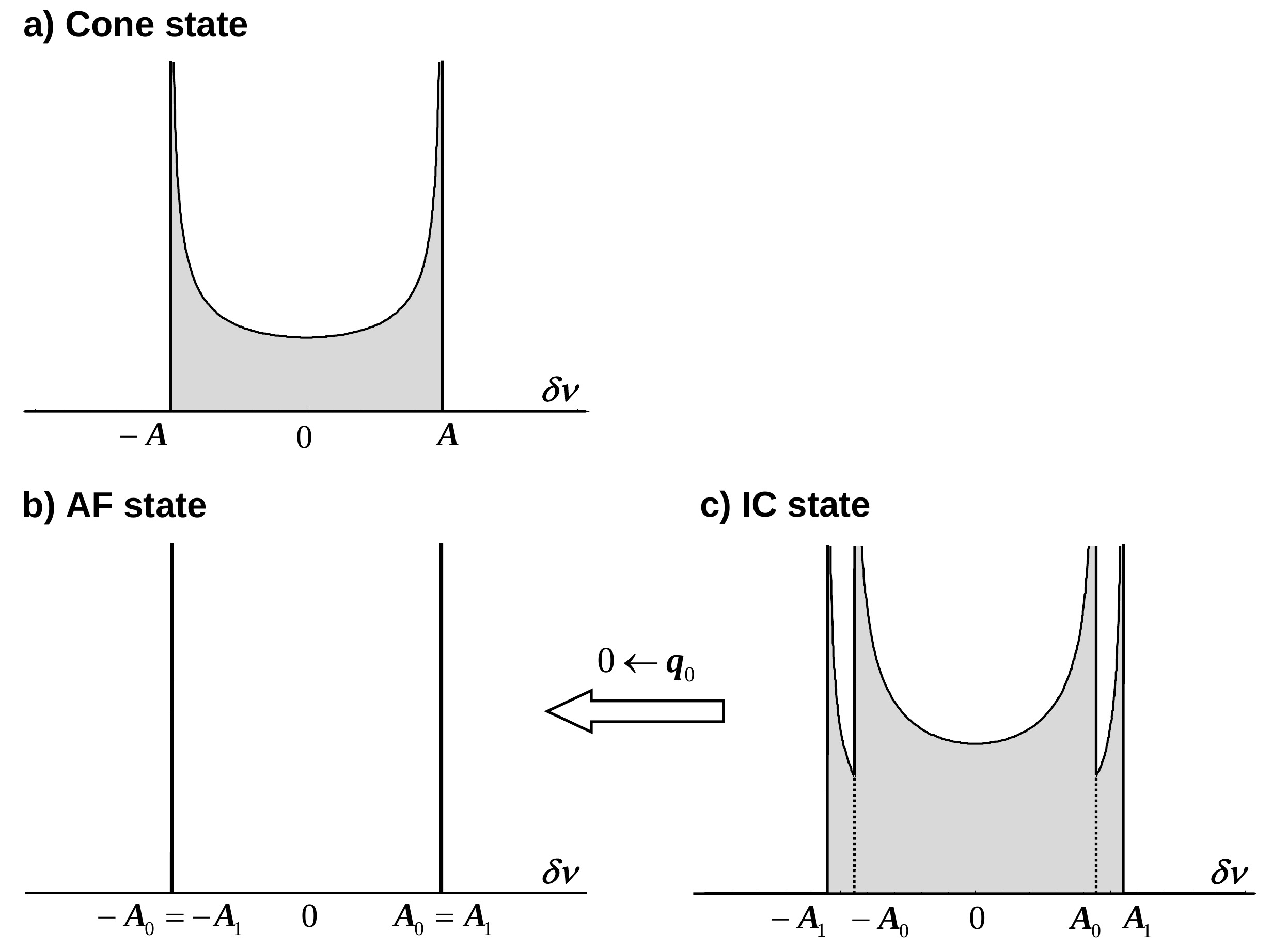}
\caption{Schematic NMR spectra in 
(a) the cone state, (b) the AF state, and (c) the IC state. 
On approaching the commensurate AF phase from the IC phase, i.e., 
$q_0 \to 0$ limit, four peaks merge pairwise as indicated by the arrow.}
\label{fig: NMR_line_shape}
\end{center}
\end{figure} 

\subsubsection{Cone state}
\label{sec:cone-state-2}

In the cone state, we can use Eq.~(\ref{eq:49a}) to evaluate
Eq.~(\ref{eq:74}).  One obtains
\begin{eqnarray}
  \label{eq:75}
   && \Delta\nu_{x+\frac{1}{2},y+\frac{1}{2},z} = \\
   \nonumber &&
 (-1)^y N \left[\kappa_2 - 2 \kappa_1 \sin \tfrac{q_0}{2}\right] \cos [(\pi +
  q_0)(x +1/2)+ \tilde\Theta_z].
\end{eqnarray}
Now the NMR lineshape reflects the {\sl distribution} of shifts,
$p(\Delta\nu) $, over all the Cs(A)
sites.  We may consider this as a sum of distributions of the shifts for
the nuclei associated with each pair of chains, i.e. ranging over $x$
for fixed $y$ and $z$.  Because $q_0$ is incommensurate, the argument of
the cosine above is distributed {\sl uniformly} over the full angular
interval from $0$ to $2\pi$.  Thus the cosine itself is distributed
between $-1$ and $+1$, and we obtain a distribution for the shift, for
fixed $y$ and $z$ with support between $\pm N |\kappa_2 - 2 \kappa_1
\sin \tfrac{q_0}{2}|$:
\begin{equation}
  \label{eq:78}
  p(\Delta \nu) = \frac{1}{\pi} \frac{1}{\sqrt{A^2 -
      (\Delta\nu)^2}}\Theta[A-|\Delta\nu|],
\end{equation}
with
\begin{equation}
  \label{eq:79}
  A = N |\kappa_2 - 2 \kappa_1\sin \tfrac{q_0}{2}|.
\end{equation}
We see that the distribution is in fact independent of $y$ and $z$, so
that the full distribution over all Cs(A) sites is identical to that for
a single pair of chains.  It has two peaks, at the edges of the
distribution, $\delta\nu=\pm A$ as shown in Fig. \ref{fig: NMR_line_shape}(a). 

\subsubsection{AF state}
\label{sec:af-state}

Applying Eq.~(\ref{eq:51}) to Eq.~(\ref{eq:74}), we obtain
\begin{eqnarray}
  \label{eq:76}
  \Delta\nu_{x+\frac{1}{2},y+\frac{1}{2},z} & = &
\sigma (-1)^{y+z}|\psi|\kappa_2 \cos[\pi(x+\sigma y/2)+\Theta]
\nonumber \\
& = & \pm \kappa |\psi|\cos\Theta.
\end{eqnarray}
One expects therefore two sharp peaks in the Cs(A) NMR spectrum,
separated by $2\kappa_2\psi|\cos\Theta|$ (see Fig. \ref{fig: NMR_line_shape}(b)).  Note that $\cos\Theta$ is
generically non-zero, as argued by symmetry in Appendix~\ref{sec:antif-af-state}.

\subsubsection{IC state}
\label{sec:ic-state}

Here we apply Eq.~(\ref{eq:52}) to Eq.~(\ref{eq:74}).  We find that the
NMR shift can be written
\begin{equation}
  \label{eq:77}
   \Delta\nu_{x+\frac{1}{2},y+\frac{1}{2},z} = A_y \cos (q_0 x + \phi_y),
\end{equation}
where
\begin{eqnarray}
  \label{eq:80}
  |A_y| & = & |\psi|\Big[ 4\kappa_1^2 \sin^2 \tfrac{q_0}{2} + \kappa_2^2
 \\
    && 
    -(-1)^y 4 \kappa_1\kappa_2 \sin \tfrac{q_0}{2} \sin (\Theta_0+\Theta_1)\Big]^{1/2},   \nonumber
\end{eqnarray}
and
\begin{eqnarray}
  \label{eq:81}
&&   \tan \phi_y  =  \\ 
&& \nonumber
\left\{ \begin{array}{cc} \frac{\kappa_1\left(\sin(q_0+\Theta_0) -
    \sin\Theta_0 \right)+ \kappa_2 \sin (\tfrac{q_0}{2}  -\Theta_{1})}{\kappa_1\left(\cos(q_0+\Theta_0) -
    \cos\Theta_0 \right)+ \kappa_2 \cos (\tfrac{q_0}{2} -\Theta_{1})} &
\textrm{$y$ even} \\
\frac{\kappa_1\left(\sin(q_0-\Theta_1) +
    \sin\Theta_1 \right)+ \kappa_2 \sin (\tfrac{q_0}{2}  +\Theta_{0})}{\kappa_1\left(\cos(q_0-\Theta_1) -
    \cos\Theta_1 \right)+ \kappa_2 \cos (\tfrac{q_0}{2}  +\Theta_{0})} &
\textrm{$y$ odd}
\end{array} \right. .
\end{eqnarray}
For each $y$, we expect from Eq.~(\ref{eq:77}) a continuum lineshape of
the form of Eq.~(\ref{eq:78}), owing to the incommensurate wavevector
$q_0$.  However, in general, $A_y$ takes two distinct values for even
and odd $y$ (owing to the $(-1)^y$ factor in Eq.~(\ref{eq:80})).  Note
that the prefactor of this term is non-vanishing since
$\sin(\Theta_0+\Theta_1)$ is generally non-zero, as argued in
Appendix~\ref{sec:incommensurate-phase}.   Hence
the full Cs(A) lineshape is expected to be the sum of both
distributions, and hence has {\sl four peaks}, at $\Delta\nu=\pm A_0,
\pm A_1$ as schematically shown in Fig. \ref{fig: NMR_line_shape}(c).  
Notably, these peaks merge pairwise as $q_0\rightarrow 0$,
i.e. on approaching the commensurate AF phase from the IC state.
Precisely such a merging of the peaks has been seen in the NMR
experiments by Takigawa and collaborators \cite{takigawa-poster}.

\subsection{Phase diagrams}
\label{sec:phase-diagrams}

In
Secs.~\ref{sec:field-along-axis},\ref{sec:field-along-b},\ref{sec:field-along-c},
we have determined (most of) the zero temperature phases for the three
major field orientations.  Here we discuss the extension of these
results to $T>0$.  

\subsubsection{Field along $a$ axis}
\label{sec:field-along-a}

This is the simplest case.  At zero temperature, the cone state extends
across the entire field range from zero up to saturation.  We have seen
that it is predominantly controlled by the DM interaction $D=D'_a$,
perturbed somewhat by the interchain exchange $J'$.  We therefore expect
a single phase boundary, $T_{\rm cone}(H)$.  One estimate for this curve
is obtained from CMFT, and is shown in Fig.~\ref{fig:dmTc} (plotted
versus magnetization $M$ rather than field).  One observes that $T_{\rm
  cone}$ at first increases with the applied field for small fields, and
then reaches a maximum, followed by a decrease to zero at the saturation
field.  

These trends can be understood simply as follows.  With increasing
magnetization, the spins become more XY-like, which decreases the
scaling dimension $\Delta_\pm$.  As a consequence, the DM interaction
becomes more relevant with increasing field, enhancing the critical
temperature.  However, on approaching saturation, the magnitude of the
transverse components of the spins, which constitute the cone order,
decrease to zero, and hence suppress the ordering temperature to zero.

As these trends are correctly captured by CMFT, we may perhaps trust the
result for the phase boundary.  However, we note that the {\sl nature}
of the phase transition is somewhat subtle, and probably not properly
described by this approximation.  Neglecting $J''$, which has a very
weak effect upon the cone state (see Eq.~\eqref{eq:66}), the system is
effectively two-dimensional, and as a consequence exhibits strong
effects of thermal fluctuations.  Since the DM terms (e.g. $D_c$) with
DM-vectors perpendicular to $a$ are also negligible here, the Hamiltonan has
approximate XY spin rotation symmetry.  As a consequence, the cone phase
is approximately a {\sl quasi-}long-range-ordered state at $T>0$, and
its thermal transition should be of Kosterlitz-Thouless (KT) type.  Obviously
the CMFT approximation does not describe the KT universality class, and
instead predicts mean-field critical behavior.  
  
It is interesting to verify that, nevertheless, the magnitude of $T_c$
obtained from CMFT agrees with an analysis based on KT theory.  For
simplicity, we will focus on the DM-dominated field range, and neglect
entirely $J'$ and $J''$ for simplicity.  In this case, the system
decouples into 2d triangular $x-y$ planes, consisting of chains
connected by the DM interaction only.  Taking the expectation value using Eq.~(\ref{eq:39}) in
Eq.~(\ref{eq:4}), one obtains the energy 
\begin{equation}
  \label{eq:85}
  H_{2d} =\sum_{y,z}  \int \! dx\, \Big\{
  \frac{v}{2}(\partial_x\vartheta_{y,z})^2 - 2D |N|^2 \cos
  \beta(\vartheta_{y,z}-\vartheta_{y+1,z}) \Big\}.
\end{equation}
In the ordered phase, one may expand the cosine, and take the continuum
limit (in $y$) for fields $\vartheta_{y,z}$ that are slowly-varying in
$y$:
\begin{equation}
  \label{eq:86}
   H_{2d} =\sum_{z}  \int \! dx dy\, \Big\{
  \frac{v}{2}(\partial_x\vartheta_z)^2 + D |N|^2 \beta^2 (\partial_y\vartheta_{z})^2 \Big\}.
\end{equation}
Now, according to KT theory, the critical temperature is proportional to
the geometric mean of the two stiffnesses, i.e. $T_{KT}\sim \sqrt{v
  D}|N|$.  We are neglecting all $O(1)$ prefactors here, as we are only
interested in the scaling behavior.  Now from scaling, or from the CMFT
calculations in Appendix~\ref{sec:cmft-T=0}, one has $|N| \sim
(D/v)^{\Delta_\pm/(2-2\Delta_\pm)} $, from which one obtains finally
$T_{KT} \sim v (D/v)^{1/(2-2\Delta_\pm)}$. 
Precisely the same
scaling is found directly from the CMFT treatment at $T>0$ in the
Appendix~\ref{sec:cmft-dm}. (Please note that the described calculation
corresponds to setting $J'=0$ the coupling $\hat\Gamma^a_{\rm cone}$, see 
\eqref{eq:cmft-dm}. That implies $q_0=0$ which, via second equation in \eqref{eq:cone-system2}, leads to 
the scaling $T_c \sim v (D/v)^{1/(2-2\Delta_\pm)}$.)
  It may appear surprising that the two approaches, which describe
the transition so differently, agree in this respect.  The reason for
the agreement is that the {\sl scale} of $T_c$ is entirely determined by
the scaling properties of the weakly perturbed one-dimensional chains.
Any approximation which respects this scaling (and both the CMFT and the
KT analysis do) will obtain the same order of magnitude answer.
Differences would appear in the prefactor, which, however, is beyond the
scope of the rough KT analysis carried out here.

\subsubsection{Field along $b$ axis}
\label{sec:field-along-b-2}

The situation in a field along $b$ is considerably more complicated.
The analysis in Sec.~\ref{sec:field-along-b} implies {\sl at least} four
phases at $T=0$: the zero field phase ``spiral'' phase, dominated by
$D$, the AF phase, the high-field cone phase, and the saturated phase.
Due to the difficulty of treating the competition between the $D$ term
and magnetic field when the two are comparable, the intervening range
between the AF and zero field phases has not been fully clarified here.
Thermodynamic measurements\cite{TokiwaPRB2006} appear to show a single
transition between the spiral and AF states, and therefore the absence
of any intermediate states.

For the ``high field'' phases (i.e. in the region where the $D$ term is
negligible), we can attempt to apply CMFT to determine the uppermost
phase boundaries, describing the transitions from the ordered to
paramagnetic states.  The key observation is that {\sl both ordered
  phases} (AF and cone) are driven by the same $J''$ interaction.  The
two states are only distinguished by the competing effects of the
weaker (at least in the renormalized sense) $J'$ and $J_2$ interactions. Thus
the upper phase boundary should be approximately continuous across this
field range up to saturation, and not very sensitive to the precise
nature (AF or cone) of the ordered phase it demarcates.   This boundary
should be similar in shape to the $T_{\rm cone}(H)$ discussed above, as
it arises from a term of the same scaling dimension as $D$ in that case,
and suffers the same reduction on approaching saturation.  

The high-field region requires one further phase boundary, between the
AF and cone states.  This should be approximately vertical, but is
expected to bend ``to the left'', as the cone state has higher entropy
than the AF one, and is thus favored with increasing temperature.  This
reasoning is based on higher order effects of other DM interactions that we have
neglected up to now (mentioned in passing in
Sec.~\ref{sec:commensurate-phase}): these sub-dominant terms are
expected to break the $U(1)$ rotational symmetry of the AF state and gap out
its Goldstone modes, while preserving the commensurability of the AF
structure. The cone state, being incommensurate, is expected to not be
affected by these small perturbations and preserve its gapless
excitations.   As a result, we expect the entropy of the cone state to be
greater than that of the AF one, and result in the mentioned bending of
the AF-cone boundary to the left.  The transition between the two states
is first order, and observables such as the ordering wavevector jump at
the critical field.

\subsubsection{Field along $c$ axis}
\label{sec:field-along-c-2}

In this, most complex field orientation, all the phases predicted for
the field along $b$ must appear, {\sl and} in addition the IC state,
taking up some territory between the AF and spiral phases.  Experiments
seem to show\cite{Coldea2001PRL,TokiwaPRB2006,takigawa-poster} even beyond these 5 states,
one or two additional ones in the regime when the $D$ term is comparable
to the Zeeman energy.  At present we have little to say about these
states. The observed linear relation between the ordering momentum
of the intermediate ``S'' state (in the notation of Ref.~\onlinecite{Coldea2001PRL})
is suggestive of an SDW phase, but at the present we do not have a good 
understanding of how the competition between the DM and the Zeeman
terms may bring out the SDW order discussed in Sec.~\ref{sec:phases}.

In the high-field region, we expect, by similar arguments to the
previous section, a rather continuous boundary between the IC, AF, and
cone states and the paramagnetic phase.  The AF-cone boundary should
appear very similar to that for this field orientation as well.  The new
feature introduced here is the IC-AF boundary, which is the location of
the CIT.  Like the cone phase, the IC phase is expected to have more
entropy than the AF state, and hence be stabilized by increasing
temperature.  Experimentally, this boundary bends quite sharply ``to the
right'', in marked contrast to the nearly vertical AF-cone transition
line.  

To understand this, consider the expression for the CIT location,
Eq.~\eqref{eq:26}.  It can be rewritten, using the expressions for
$g_{\rm bq}$, Eq.~\eqref{eq:15}, and for $d_c$ (given in the text following
Eq.~\eqref{eq:21}), as 
\begin{equation}
  \label{eq:82}
  |\psi|^2 \sim \frac{v D_c}{\beta J J'},
\end{equation}
where we drop all $O(1)$ factors.  A simple treatment, which takes into
account some of the one-dimensional fluctuations, but not the
high-dimensional ones, is to simply apply Eq.~\eqref{eq:26}, but
replacing $|\psi|^2$ with its reduced value {\sl at $T=T_{\rm CIT}$}.  In general,
this is difficult to actually calculate analytically, e.g. with CMFT,
but the detailed form is not important to our argument.  Scaling implies
that it can be written
\begin{equation}
  \label{eq:83}
  |\psi|^2(T) = |\psi_0|^2 {\mathcal F}(T/T_c),
\end{equation}
where $T_c$ is the critical temperature at which $\psi$ vanishes, and
the scaling function $\mathcal F$ is smooth and obeys ${\mathcal
  F}(0)=1$ and ${\mathcal F}(1)=0$.  For illustrative purposes, we can
take the simple approximation $\mathcal{F}(t) = 1-t^2$.  Using this
form, one finds
\begin{equation}
  \label{eq:84}
  \frac{T_{\rm CIT}}{T_c} = \sqrt{1- \frac{v D_c}{\beta JJ'|\psi_0|^2}}.
\end{equation}
By construction, the right-hand side vanishes at the zero temperature
CIT, where $T_{\rm CIT}=0$.  Both $\psi_0$ and $\beta$ are rather weak
functions of magnetic field.  However, the velocity $v$ varies
considerably (on the scale of $J$) with field, indeed vanishing as
saturation is approached.  Hence, the right-hand-side increases rather
quickly with field, leading to rapid variation of $T_{\rm CIT}$ with an
approximate square-root form, consistent with experiments.  

It is interesting to note that, in
experiment,\cite{TokiwaPRB2006,takigawa-poster} the AF-IC and AF-cone
boundaries are observed to approach each other very closely with
increasing temperature, leading to an extremely narrow range of
transition directly from the paramagnet to the AF phase. This suggests
some physical mechanism which ``avoids'' this transition.  In fact, one
can argue that,according to Landau theory, a {\sl continuous}
AF-paramagnetic transition is {\sl forbidden} for this field
orientation.  To do so, consider the Landau expansion of the free
energy $F$ in the ``order parameters'' $\psi_{y,z}(x) \equiv \langle
\mathcal{S}_{y,z}^-(x)\rangle$.  Such an expansion, in powers of
$\psi_{y,z}$, is valid near any putative continuous transition.  We
presume $\psi_{y,z}(x)$ to be a slowly-varying function of $x$.  The
Landau expansion has the form $F=F_2+F_4+\cdots$, where 
\begin{eqnarray}
  \label{eq:108}
  F_2 & = & \sum_{y,z}\int\! dx\, \Big\{
  \frac{\bar{v}}{2}|\partial_x\psi_{y,z}|^2 - i \bar{\gamma}_c
  \left(\psi_{y,z}^* \partial_x \psi^{\vphantom*}_{y+1,z} + {\rm c.c.}\right) \nonumber \\
& &  -i\bar{d}_c
(-1)^y \psi_{y,z}^* \partial_x \psi^{\vphantom*}_{y,z} + \bar{\gamma}_2\left(
\psi_{y,z}^* \psi_{y+2,z}^{\vphantom*} + {\rm h.c.}\right)\nonumber \\
& & + \bar{\gamma}''_\pm \left( \psi_{y,z}^* \psi_{y,z+1}^{\vphantom*} +
  {\rm h.c.}\right) + r |\psi_{y,z}|^2\Big\}
\end{eqnarray}
contains quadratic terms in the order parameter, and
\begin{eqnarray}
  \label{eq:109}
  F_4 & = & \sum_{y,z}\int\! dx\, \Big\{ u |\psi_{y,z}|^4 \nonumber \\
  & & +\bar{\gamma}_{\rm bq} \left[ \left(\psi_{y,z}^* \psi_{y+1,z}^{\vphantom*} \right)^2+
  {\rm h.c.}\right] \Big\}
\end{eqnarray}
is quartic.  In Eqs.~(\ref{eq:108},\ref{eq:109}), the couplings with
overlines on them are analogous to the corresponding couplings in the
bosonized Hamiltonian, as can be seen if one assumes
$\psi_{y,z}=|\psi|e^{i\vartheta_{y,z}}$.  They are, however, from the
present point of view, phenomenological coefficients which are at best
proportional to those microscopic couplings.

Let us consider possible continuous transitions from the paramagnetic
state.  In this case, we may assume $|\psi_{y,z}|$ is arbitrarily small,
and thus $F_4$ is a small perturbation to $F_2$.  The transition occurs
on decreasing $r$ from large positive values, at the point at which the
smallest eigenvalue of the quadratic form in $F_2$ vanishes.  Fourier
transforming into the two-site unit cell, $\psi_{y,z}(x) = \int \!
d^3k/(2\pi)^3 \, \psi_{a,k} e^{i k_x x + i k_y y+ i k_z z}$, with $a=0$
for $y$ even and $a=1$ for $y$ odd, one obtains
\begin{eqnarray}
  \label{eq:110}
  F_2 & = & \int \! \frac{d^3k}{(2\pi)^3} \, \psi_{a,k}^*
  \mathcal{F}_{ab}(k) \psi_{b,k},
\end{eqnarray}
where the matrix $\mathcal{F}(k)$ can be decomposed into the identity
matrix, ${\bm I}$, and the Pauli matrices, ${\boldsymbol \sigma}_\mu$,
according to
\begin{equation}
  \label{eq:112}
  \mathcal{F}(k) = \mathcal{F}_0 {\bm I} + \mathcal{F}_x {\boldsymbol
    \sigma}_x + \mathcal{F}_z {\boldsymbol \sigma}_z,
\end{equation}
with
\begin{eqnarray}
  \label{eq:113}
  \mathcal{F}_0 & = &  \frac{\bar{v}}{2}k_x^2 + 2 \bar{\gamma}''_\pm \cos k_z +
    2\bar{\gamma}_2 \cos 2k_y + r, \nonumber \\
    \mathcal{F}_x & = & 2\bar{\gamma}_c  k_x \cos k_y, \nonumber \\
    \mathcal{F}_z & = & \bar{d}_c k_x.
\end{eqnarray}
One immediately concludes that the minimum eigenvalue of $\mathcal{F}$
is
\begin{eqnarray}
  \label{eq:114}
  \mathcal{F}_{\rm min} & = & \mathcal{F}_0 - \sqrt{\mathcal{F}_x^2 +
    \mathcal{F}_z^2} \nonumber \\
   & = & \mathcal{F}_0(k) - |k_x|\sqrt{4\bar{\gamma}_c^2 \cos^2 k_y + \bar{d}_c^2} .
\end{eqnarray}
This, in turn, should be minimized over $k$.  Minimization over $k_x$
and $k_z$ is simple: the minimum occurs at
\begin{equation}
  \label{eq:116}
  |k_x|=\sqrt{4\bar{\gamma}_c^2 \cos^2 k_y + \bar{d}_c^2}/\bar{v}
\end{equation}
and
$k_z=\pi$.  Then
\begin{eqnarray}
  \label{eq:115}
   \mathcal{F}_{\rm min}(k_y) & = &
  r - 2 \bar{\gamma}''_\pm - \frac{\bar{d}_c^2}{2\bar{v}}
   -\frac{2\bar{\gamma}_c^2 \cos^2 k_y}{\bar{v}}  + 2 \bar{\gamma}_2
   \cos 2k_y\nonumber \\
   & = & r- 2 \bar{\gamma}''_\pm - \frac{\bar{d}_c^2+ 2
     \bar{\gamma}_c^2}{2\bar{v}} + (2\bar{\gamma}_2 -
   \frac{\bar{\gamma}_c^2}{\bar{v}}) \cos 2k_y.\nonumber \\
\end{eqnarray}
From here we immediately see that the minimum free energy is obtained
for $k_y=0$ when $\bar{\gamma}_c^2/\bar{v}> 2\bar{\gamma}_2$ and for
$k_y=\pi/2$ otherwise.   In either case, we see from Eq.~(\ref{eq:116})
that $k_x \neq 0$.  The two cases therefore correspond to the cone and
IC states, respectively.  Noteably, the commensurate AF state is {\sl
  absent}.  This is easily understood since it is stabilized by the
biquadratic coupling $g_{\rm bq}$, which in Landau theory corresponds to
the {\sl quartic} interaction $\bar{\gamma}_{\rm bq}$.  Since this
becomes parametrically small relative to the quadratic terms as
$|\psi|\rightarrow 0$, it cannot stabilize a commensurate phase in this
limit.  Thus we conclude that {\sl if} there is a continuous transition
from the paramagnet to an ordered state for this field orientation, it
can only be to the cone or IC phases, and {\sl not} to the AF state.
Conversely, if there is a direct transition between the paramagnet and
AF states, it must be first order.  This latter scenario appears to be
the case in experiment.\cite{TokiwaPRB2006,takigawa-poster}  We note that for
fields along the $b$ axis, where $d_c=0$, a direct continuous transition to the AF
state is possible, since in that case  $k_y=\pi/2$ and $k_x$ vanishes
from Eq.~(\ref{eq:116}).

\subsection{\ccb}
\label{sec:cscubr}

It is instructive to compare the case of \ccc\ extensively reviewed here
with that of its isostructural equivalent \ccb. The latter material is
more two-dimensional, with $J'/J \approx 0.75$ as estimated in
Ref. \onlinecite{ono2005} by comparing the observed momentum of magnetic
Bragg reflections, ${\bm q}_0 = (0, 0.575, 0)$, with the result of the
series expansion calculations in Ref. \onlinecite{weihong1999}.  This
estimate should be taken with some caution, as the theory neglects DM
coupling, which is clearly present in experiment (as witnessed by the
distinct differences between the behavior in a field along $a$ and
perpendicular to it).  Ref.~\onlinecite{ono2005} argued that the
inter-plane coupling in \ccb\ is weaker than in \ccc, since the ratio of
the saturation field to the N\'eel temperature is approximately $1.5$
times larger in \ccb\ than in \ccc, and usually the N\'eel temperature
in quasi-2d systems is expected to be determined by inter-plane
coupling.  In principle, this need not be the case when DM interactions
are strong, but we believe the conclusion is probably correct.  Thus,
compared to \ccc, we surmise that $J'/J$ is increased and $J''/J$ is
decreased.  This behavior is in agreement with the estimate based on
the band-structure calculation of material's microscopic parameters in Ref. 
\onlinecite{valenti2009}.

We believe that reduced three-dimensional coupling is the primary reason
for the observed cascade of phase transitions in \ccb\ subject 
to magnetic field in $b$-$c$ plane.
\cite{ono2004,ono2005,fujii2007,tsujii2007,fortune2009} Particularly striking is
the observation of a robust $M=\frac{1}{3} M_{\rm sat}$ magnetization plateau as well as
a hint of possible second plateau, at or near $2/3$ of the saturation magnetization.

In the quasi-one-dimensional approach adopted here, as discussed in
Section~\ref{sec:plateau}, the existence of the SDW state is a necessary
condition for the plateau. Given that inter-plane exchange $J''$
strongly favors cone state over the SDW one, we understand that \ccb\
with its small inter-plane coupling is indeed a good candidate for the
magnetization plateau.  One must remember that this argument is based on
one-dimensional reasoning, the validity of which in
\ccb\ ($J'/J = 0.75$) is questionable. However, the final outcome of this
-- that the magnetization plateau is stable in the full range of $J'/J
\leq 1$ ratio -- is completely consistent with two recent studies
\cite{alicea2009,Tay_Motrunich} which approached the problem as a
spatially-deformed two-dimensional one.

It is worth noting that abrupt variation of the SDW ordering momentum
$Q$ on approaching the plateau value, described by  Eq.~\eqref{eq:umk14},
can be clearly seen in Figure 9 of Ref.~\onlinecite{ono2005}. While this strong feature
was interpreted there as an indication of a first order transition, our
theory would predict very similar behavior from a continuous
two-dimensional C-IC transition.  

It is interesting to contrast the physical scenario emerging from the
quasi-one-dimensional approach to what is expected based on
semi-classical physics and the more isotropic regime.\cite{alicea2009}\
Notably, the phases immediately bordering the $1/3$ magnetization
plateau in the latter case {\sl are not} collinear SDW states.  Instead,
Ref.~\onlinecite{alicea2009} finds commensurate planar or non-coplanar
incommensurate distorted umbrella states.  These states are connected
to the uud plateau state by continuous phase transitions which however
can be driven first order by residual DM
interactions,\cite{alicea2009} which generally allow for cubic terms in free energy expansion.
NMR measurements\cite{fujii2007} find that the states below and above the
plateau are incommensurate, but cannot distinguish SDW from distorted
umbrella states.  These experiments and others\cite{tsujii2007} also
find some hysteresis at the plateau edges, which suggests first order
transitions there.  As we have discussed, one expects second order
transitions in the SDW case, so this probably suggests that SDW state
does not occur in \ccb.  
This is also supported by the neutron
scattering experiments,\cite{ono2004} which observe a smooth evolution
of the scattering intensity from zero field up to the plateau edge.

Despite the evident absence of SDW physics in \ccb, it is still
interesting to consider the predictions of our theory for the plateaux
themselves.  Apart from the persistence of the $1/3$ plateau to small
$J'/J$, the most striking outcome of our theory is probably that the
second ``strongest'' candidate plateau is {\sl not} at $2/3$ of
saturation but at $3/5$ of it.  This feature should be taken as another
definite prediction of our work.

\section{Discussion}
\label{sec:conclusions}

\subsection{Resume}
\label{sec:resume}

In this paper, we have presented a fairly thorough analysis of the low
temperature phases of \ccc, obtained from a quasi-one-dimensional
approach.  The results explain most of the specific heat, magnetization,
NMR, and neutron data available.  Several aspects are particularly
remarkable.  First, contrary to the popular view of this material as an
``anisotropic triangular lattice antiferromagnet'', for magnetic fields
within the XY plane, the strongest two-dimensional ground state
correlations are within the $a$-$b$ planes, {\sl perpendicular} to the
nominal triangular ($b$-$c$) layers!  Second, in establishing the phase
diagram, we have argued for the critical importance of {\sl four}
different very weak interactions ($D$, $J''$, $D_c$ and $J_2$), only two
of which have been generally recognized ($D$ and $J''$) in prior work.
It is remarkable that interactions of a magnitude of only a few percent
of the largest exchange can induce entirely new phases.  Finally, we
have discovered an heretofore unnoticed commensurate-incommensurate
transition in this material, and located its telltale signature in NMR
experiments.

\subsection{Relation to previous work}
\label{sec:relat-prev-work}

The subject of quantum antiferromagnetism on the triangular lattice is
long and storied.  Here, we will review various aspects of the problem
discussed in the literature which relate to this paper.  
Some of the earliest work\cite{bocquet2001a,bocquet2001} applied the
random phase approximation (RPA), using bosonization results for
one-dimensional Heisenberg chains, to study the susceptibility in the
paramagnetic phase, and estimate critical temperatures. Indeed, the RPA
is equivalent to the CMFT used here, as far as
predictions of the critical temperature are concerned, provided the same
interactions are taken into account.  At a more general level, this
early work correctly emphasized the importance of the one-dimensional
regime.  However, the analysis here (and in Ref.~\onlinecite{Starykh2007}) is
much more complete in a number of significant ways.  It treats the
ordered phases below $T_c$, takes full account of anisotropic DM
couplings, and includes fluctuation-generated interactions which are
ignored within RPA.  These effects rather dramatically alter the phase
diagram of \ccc\ from the expectations of
Refs.~\onlinecite{bocquet2001} and~\onlinecite{bocquet2001a}. 

Much of the theoretical work motivated by \ccc\ focused on the inelastic
neutron structure factor, addressing experiments\cite{Coldea2003PRB}
which observed very broad lineshapes and extracted dispersion relations
for putative ``magnon'' peaks.  Several groups applied spin-wave
theory\cite{PhysRevB.72.134429,PhysRevB.73.184403,merino1999heisenberg,PhysRevB.60.2987}\
to study the ground state (staggered) magnetization and the structure
factor, including higher order corrections in $1/S$.  The low energy
dispersion of the zero field magnon peak is reasonably well reproduced
by this approach, while higher energy portions are not.  To fit them,
requires ``renormalizing'' the exchange couplings by hand, in a manner
inconsistent with other measurements (e.g. at high fields).  Moreover,
the large continuum scattering is not obtained in this approach.
Another series of works study the excitation spectrum of ``magnons''
using series expansions.\cite{PhysRevB.75.174447,PhysRevB.74.224420,PhysRevLett.96.057201,PhysRevB.71.134422}
In our opinion, because the ground state {\sl is} ordered, and the
series are constructed from such a starting point, they are fairly
reliable in determining the energies of magnon-type excitations of the
system.  (Although they do miss important finite lifetime effects which
can be quite large in non-collinear spin configurations.\cite{chernyshev2009})
Indeed, the results compare well to the dispersion of the peaks
of intensity in experiment.\cite{PhysRevB.75.174447}\ This method
provides a useful computational tool, especially helpful in estimating
exchange couplings.  However, it does not elucidate the {\sl mechanism}
of magnetic ordering or provide a full description of {\sl all} the
excitations.  Thus it is much less useful if the ground state is not
known (as in much of the non-zero field experiments), and it does not
address the dominant continuum portion of the experimental spectra.  
Several theories approached the excitation spectrum from more exotic
perspectives,\cite{IsakovPRB2005PRB,PhysRevB.73.174430,AliceaPRL2005,PhysRevB.72.064407,PhysRevLett.92.157003}
postulating proximity to quantum spin liquid phases and/or quantum
critical points.  We believe there is little support for such proximate
exotic phases from experiments on \ccc.  Instead, the most compelling
explanation for the neutron experiments comes from a
theory\cite{kohno2009dps,PhysRevLett.103.197203,kohno07:_spinon_and_tripl_in_spatial}
in which the excited states are constructed from superposition of a
small number of elementary ``spinon'' excitations of the individual
Heisenberg chains.  This approach  quantitatively and qualitatively
explains the main features of experiment, with no adjustable
parameters.  Its success is a strong argument in favor of the
quasi-one-dimensional approach adopted here.  

Several works address the ground states of spatially anisotropic
triangular antiferromagnets, and \ccc\ in particular.  In zero magnetic
field, the ideal problem (discussed here in
Sec.~\ref{sec:ideal-2d-model}) has been heavily
studied.\cite{weihong1999,Heidarian_Sorella_PRB2009,chung2001large,DN_Sheng_PRB2006,
bishop2009,Pardini_Singh_PRB2008,Starykh2007}
Many approaches find simply that, in the quasi-1d limit, the
correlations between chains are extremely weak, and either regard this
small $J'/J$ region as a ``spin liquid'' or are inconclusive as to the
actual ground
state.\cite{weihong1999,Heidarian_Sorella_PRB2009,DN_Sheng_PRB2006}  The
most recent series expansion calculations of
Ref.~\onlinecite{Pardini_Singh_PRB2008} favor a spiral state, but do not make a
definitive conclusion.  The approach described here, applied to this
case in Ref.~\onlinecite{Starykh2007}, predicts definitively a collinear ground
state, arising from a rather subtle fourth order fluctuation effect.
Very recently, a numerical coupled cluster method \cite{bishop2009} also obtained this
state.  Unfortunately, because of the very weak fourth order nature of
the stabilization of this phase, we expect this to be a somewhat
academic result.  The DM interactions in \ccc\ (and likely in many other
anisotropic triangular systems) completely overwhelm the fluctuation
effect and result in a spiral state, as decribed in
Sec.~\ref{sec:ideal-2d-model}.  

In a non-zero applied field, there has been less
effort.\cite{veillette2005incomm,veillette2006commensurate,Starykh2007,PhysRevB.73.174430}
Spin wave theory, applied to the ``standard'' model of \ccc, has
considerable success in reproducing many of the features of the ground
state phase diagram.\cite{veillette2006commensurate,veillette2005incomm}
It does not, however, explain the broad regions of AF and IC phases
appearing for fields in the $b$-$c$ plane.  An explanation of the region
corresponding to the AF phase found here was given in
Ref.~\onlinecite{veillette2006commensurate}, based on a dilute spin-flip
approximation.  However, the state obtained there differs from our AF
state.  It is a non-coplanar commensurate state, with the spin
components transverse to the field lying in orthogonal directions on
neighboring chains.  Such a state would obtain for {\sl negative}
biquadratic coupling $g_{\rm bq} < 0$ in \eqref{eq:14}), and would
probably have distinct signatures in NMR measurements.  In our opinion,
the sign $g_{\rm bq}>0$, found here, is rather more natural, and more
consistent with the usual expectation that fluctuations (``order by disorder'') favor more
collinear states.\cite{Henley_JAP}   The IC phase and CIT discussed here are entirely new,
and could not possibly have been obtained in previous works, all of
which assume rotational symmetry of the exchange interactions in the
$b$-$c$ plane.  As a general point, it is not too surprising that
spin-wave based approximations can capture most of the phases in an
applied field.  This is because we have found that these have
``classical'' order parameters (non-zero $\langle {\mathcal
  S}^\pm_{y,z}(x)\rangle$), and are thus adiabatically connected to
mean-field states.  Such approaches may, however, have large
quantitative errors, and furthermore, may miss states where
fluctuation-induced interactions are important (such as the AF state).
The problems with spin-wave theory are most extreme actually in the
ideal model, which we have found displays a wide range of SDW state,
which is entirely non-classical and difficult to obtain from spin-waves.
The SDW state is very naturally related to magnetization plateaus, which
have been widely discussed, mostly in the spatially isotropic regime.
This is discussed in detail in Sec.~\ref{sec:cscubr}.

\subsection{Open Questions and Parting Remarks}
\label{sec:parting-remarks}

Our study has resolved most of the main questions regarding the phase
diagram of \ccc.  However, there are still some smaller details which
remain to be understood.  We have not addressed the regime of magnetic
fields $h$ of order $D$, for fields in the triangular plane.  There, the
ground state should involve a non-trivial balance between $D$ and
inter-chain exchanges.  Evidently, this gives rise to additional phases
for fields along the $c$ axis.  There is also a discrepancy between the
measured incommensurability ($q_0$) in the high-field cone state for
this field orientation\cite{veillette2005incomm} and theoretical
expectations.  To check whether this discrepancy might be explained by
the additional DM interactions considered in this paper, we evaluated in
Appendix~\ref{sec:swa} the single-magnon spectrum in the fully-polarized
state, including these interactions.  We have not found any set of
parameters which appear consistent with the measured incommensurability,
which is {\sl reduced} compared to the expected one.  Indeed, the
theoretical result in Eq.~(\ref{eq:90}) shows that DM interactions only
{\sl increase} the incommensurability, making the problem worse.  This
might be a point that warrants more extensive experimental
investigation.

While we have focused on the application of our methods to \ccc\ (and to
a lesser extent, \ccb), the analysis can be applied to other quasi-1d
materials.  Indeed, it appears that the commensurate AF state described
here has been observed\cite{private-coldea} in the another triangular
lattice material, Cs$_2$CoCl$_4$, in Ref.\onlinecite{coldeaXY}. This
material is a spin-$1/2$ XY-like antiferromagnet forming a spatially
anisotropic triangular lattice. In contrast with the na\"ive expectation
of incommensurate spiral order along the chains,
experiment\cite{coldeaXY} finds commensurate antiferromagnetic ordering
in the absence of an applied magnetic field. This finding matches nicely
our description here, as in Cs$_2$CoCl$_4$ pronounced easy-plane
anisotropy (the estimate\cite{coldeaXY} is $J^z/J^{x,y} = 0.25$, where
$J^a$ represents in-chain exchange between $S^a$ components of the
nearest spins) plays a role quite similar to the external magnetic
field: it enhances XY spin correlations at the expense of Z ones (which,
however, remain commensurate with the lattice).  In fact, the behavior
of the compactification radius as function of XY
anisotropy\cite{affleck1999field} is not very different from that in a
magnetic field, discussed in Appendix~\ref{sec table}: $2\pi R^2 = 1 -
\arccos[J^z/J^{x,y}]/\pi$. Thus the results of
Sec.~\ref{sec:field-along-b} provide a natural theoretical explanation of
the observed commensurate AF phase in Cs$_2$CoCl$_4$.

To facilitate further application of our methods such as this one, we
have described them here in sufficient detail that they could be readily
applied to other problems.  We hope that the level of detail exposed
here serves to amplify the tremendous power of the quasi-one-dimensional
approach, which allows real quantitative contact betweem the microscopic
spin Hamiltonian and universal long-wavelength physics, while including
at the same time strong fluctuations and frustration.


 


\section*{Acknowledgments}
We thank J. Alicea, A. Chubukov, R. Coldea, V. Mitrovic, 
O. Motrunich, M. Takigawa 
and Y. Takano for stimulating discussions.  Much of this work was
carried out using resources provided by the KITP through NSF grant
PHY05-51164.  OAS is supported by the 
National Science Foundation through grant DMR-0808842.
LB was supported by the Packard Foundation, and the
National Science Foundation through grant DMR-0804564.  HK is partly
supported by the JSPS Postdoctoral Fellowships for Research Abroad.

\begin{figure}
\begin{center}
\vspace{.5cm}
\hspace{-.0cm}\includegraphics[width=0.9\columnwidth]{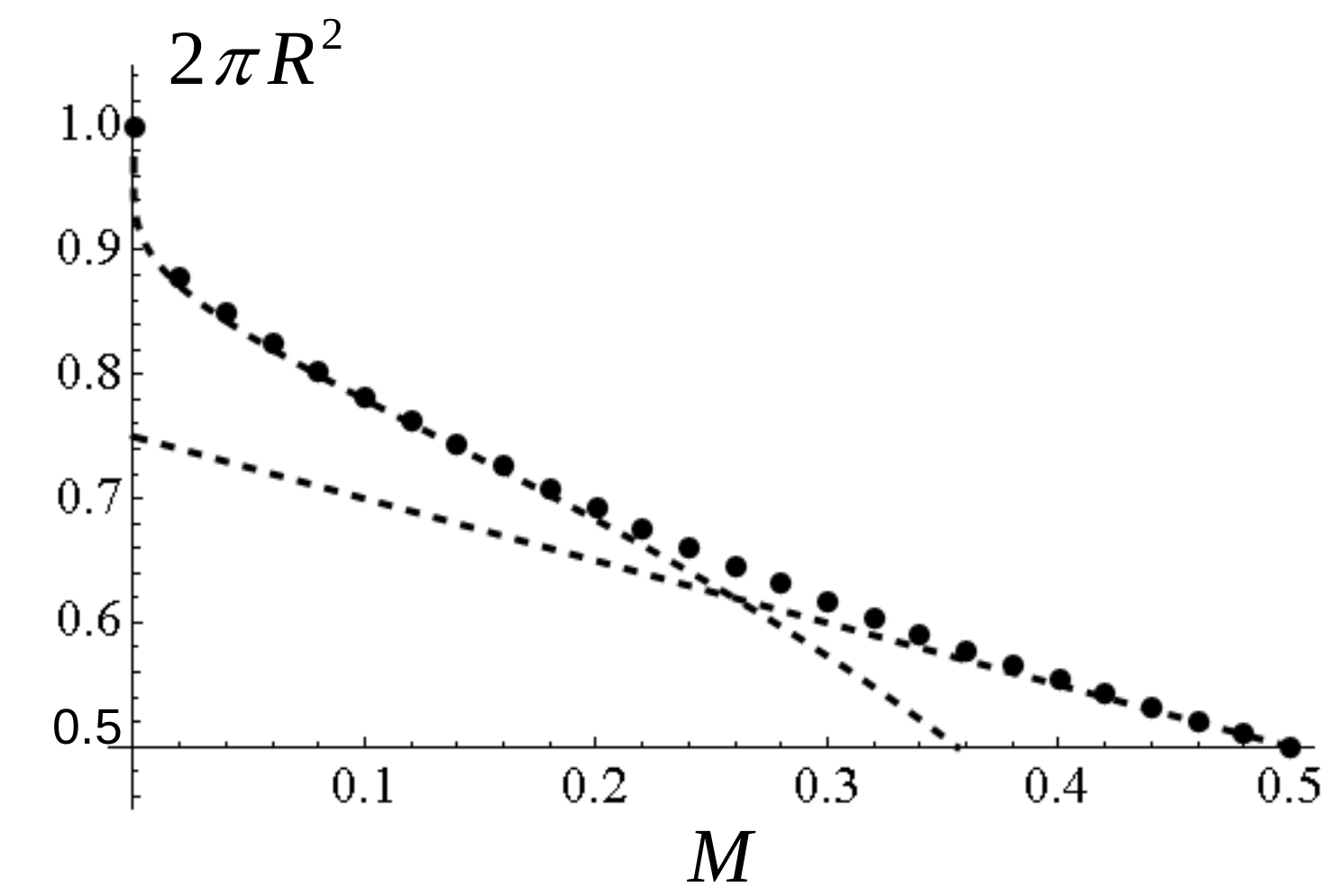}
\vspace{.0cm}
\caption{Parameter $2\pi R^2$ as a function of the magnetization $M$. Numerical solutions in 
Ref. \onlinecite{Essler-Hikihara-Furusaki} shown by the dots are compared with 
asymptotic solutions Eq. (\ref{eq: sat-field}) and Eq. (\ref{eq: zero-field}).}
\label{fig:Beta_curve}
\end{center}
\end{figure}

\appendix
\begin{widetext}
\section{Notations and parameters}
\label{sec table}
Throughout the paper the following conventions are used: coupling constants of quantum 
Hamiltonians written in terms of spin densities ${\cal S}^{z,\pm}$ are denoted as $\gamma$.
When these terms are expressed in terms of bosonic fields $\phi$ and $\theta$, the corresponding
couplings change into $\tilde{\gamma}$. Coupling constants of various  interaction terms
of the effective two- and three-dimensional Hamiltonians, expressed in terms of the {\em classical} phase
$\vartheta$, are denoted as $g$.

Table \ref{tab: notation} contains a list of notations in which several perturbations are summarized 
along with their scaling dimensions. The scaling dimensions are functions of $2\pi R^2$. 
\begin{table}[h]
\begin{center}
\vspace{0.1in}
\begin{tabular}{||c||c|c|} \hline\hline
Hydrodynamic rep. & Bosonized rep. & Scaling dimension \\
\hline\hline
$\vectorize{\gamma_{\rm sdw}({\cal S}^z_{y,z;\pi-2\delta} {\cal S}^z_{y+1,z;\pi+2\delta}+{\rm h.c.})}{\gamma_{\rm sdw}=2J'\sin \delta}$ & $\vectorize{{\tilde \gamma}_{\rm sdw}\cos [\frac{2\pi}{\beta} (\phi_{y,z}-\phi_{y+1,z})]}{{\tilde \gamma}_{\rm sdw}=J'A^2_1 \sin \delta}$ & $\frac{1}{2\pi R^2}$ \\
\hline
$\vectorize{-i\gamma_{\rm cone}{\cal S}^+_{y,z;\pi}\partial_x {\cal S}^-_{y+1,z;\pi}+{\rm h.c.}}{\gamma_{\rm cone}=J'/2}$ & 
$\vectorize{-{\tilde \gamma}_{\rm cone}(\partial_x \theta_{y,z}+\partial_x \theta_{y+1,z}) \cos[\beta (\theta_{y,z}-\theta_{y+1,z})]}
{{\tilde \gamma}_{\rm cone}=J'A^2_3 \beta/2}$ & $2\pi R^2+1$ \\
\hline
$\vectorize{\gamma''_{\pm}({\cal S}^+_{y,z;\pi}{\cal S}^-_{y,z+1;\pi}+{\rm h.c.})}{\gamma''_{\pm}=J''/2}$ & 
$\vectorize{{\tilde \gamma}''_{\pm} \cos[\beta (\theta_{y,z}-\theta_{y,z+1})]}{{\tilde \gamma}''_{\pm}=A^2_3 J''}$ & $2\pi R^2$ \\
\hline
$\vectorize{\gamma_2 {\cal S}^+_{y,z;\pi} {\cal S}^-_{y+2,z;\pi}+{\rm h.c.}}{\gamma_2=J_2/2}$ &
$\vectorize{{\tilde \gamma}_2 \cos[\beta (\theta_{y,z}-\theta_{y+2,z})]}{{\tilde \gamma}_2=A^2_3 J_2}$ 
& $2\pi R^2$ \\
\hline
$\vectorize{\rm biquadratic ~fluctuation-generated}{\rm coupling}$ & $\vectorize{-{\tilde \gamma}_{\rm bq}\cos[2 \beta (\theta_{y,z}-\theta_{y+1,z})]}{g_{\rm bq} = {\tilde \gamma}_{\rm bq}(\ell'')\sim (J')^2 |\psi|^4/v}$ & $8\pi R^2$ \\
\hline
$D_c(-1)^y {\cal J}^z_{y,z}$ & $\vectorize{d_c(-1)^y \partial_x \theta_{y,z}}{d_c=vD_c/(\beta J)}$ & $1$ \\
\hline \hline
\end{tabular}
\caption{List of notations.} \label{tab: notation} 
\end{center}
\end{table}
\end{widetext}
The parameter $2\pi R^2$ as a function of the magnetization $M$ is obtained 
by solving the Bethe ansatz integral equations.\cite{Bogoliubov-Izergin-Korepin, Qin-Fabrizio, Cabra-Honecker-Pujol}  
Figure \ref{fig:Beta_curve} shows the numerical data obtained from $\beta=2 \pi R$ in Ref. \onlinecite{Essler-Hikihara-Furusaki}.
Near the saturation magnetization, $M\sim 1/2$, one can solve the integral equation analytically and obtain
\begin{equation}
\label{eq: sat-field}
2 \pi R^2 = \frac{3}{4}-\frac{M}{2}. 
\end{equation}
In the opposite limit of zero magnetization, $M\sim 0$, $2\pi R^2$ is well fitted by the following function:
\begin{equation}
\label{eq: zero-field}
2\pi R^2=1-\frac{1}{2 \ln(M_0/M)},
\end{equation}
where $M_0 = \sqrt{8/(\pi e)}$. An abrupt, inverse-log, deviation from the $SU(2)$ value is due
to the marginally irrelevant current-current interaction typical for the spin-1/2 Heisenberg chain.\cite {affleck1999field}

The relation between the magnetization $M$ and the magnetic field $h$
has been discussed previously in Ref.\onlinecite{Starykh2007} and we
briefly describe the result here. As discussed below \eqref{eq:H1dec},
the interchain interaction $J'$ increases the energy of the system of coupled
chains by $2 J' M^2$ which results in suppression of the {\em
  two-dimensional} $M(h)$ curve with respect to the one-dimensional
$M_1(h)$ curve for a single spin chain, at fixed external field
$h$. Using the relation $h = -\frac{\partial E}{\partial M}$, we observe
that the field $h$ naturally decomposes into a sum of
``one-dimensional'' field $h_1 = -\frac{\partial E_1}{\partial M}$ for a
single magnetized chain with energy $E_1(M)$ and the inter-chain
contribution $- 4 J' M$. As a result, we arrive at a self-consistent
equation where magnetization $M(h)$ of the system of coupled chains is
approximated by that of the single chain, but evaluated at a shifted
field $h - 4 J' M$:
\begin{equation}
M(h) = M_1(h-4 J' M).
\label{eq:Mh}
\end{equation}
This equation is easily solved numerically using data of Ref. \onlinecite{Essler-Hikihara-Furusaki}
and results in $M(h)$ curve going below $M_1(h)$ for all $h$. An essentially identical result is
obtained if one uses the interpolating formula $M_1(h) = \pi^{-1} \arcsin[1/(1 - \pi/2 + \pi/h)]$,
suggested in Ref.\onlinecite{TokiwaPRB2006}. This approximation predicts saturation field
$h_{\rm sat} = 2J + 2J'$ which is very close to the exact 2d result 
$h_{\rm sat}^{\rm exact} = 2 J + 2 J' + (J')^2/(2 J)$. It is also worth noting that while $M_1$
approaches saturation with an infinite slope, $\partial M_1/\partial h \sim (2 - h)^{-1/2}$, the
two-dimensional curve is characterized by the finite slope $1/(4 J')$, so that
$M(h \approx h_{\rm sat}) = 1/2 + (h - h_{\rm sat})/(4 J')$.
\begin{figure}
\begin{center}
\vspace{.5cm}
\hspace{-.0cm}\includegraphics[width=0.9\columnwidth]{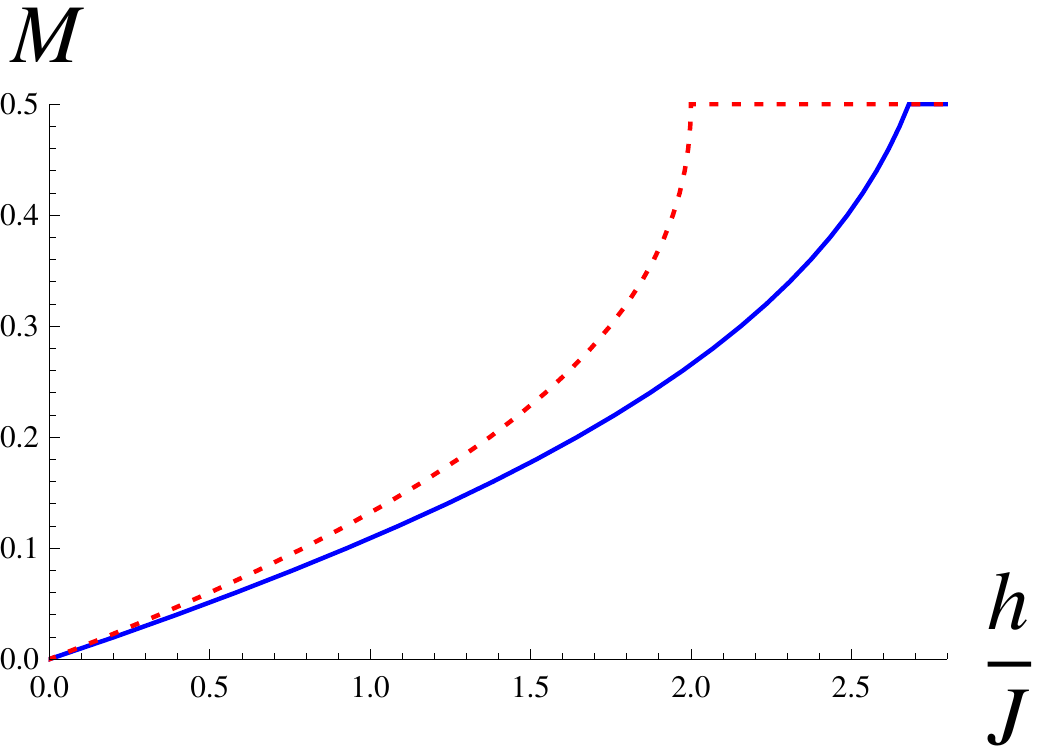}
\vspace{.0cm}
\caption{(Color online) Magnetization $M$ versus magnetic field $h$. Dashed (red) curve
shows magnetization of a single Heisenberg chain, based on the data in 
Ref. \onlinecite{Essler-Hikihara-Furusaki}. Solid (blue) line shows mean-field result \eqref{eq:Mh}.}
\label{fig:Mh}
\end{center}
\end{figure}

\section{Symmetry analysis}
\label{sec:dmv}

Here we consider the full symmetries of \ccc, and some of their
consequences.  Most importantly, we derive the possible DM vectors of
the on-chain and diagonal bonds. The direction of the DM vectors are
constrained by the space group symmetry in the crystal.  In the ideal
triangular lattice, the DM vectors must be perpendicular to the plane
and hence is parallel to the $a$ axis. However, this is not true in the real
crystal due to the lower symmetry.

\subsection{Crystal structure and symmetry generators}
\label{sec:cryst-struct-symm}

\begin{figure}[h]
\begin{center}
\vspace{.5cm}
\includegraphics[width=\columnwidth]{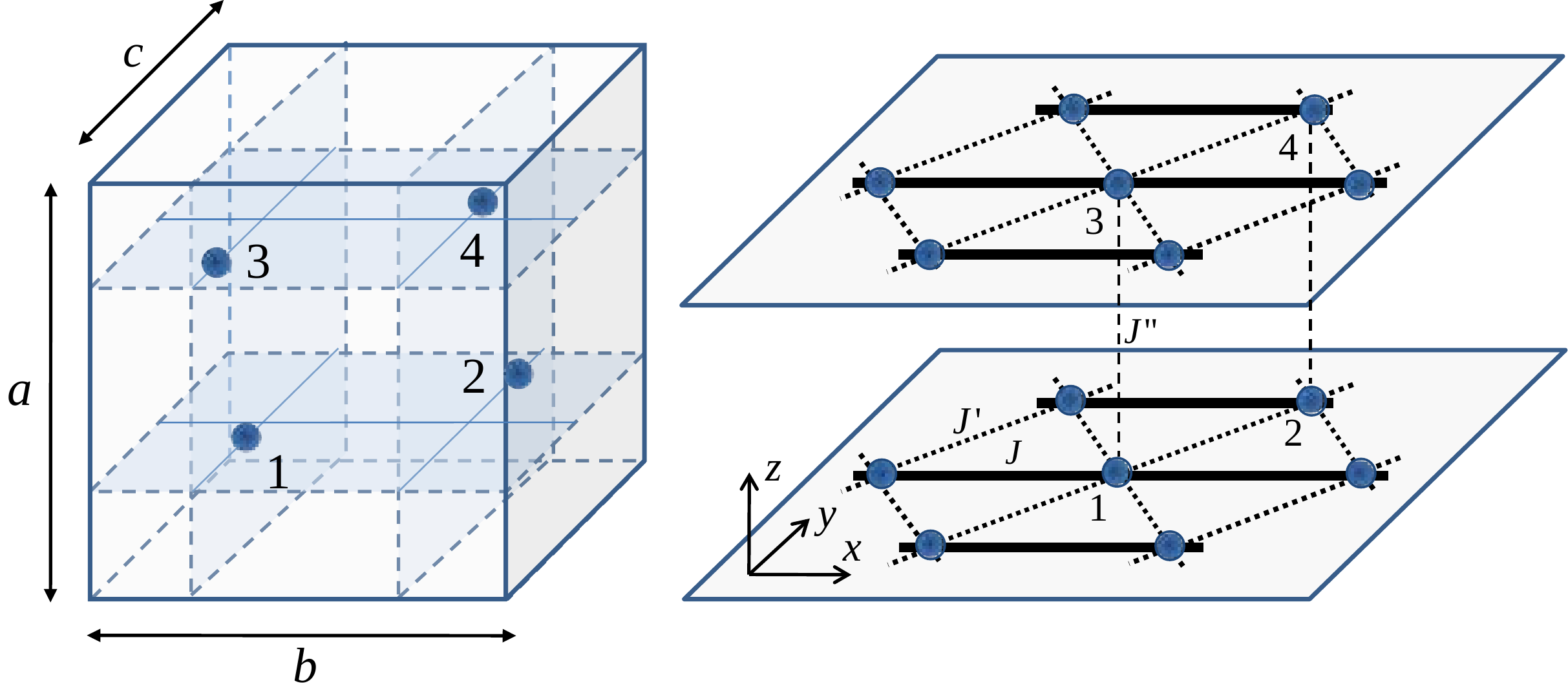}
\caption{Left: Four independent Cu spins (labeled by 1,2,3, and 4) in the unit cell of \ccc. The spins 1(2) and 3(4) lie in the same $a$-$c$ plane. 
Right: Layered-triangular lattice formed by magnetic sites. Thick, dotted, and broken lines indicate $J$, $J'$, and $J''$ exchange couplings, respectively.}
\label{fig: Pnma_crystal}
\end{center}
\end{figure}

\ccc\ has an orthorhombic crystal structure with space group
$Pnma$.\cite{Bailleul} The lattice parameters are $a=9.65$\AA,
$b=7.48$\AA, and $c=12.35$\AA~at 0.3 K.  The unit cell contains four
independent Cu$^{2+}$ ions as shown in Fig. \ref{fig: Pnma_crystal}.
The locations of the ions are given by ${\vs R}+{\vs \delta}_\alpha$ ($\alpha=1,2,3$, and $4$), where
\begin{equation}
{\vs R}=l {\hat a}+m {\hat b}+n{\hat c}
\label{crystal_coordinate}
\end{equation}
denotes the location of the unit cell and 
\begin{eqnarray}
{\vs \delta}_1 &=& z_0{\hat a}+\frac{1}{4}{\hat b}+y_0{\hat c}, \nonumber \\
{\vs \delta}_2 &=& \left( \frac{1}{2}-z_0 \right){\hat a}+\frac{3}{4}{\hat b}+\left( \frac{1}{2}+y_0 \right){\hat c}, \nonumber \\
{\vs \delta}_3 &=& \left(\frac{1}{2}+z_0 \right){\hat a}+\frac{1}{4}{\hat b}+\left(\frac{1}{2}-y_0 \right){\hat c}, \nonumber \\
{\vs \delta}_4 &=& \left( 1-z_0 \right){\hat a}+\frac{3}{4}{\hat b}+ 
\left(1-y_0 \right){\hat c}, \nonumber
\end{eqnarray}
with $z_0=0.23$ and $y_0=0.42$. 
In the simplified notation used in Eq. (\ref{standard_hamiltonian}), the spins 1,2,3, and 4 correspond to ${\bm S}_{x,y,z}$, ${\bm S}_{x+\frac{1}{2},y+1,z}$, ${\bm S}_{x, y, z+1}$, and ${\bm S}_{x+\frac{1}{2},y+1,z+1}$, respectively. 
Note that $m$ in Eq. (\ref{crystal_coordinate}) is the coordinate along the chains, i.e., $x$. 
For fixed $l$, 1 and 2 spins constitute one triangular layer while 3 and 4 another layer. On the other hand, 1 and 3 spins lie in the same $a$-$c$ plane which is a mirror plane through the midpoints of the on-chain bonds (see Fig. \ref{fig: Pnma_crystal}). The same thing holds for 2 and 4 spins. Therefore, according to Moriya's rule, the DM vector on the chain bonds must lie in the $a$-$c$ plane.\cite{Moriya_rule} In contrast to this, there is no symmetry constraint on the DM vectors on diagonal bonds. 

\begin{figure}[h]
\begin{center}
\vspace{.5cm}
\includegraphics[width=\columnwidth]{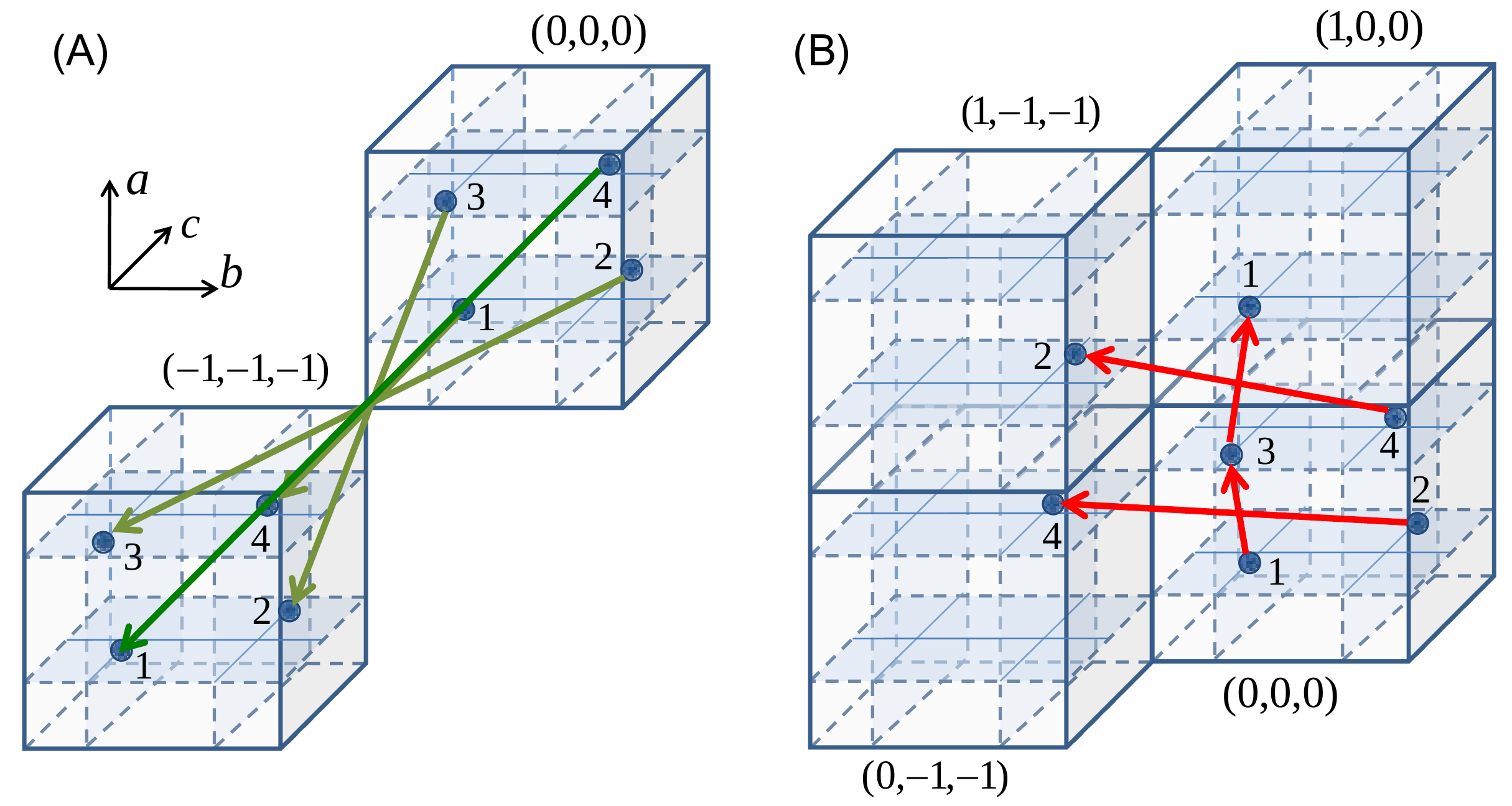}
\caption{Symmetry operations $A$ and $B$. The locations of unit cells are indicated by $(l,m,n)$.}
\label{fig: Pnma_sym}
\end{center}
\end{figure}

We shall next determine the pattern of relative signs of the DM vectors, which requires a more elaborate argument of symmetry. 
The $Pnma$ space group has three kinds of symmetry operations, $A$, $B$, and reflection apart from Bravais lattice translations. We consider the transformation property of spins under $A$ and $B$. Let us denote by ${\bm S}_\alpha (l,m,n)$ the spin at the position ${\vs R}+ \vs {\delta}_\alpha$. 
The spins are transformed under $A$ as follows:
\begin{equation}
A: \left\{ \begin{array}{l}
S^\mu_{1} (l,m,n) \leftrightarrow S^\mu_{4} (-l-1,-m-1,-n-1) \\
S^\mu_{2} (l,m,n) \leftrightarrow S^\mu_{3} (-l-1,-m-1,-n-1)
\end{array} \right. , 
\label{eq: sym_A}
\end{equation}
where $\mu=a,b,c$. 
Since $A$ is the inversion operation as shown in Fig. \ref{fig: Pnma_sym}, the spins do not change sign.  
The symmetry operation $B$, graphically shown in Fig. \ref{fig:
  Pnma_sym}, corresponds to a $\pi$ rotation about an axis parallel to
$a$. Under it, two of three components of the spin change
sign. Therefore, one obtains
\begin{equation}
B: \left\{\begin{array}{l}
S^\mu_{1} (l,m,n) \leftrightarrow \gamma_\mu S^\mu_{3} (l,-m,-n) \\
S^\mu_{2} (l,m,n) \leftrightarrow \gamma_\mu S^\mu_{4} (l,-m-1,-n-1)
\end{array}\right. ,
\label{eq: sym_B}
\end{equation}
where $\gamma_b=\gamma_c=-1$ and $\gamma_a=1$. 

\subsection{DM vectors}
\label{sec:dm-vectors}

We can determine the relative signs of the DM vectors using Eqs. (\ref{eq: sym_A}) and (\ref{eq: sym_B}). 
Let us first consider components of the DM interactions on the on-chain bonds:
\begin{equation}
\sum^4_{\alpha=1} (D_{\alpha,a} {\hat a} + D_{\alpha,c} {\hat c}) \cdot {\bm S}_\alpha (l,m,n) \times {\bm S}_\alpha (l,m+1,n)
\end{equation}
Applying $A$ and translations, we find 
\begin{eqnarray}
D_{1,a} = -D_{4,a},~~~~~ D_{2,a}=-D_{3,a} \\
D_{1,c} = -D_{4,c},~~~~~ D_{2,c}=-D_{4,c}
\end{eqnarray} 
while applying $B$ and translations, we find 
\begin{equation}
D_{1,a}=-D_{3,a},~~~~~D_{1,c}=D_{3,c}.
\end{equation}  
These six relations determine the relative signs of $D_a$ and $D_c$ as in Eq. (\ref{DM_chain}).  
Let us apply the same technique to analyze the DM vectors on the diagonal bonds. 
Since each site lies in the mirror plane, we can assume 
the coupling of the following form:
\begin{eqnarray}
&& {\bm D}^+_{12} \cdot {\bm S}_1(l,m,n) \times {\bm S}_2 (l,m,n) \nonumber \\
&+& {\bm D}^-_{12} \cdot {\bm S}_1(l,m,n) \times {\bm S}_2(l, m-1,n) \nonumber \\
&+& {\bm D}^+_{21} \cdot {\bm S}_2(l,m,n) \times {\bm S}_1(l,m+1,n+1) \nonumber \\
&+& {\bm D}^-_{21} \cdot {\bm S}_2(l,m,n) \times {\bm S}_1(l,m,n+1) \nonumber \\
&+& (1 \rightarrow 3, 2 \rightarrow 4)
\end{eqnarray}
with
\begin{equation}
{\bm D}^{\pm}_{\alpha \beta}=\pm (D'_{\alpha \beta,a}) {\hat a}+(D'_{\alpha \beta,b}) {\hat b} \pm D'_{\alpha \beta,c} {\hat c}.
\end{equation}
Here we have used the fact that $S^a_\alpha$ and $S^c_\alpha$ change signs while $S^b_\alpha$ does not under reflection through a mirror plane. Applying $A$ and translations, we find
\begin{equation}
{\bm D}^\pm_{12}=-{\bm D}^\pm_{34},~~~~~{\bm D}^\pm_{21}=-{\bm D}^\pm_{43}.
\label{eq: DM_diagonal1}
\end{equation}
On the other hand, applying $B$ with translations, we obtain
\begin{equation}
D'_{12,a} = -D'_{43,a},~~
D'_{12,b} = D'_{43,b},~~
D'_{12,c} = D'_{43,c}.
\label{eq: DM_diagonal2}
\end{equation}
Using Eqs. (\ref{eq: DM_diagonal1}) and (\ref{eq: DM_diagonal2}), 
one can show that relative signs are given by Eq. (\ref{DM_diagonal}),
which is shown in Fig. \ref{fig: DM_distribution}. 

\subsection{Symmetry with in-plane magnetic field}
\label{sec:symmetry-c-axis}

\begin{table}
  \centering
  \begin{tabular}{c||c|c|c||c|c|c}
    \# & $x'_a$ & $x'_b$ & $x'_c$ & $S^{a'}$ &$S^{b'}$ & $S^{c'}$ \\
\hline
0 & $x_a$ & $x_b$ & $x_c$ & $S^a$ & $S^b$ & $S^c$ \\
1 ($A$) & $-x_a$ & $-x_b$ & $-x_c$ & $S^a$ & $S^b$
& $S^c$ \\
2 & $\tfrac{1}{2}-x_a$ & $-x_b$ & $\tfrac{1}{2}+ x_c$ & $-S^a$ & $-S^b$
& $S^c$ \\
3 ($D$) & $\tfrac{1}{2}+x_a$ & $x_b$ & $\tfrac{1}{2}- x_c$ & $-S^a$ & $-S^b$
& $S^c$ \\
4 & $-x_a$ & $\tfrac{1}{2}+x_b$ & $-x_c$ & $-S^a$ & $S^b$
& $-S^c$ \\
5 ($C$) & $x_a$ & $\tfrac{1}{2}-x_b$ & $x_c$ & $-S^a$ & $S^b$
& $-S^c$ \\
6 ($B$) & $\tfrac{1}{2}+x_a$ & $\tfrac{1}{2}-x_b$ & $\tfrac{1}{2}- x_c$ & $S^a$ & $-S^b$
& $-S^c$ \\
7 & $\tfrac{1}{2}-x_a$ & $\tfrac{1}{2}+x_b$ & $\tfrac{1}{2}+ x_c$ & $S^a$ & $-S^b$
& $-S^c$ 
  \end{tabular}
  \caption{Point group operations in the $Pnma$ space group.  }
  \label{tab:Pnma}
\end{table}

To fully determine the spin structures, it is sometimes useful to have a
more detailed understanding of the residual symmetry in a field.  The
uniform applied field obviously preserves the translational symmetry of
the lattice, but breaks time reversal symmetry.  Its effects upon the
point group operations are less obvious.  In general, the $Pnma$ space
group contains 7 non-trivial (and one trivial) point operations, which
are not all independent (see Table~\ref{tab:Pnma}).  Of these 7
operations, 3 preserve any one component of the magnetization,
and of these 3, two are independent.  Therefore the point group symmetry
in the presence of the magnetic field is generated by just two
operations, which have Ising character.  The first operation is simply
the inversion tranformation $A$, from Eq.~(\ref{eq: sym_A}) in
Sec.~\ref{sec:cryst-struct-symm}.  In the notation of the main text,
this operation reads
\begin{equation}
\label{eq:55}
  A: \; {\bf S}_{x,y,z} \rightarrow {\bf S}_{\frac{1}{2}-x,1-y,1-z}.
\end{equation}
Clearly, this operation preserves {\sl all} components of the uniform
magnetization, and hence is a symmetry for an arbitrary applied magnetic
field.  

For a magnetic field along $a$, the $B$ operation given in the previous
subsection, Eq.~(\ref{eq: sym_B}), can be chosen as the other symmetry
generator.  It can be written as
\begin{equation}
  \label{eq:68}
  B: \; S^\mu_{x,y,z} \rightarrow \upsilon_\mu S^\mu_{-x,-y,z+1},
\end{equation}
with $\upsilon_a=1$, $\upsilon_b=\upsilon_c=-1$.

However, operation $B$ does not keep either the $b$ or the $c$-axis
magnetization invariant.  The second independent operation should be
chosen differently for these field orientations.  For a magnetic field
along $b$, it can be taken as a reflection in an $a$-$c$ plane:
\begin{equation}
\label{eq:57}
  C: \, S^\mu_{x,y,z}\rightarrow \zeta_\mu S^\mu_{-x,y,z},
\end{equation}
where $\zeta_a=\zeta_c=-1$ and $\zeta_b=1$.  

For a field along $c$, it can be taken as a reflection in an
$a$-$b$ plane:
\begin{equation}
\label{eq:56}
  D: \, S^\mu_{x,y,z}\rightarrow \eta_\mu S^\mu_{x,-y,z},
\end{equation}
where $\eta_a=\eta_b =-1$ and $\eta_c=1$.  

Using the bosonization formulae, Eq.~(\ref{eq:2}), one can deduce the
transformation of the boson field $\theta_{y,z}(x)$ under these
symmetries.  One finds, for the inversion operation,
\begin{eqnarray}
\label{eq:24}  
  A: \theta_{y,z}(x) & \rightarrow & \theta_{1-y,1-z}(\tfrac{1}{2}-x)
  + \frac{\pi}{2\beta} (-1)^y
\end{eqnarray}
For the field along $b$, the reflection gives
\begin{equation}
  \label{eq:59}
  C: \theta_{y,z}(x) \rightarrow \theta_{y,z}(-x) - \frac{\pi}{2\beta}[1-(-1)^y].
\end{equation}
For the field along $c$, the corresponding reflection instead gives
  \begin{equation}
    \label{eq:58}
    D: \theta_{y,z}(x) \rightarrow  \theta_{-y,z}+ \frac{\pi}{\beta}.
  \end{equation}

\section{Breaking of $U(1)$ spin rotation symmetry}
\label{sec:select-discr-ground}

In the discussion of
Secs.~\ref{sec:field-along-b}-\ref{sec:field-along-c}, the {\sl overall}
phase angle of the spins in the plane perpendicular to the magnetic
field remained arbitrary.  This reflects the $U(1)$ symmetry of spin
rotations about the field axis, which is present in the effective
Hamiltonian having dropped DM terms with DM-vectors perpendicular to the
field.  While we expect this to be an excellent approximation, it is not
exact, and the weak effects which we have neglected should remove this
artificial invariance.  This is appropriate for a crystalline system
with only discrete symmetries.  In this Appendix, we use symmetry
analysis to determine how this symmetry breaking occurs, for different
field orientations.  We will not attempt to determine the microscopic
origin of these effects here, which might be, for instance, symmetric
exchange anisotropy, or fluctuation-generated interactions.  Instead, we
ask what terms might arise in the energy as a function of the remaining
parameters describing the orientation of the spins in the plane
perpendicular to the field.  To do so, we must consider the {\sl
  reduced} symmetries of the system in the presence of the magnetic
field.

\subsection{Cone state}
\label{sec:cone-state-1}

In Sec.~\ref{sec:field-along-axis}, we obtained a cone state for
arbitrary magnetic fields (below saturation) along the $a$ axis.
Considering the standard model, this incommensurate state has both a
$U(1)$ spin-rotational degeneracy and a pseudo-$U(1)$ phason
degeneracy.  The latter degeneracy is protected by translational
invariance, but the $U(1)$ rotational symmetry is an artifact, violated
for instance by the $D_c$ term and other DM interactions with DM-vectors
along $b$ or $c$.  To study the breaking of the $U(1)$ spin rotational
symmetry, it is useful to consider the combinations $\Theta_\pm =
\Theta_0\pm \Theta_1$.  The $\Theta_-$ field describes the phason mode.  We
focus instead of $\Theta_+$.

Apart from translation invariance, the residual symmetries in the field
along $a$ are $A$ and $B$.  Under these operations, we find
\begin{eqnarray}
  \label{eq:69}
  A: \; \Theta_+ & \rightarrow & -\Theta_+ + 2\pi , \\
  B: \; \Theta_+ & \rightarrow & -\Theta_+ .
\end{eqnarray}
One sees from this that the effective potential should be an even,
$2\pi$-periodic function of $\Theta_+$.    {\sl There are no further
  symmetry restrictions on this potential.}   In the simplest situation,
e.g. $V_{\rm eff}(\Theta_+) = V \cos \Theta_+$, this potential has a
unique minimum ($\Theta_+=0$ or $\pi$, depending upon the sign of $V$).
Thus one expects the artifical $U(1)$ rotational symmetry about the $a$
axis to be completely lifted, and the only degeneracy of the cone state
to be that associated with the phason mode.

\subsection{Antiferromagnetic (AF) state}
\label{sec:antif-af-state}

The breaking of the spin-rotation symmetry about the field axis is
crucial for determining the precise nature of the spin structure in the
AF phase.  Since this phase is commensurate, and the Hamiltonian has only
discrete symmetries, we expect only a discrete ground state degeneracy.
This will fix the angles of the spin projections into the plane
transverse to the magnetic field.  We want to construct an effective
potential which depends upon the parameters of the AF state, $\Theta$
and $\sigma$.

\subsubsection{field along $b$ axis}
\label{sec:field-along-b-1}

For a field along the $b$ axis, the remaining symmetries are the
inversion and reflection operations, denoted $A$ and $C$, given in
Eqs.~(\ref{eq:55},\ref{eq:56}), in Appendix~\ref{sec:symmetry-c-axis}.
Using the formula in the appendix, and Eq.~(\ref{eq:25}), we find that
under these operations,
\begin{eqnarray}
  \label{eq:60}
  A: \sigma\rightarrow \sigma, & \qquad & \Theta \rightarrow \Theta +
  \frac{\pi}{2}(1+\sigma), \\
  C: \sigma\rightarrow - \sigma, & \qquad & \Theta \rightarrow \Theta . \label{eq:62}
\end{eqnarray}
In addition to these symmetries, under translations by one unit along $x$, one has
\begin{equation}
  \label{eq:61}
  T: \sigma \rightarrow \sigma \qquad \Theta \rightarrow \Theta + \pi.
\end{equation}
From this, we may construct the simplest energy function which depends
upon $\Theta$ and $\sigma$, in the spirit of Landau theory.  From
$C$, Eq.~(\ref{eq:62}), we see that it must be independent of $\sigma$.
From $A$ and $T$, Eqs.~(\ref{eq:60},\ref{eq:61}), it must be periodic in
$\Theta$ with period $\pi$.  Importantly, {\sl there are no other
  symmetry restrictions}.  Hence the most general energy is of the form
\begin{equation}
  \label{eq:63}
  V(\Theta) = \sum_{n=1}^\infty a_n \cos (2n\Theta - \alpha_n).
\end{equation}
Generically, since there are no restrictions on $\alpha_n$, such a
potential $V(\Theta)$ has two inequivalent minima (e.g. if we take the
simplest form with $a_n=0$ for $n>1$, located at $\Theta=\alpha_1/2,
\alpha_1/2+\pi$) located at points of no particular symmetry.  This
means that the transverse (to the magnetic field) components of spins do
not lie parallel to {\sl either} the $a$ or $c$ axes.  In total, one
therefore expects 4 ground states, with $(\sigma,\Theta)=(\pm
1,\Theta_0), (\pm1,\Theta_0+\pi)$.  Physically, the four states are
obtained from one another by the arbitrary choice of global sign for the
transverse components of the spins on the even and odd chains,
separately.

\subsubsection{field along $c$ axis}
\label{sec:field-along-c-1}

In this field orientation, the good symmetry operations are $A$, $T$ and $D$,
given in Eqs.~(\ref{eq:55},\ref{eq:56}) of Appendix~\ref{sec:dmv}.  $A$
and $T$ continue to act as in Eqs.~(\ref{eq:60},\ref{eq:61}), while $D$
leads instead to
\begin{equation}
  \label{eq:64}
  D: \sigma\rightarrow -\sigma, \qquad \Theta \rightarrow \Theta+\pi.
\end{equation}
Combining the $D$ and $T$ operations, one sees that the energy must be
independent of $\sigma$, and then using $A$ or $D$, one obtains again the
effective potential for $\Theta$ in form of Eq.~(\ref{eq:63}).  Thus,
the ground state degeneracy (four-fold) is the same as in the AF state
for fields along $b$, and the transverse components of the spins do not
point along the $a$ or $b$ axes.

\subsection{Incommensurate Phase}
\label{sec:incommensurate-phase}

Here we must reconsider the valid symmetry operations -- $A$, $T$, and $D$ --
when acting upon the spin structure in Eq.~(\ref{eq:52}).  This
structure is parametrized by two angles, $\Theta_0$ and $\Theta_1$,
which describe the spins in the even and odd chains, respectively.
Equivalently, we can choose the combinations $\Theta_\pm = \Theta_0\pm
\Theta_1$.  We find
\begin{eqnarray}
  \label{eq:65}
  & & A : \Theta_+ \rightarrow \Theta_++2\pi , \hspace{0.2in} \Theta_-
  \rightarrow -\Theta_--q_0, \\
  & & D : \Theta_+ \rightarrow \Theta_+ + \pi,  \hspace{0.3in} \Theta_- \rightarrow
  \Theta_-+\pi, \\
  & & T: \Theta_+ \rightarrow \Theta_++2\pi, \hspace{0.25in} 
  \Theta_-\rightarrow\Theta_- + 2q_0.
\end{eqnarray}

Note that the transformations of the $\Theta_-$ field under $A$ and $T$ involve shifts by
multiples of the incommensurate wavevector $q_0$.  Under multiple
actions of such shifts, any value of $\Theta_-$ can be approached
arbitrarily closely (due to the $2\pi$ periodicity).  Thus there is {\sl
  no} potential which can pin the values of $\Theta_-$.  This is the
reason for the gapless ``phason'' mode in the IC phase.   

By contrast, the effective Hamiltonian can certainly depend upon
$\Theta_+$, reflecting the lack of rotational invariance about the $c$
axis.  In general, due to the action of $D$ (which is most constraining),
the effective potential must be a $\pi$-periodic function of
$\Theta_+$.  There are, however, no other constraints.  Since shifts of
either $\Theta_0$ or $\Theta_1$ by $2\pi$ have no physical significance,
$\Theta_+$ is itself defined only up to $2\pi$.  This implies that there
should be two discrete sets of IC solutions, described by $\Theta_+ =
\Theta_0, \Theta_0+\pi$, and $\Theta_0$ should generically take an
incommensurate value with no special symmetry.

\section{Chain mean field theory}
\label{sec:cmft}

\subsection{SDW order}
\label{sec:cmft-sdw}
We start by applying CMFT to the ideal 2d model Eq.~\eqref{eq:Hsg1} and consider first 
SDW order at finite temperature. Thus we write 
\begin{equation}
H_{1,\rm{sdw}} = \sum_{y,z} \int\! dx\, {\tilde \gamma}_{\rm{sdw}} 
\cos [2\pi (\phi_{y,z}-\phi_{y+1,z})/\beta ].
\label{eq:H1sdw}
\end{equation}
It is convenient to shift $\phi$ fields slightly, $\phi_{y,z} \to \phi_{y,z} + (-1)^y \beta/4$, 
so as to change the sign of the interaction in the equation above. In terms of the shifted
fields the minimum corresponds to a uniform configuration $\phi_{y,z} = \phi_0(z)$
for each $z$, since the layers are decoupled.
It is clear that the inter-chain interaction can next be written as ${\bm \sigma}_{y,z} \cdot {\bm \sigma}_{y+1,z}$
where ${\bm \sigma}_{y,z} = (\cos[2\pi \phi_{y,z}/\beta], \sin[2\pi \phi_{y,z}/\beta])$ is the SDW order parameter 
vector describing chain $(y,z)$. Chain mean-field consists in a self-consistent assumption
that SDW order spontaneously develops at some critical temperature $T_{\rm{sdw}}$, below
which the SDW order parameter acquires a finite value along some arbitrary direction
in the SDW plane. For concreteness we choose this direction to be along $x$-axis,
$\langle {\bm \sigma}_{y,z}\rangle = (\tilde\psi, 0)$. This choice corresponds to $\alpha_z =0$ 
in Eq.~\eqref{eq:44}. With these approximations we have
\begin{equation}
H_{1,\rm{sdw}} \to H_{1,\rm{sdw}}^{\rm{mf}} = -2 {\tilde \psi} {\tilde \gamma}_{\rm{sdw}} \sum_{y,z} \int\! dx\, 
\cos [\frac{2\pi}{\beta} \phi_{y,z} ],
\end{equation}
where the factor of $2$ arises from the coordination number of chain $y$ in the layer $z$.
The self-consistent condition reads
\begin{equation}
\tilde\psi = \langle \cos [\frac{2\pi}{\beta} \phi_{y,z}] \rangle_{\rm{sdw}} ,
\label{eq:cmft-sdw1}
\end{equation}
where angular brackets denote finite-temperature average with the sine-Gordon
Hamiltonian $H^{\rm{mf}}_{\rm{sdw}}$ of the single chain:
\begin{equation}
H^{\rm{mf}}_{\rm{sdw}} = \int\! dx\, \frac{v}{2} \left( (\partial_x \theta_{y,z})^2 + (\partial_x \phi_{y,z})^2 \right) 
+ H_{1,\rm{sdw}}^{\rm{mf}} .
\label{eq:H-mf-sdw}
\end{equation}
The right-hand side of Eq.~\eqref{eq:cmft-sdw1} is evaluated perturbatively in powers of vanishing
$\tilde\psi$ and the leading order result is
\begin{equation}
\tilde\psi = 2\tilde\psi {\tilde \gamma}_{\rm{sdw}} \chi_{\Delta_{\rm sdw}}(q=0,\omega_n=0;T).
\label{eq:cmft-sdw2}
\end{equation}
Here we defined the momentum and frequency dependent {\sl
  susceptibility}, $\chi_{\Delta}(q,\omega_n;T)$, of the vertex operator
${\mathcal O}_\Delta=\cos(\sqrt{4\pi\Delta}\phi)$ (or ${\mathcal
  O}_\Delta=\cos(\sqrt{4\pi\Delta}\theta)$, which gives identical
results) at temperature $T$, in the canonical free boson theory,
Eq.~(\ref{eq:H-bos-0}):
\begin{eqnarray}
  \label{eq:91}
  \chi_{\Delta}(q=0,\omega_n=0;T) & = & \int_{-\infty}^{\infty} \! dx\,
  \int_0^{1/T} d\tau\, e^{i q x + i \omega_n \tau} \nonumber \\
  && \times \left\langle {\mathcal O}_\Delta(x,\tau) {\mathcal O}_\Delta(0,0)\right\rangle_0.
\end{eqnarray}
The subscript $0$ reminds us that that this is evaluated in the free
theory.  This susceptibility, in various limits, plays a central role in
the determination of critical temperatures within CMFT.  It is evaluated
at the end of this appendix, Sec.~\ref{sec:Fcos}.  Here, we need the SDW
susceptibility, for which $\Delta_{\rm sdw} = 1/4\pi R^2$.   At the critical temperature
Eq.~\eqref{eq:cmft-sdw2} acquires a non-trivial solution, $\tilde\psi\neq 0$,
resulting in the implicit equation for $T_{\rm sdw}$:
\begin{equation}
1 = 2 {\tilde \gamma}_{\rm{sdw}}  \chi_{\Delta_{\rm sdw}}(q=0,\omega_n=0;T_{\rm sdw}).
\label{eq:cmft-sdw3}
\end{equation}
Using Eq.~\eqref{eq:A} from Sec.~\ref{sec:12delta1}, we obtain
\begin{eqnarray}
&&\Big(\frac{2\pi T_{\rm{sdw}}}{v}\Big)^{2-2\Delta_{\rm sdw}} = \lambda_{\rm{sdw}}
\frac{\Gamma(1-\Delta_{\rm sdw}) \Gamma^2(\Delta_{\rm sdw}/2)}
{\Gamma(\Delta_{\rm sdw}) \Gamma^2(1-\Delta_{\rm sdw}/2)} \\
&&\times\Big[1 + \lambda_{\rm{sdw}}\Gamma(\Delta_{\rm sdw}-1/2)/
(\sqrt{\pi} \Gamma(\Delta_{\rm sdw}) (1-\Delta_{\rm sdw}))\Big]^{-1} ,\nonumber
\end{eqnarray}
where $\lambda_{\rm{sdw}} = \pi \tilde{\gamma}_{\rm sdw}/v = \pi A_1^2 \sin(\delta) J'/v$.
This result, $T_{\rm{sdw}}$ as a function of magnetization $M$, is plotted in Fig. \ref{fig:RG1}.

\subsection{Cone order}
\label{sec:cmft-cone}

Here we consider the cone (twist) ordering instability of the ideal 2d model Eq.~\eqref{eq:Hsg1}.
The cone Hamiltonian is given by 
\begin{eqnarray}
H_{1,\rm{cone}} &=&   - {\tilde \gamma}_{\rm{cone}} \sum_{y,z} \int\! dx\, 
  (\partial_x \theta_{y,z}+ \partial_x \theta_{y+1,z}) \nonumber\\
&&\times  \cos [\beta(\theta_{y,z} - \theta_{y+1,z})] 
\label{eq:H1cone}
\end{eqnarray}
where ${\tilde \gamma}_{\rm{cone}}=J' A_3^2 \beta/2$. The spatial derivatives
in Eq.~\eqref{eq:H1cone} require a generalization of the procedure
described in subsection~\ref{sec:cmft-sdw}.  We begin by shifting the
$\theta$ fields by a position-dependent phase corresponding to a
wavevector shift $q_0$, the magnitude of which
is to be determined later self-consistently. Thus
\begin{equation}
\theta_{y,z}(x) = q_0 x/\beta + \tilde{\theta}_{y,z}(x) .
\label{eq:theta-shift}
\end{equation}
This shift transforms Eq.~\eqref{eq:H1cone} into
\begin{equation}
\tilde{H}_{1,\rm{cone}} \approx - 2 \frac{q_0 {\tilde \gamma}_{\rm{cone}}}{\beta} \sum_{y,z} \int\! dx\, 
 \cos [\beta(\tilde{\theta}_{y,z} - \tilde{\theta}_{y+1,z})] ,
 \label{eq:H1cone-shifted}
\end{equation}
where we have neglected as subleading $\partial_x\tilde{\theta}$ terms.
The transformed Hamiltonian Eq.~\eqref{eq:H1cone-shifted} is now of the form Eq.~\eqref{eq:H1sdw}
and can be manipulated similarly. However, the shift Eq.~\eqref{eq:theta-shift}
has modified the free boson Hamiltonian  Eq.~\eqref{eq:H-bos-0} into
\begin{equation}
\tilde{H}_0 = \sum_{y,z} \int\! dx\, \frac{v}{2} \left((\partial_x \tilde{\theta}_{y,z} + q_0/\beta)^2 + (\partial_x \phi_{y,z})^2 \right) .
\label{eq:H0-tilde}
\end{equation}
Introducing the order parameter 
\begin{equation}
\tilde\psi = \langle \cos [\beta \tilde{\theta}_{y,z}] \rangle_{\rm{cone}} ,
\label{eq:cmft-cone1}
\end{equation}
where the average is over the Hamiltonian 
\begin{equation}
H^{\rm{mf}}_{\rm{cone}} = \tilde{H}_0 - 2 \tilde\psi {\hat \gamma}_{\rm{cone}} \sum_{y,z}\int\! dx\, 
\cos[\beta \tilde{\theta}_{y,z}] ,
\label{eq:H-mf-cone}
\end{equation}
and ${\hat \gamma}_{\rm{cone}} = 2 q_0 {\tilde \gamma}_{\rm{cone}}/\beta = q_0 J' A_3^2$.
As before, expanding the expectation value in Eq.~\eqref{eq:cmft-cone1} to leading order in
$\hat\gamma$, and assuming $\tilde\psi\neq 0$, gives the condition for the
critical temperature
\begin{equation}
1 = 2 {\hat \gamma}_{\rm{cone}} \tilde\chi_{\Delta_\pm}(q=0,0;T_c) \
\label{eq:cmft-cone3}
\end{equation}
where {\em tilde} on the susceptibility indicates that it is to be
calculated using the free but {\sl shifted} Hamiltonian
Eq.~\eqref{eq:H0-tilde}.  However we can now undo the shift
Eq.~\eqref{eq:theta-shift} and transform back to the original $\theta$
fields.  As a result, one obtains the identity
\begin{equation}
\tilde\chi_{\Delta_\pm}(q=0,0;T_c) =\chi_{\Delta_\pm}(q_0,0;T_c) .
\end{equation}
Then Eq.~\eqref{eq:cmft-cone3} becomes
\begin{equation}
1=2 {\hat \gamma}_{\rm{cone}} \chi_{\Delta_\pm}(q_0,0;T_c).
\label{eq:cmft-cone2}
\end{equation}
This wavevector dependent static susceptibility is well-known, and given
in Eq.~\eqref{eq:95}.  Using it, Eq. \eqref{eq:cmft-cone2} can be solved
by maximizing the right-hand-side with respect to $q_0$ at given $T$,
and then finding the maximum $T_c=T_{\rm{cone}}$ (this is equivalent to
choosing the wavevector $q_0$ for which $T_c$ is maximum).   Expressing
all quantities in terms of
dimensionless variables, $r=v q_0/(4\pi T_{\rm{cone}})$ and $s=2\pi
T_{\rm{cone}}/v$, we obtain a system of two equations
\begin{eqnarray}
4~ {\rm Im} \Psi(\frac{\Delta_\pm}{2} + i r) &=& \frac{2\pi \sinh[2\pi r]}{\cosh[2\pi r] - \cos[\pi \Delta_\pm]} + \frac{1}{r},\nonumber\\
s^{1-2\Delta_\pm} &=& \frac{2\tilde\gamma_{\rm cone}}{\pi\beta v} \frac{\Gamma(1-\Delta_\pm)}{\Gamma(\Delta_\pm)}
r \Big|\Gamma(\frac{\Delta_\pm}{2} + i r)\Big|^4 \nonumber\\
&&\times (\cosh[2\pi r] - \cos[\pi \Delta_\pm]) .
\label{eq:cone-system}
\end{eqnarray}
The resulting $T_{\rm{cone}}$ is plotted in Fig.~\ref{fig:RG1}.

It is worth mentioning here that the outlined calculation can be done by keeping
track of lattice as well, so that spatial derivatives in Eq.~\eqref{eq:H1cone} become 
lattice differences. Following this route (see for example Ref.~\onlinecite{bocquet2001}) one again
arrives at Eq.~\eqref{eq:cmft-cone2} but with the coupling constant given by 
${\hat \gamma}_{\rm{cone,lattice}} = 2 J' A_3^2 \sin[q_0/2]$. This
difference, $q_0 \to 2\sin[q_0/2]$, does
not affect the outcome as in the regime where CMFT is applicable the ordering vector remains
small, $q_0 \sim J'/J \ll 1$. See section \ref{sec:cmft-limits} for more discussion.

\subsection{Inter-layer interaction $J''$}
\label{sec:cmft+Jpp}
Here we consider the fate of SDW and cone orders in the presence of inter-layer coupling $J''$.
As discussed in Section~\ref{sec:field-along-axis}, the inter-layer interaction 
is a strongly relevant perturbation which should be accounted for in CMFT.

\subsubsection{SDW order}
\label{sec:sdw-phase}

\begin{figure}[h]
  \centering
  \includegraphics[width=3.4in]{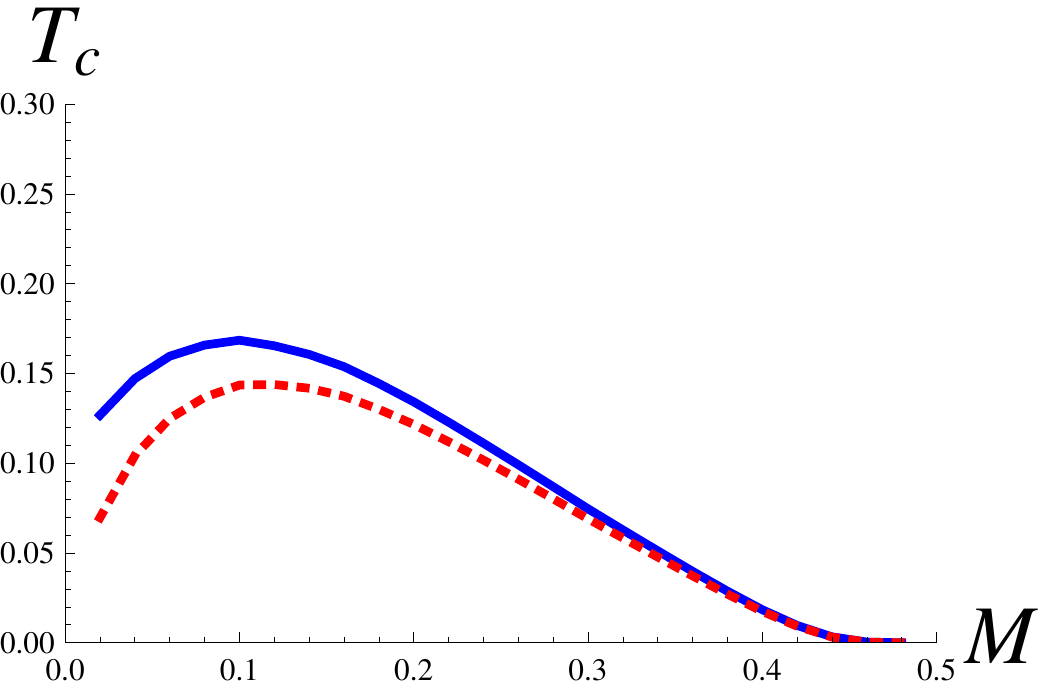}
  \caption{(Color online) SDW ordering temperature $T_{\rm SDW}$ (dotted (red) line) of the ideal 2d model, as obtained from Eq.~\eqref{eq:cmft-sdw3}.
  Solid (blue) line: the same but with inter-layer $J''$ accounted for, see Eq.~\eqref{eq:96}.} 
  \label{fig:sdwTc}
\end{figure}

Consider the SDW channel first.  Eq.~\eqref{eq:H1sdw} should now be complimented by 
\begin{eqnarray}
H'_{2,\rm{sdw}} &=& J'' \sum_{y,z} \int\! dx\, ( \mathcal{S}^z_{y,z;\pi-2\delta}
    \mathcal{S}^z_{y,z+1;\pi+2\delta} + \nonumber\\
    && + \mathcal{S}^z_{y,z;\pi+2\delta} \mathcal{S}^z_{y,z+1;\pi-2\delta}) .
\end{eqnarray}
This is simply a rewriting of the $\gamma''_z$ term in Eq.~\eqref{eq:3}.
Its bosonized form is
\begin{equation}
H'_{2,\rm{sdw}} = \frac{1}{2} A_1^2 J'' \sum_{y,z} \int\! dx\,
\cos [2\pi (\phi_{y,z}-\phi_{y,z+1})/\beta ].
\label{eq:H2sdw}
\end{equation}
The total SDW Hamiltonian is obtained by adding Eq.~\eqref{eq:H1sdw} and Eq.~\eqref{eq:H2sdw}.
Both terms can be made negative by a shift $\phi_{y,z} \to \phi_{y,z} + ((-1)^y + (-1)^z)\beta/4$.
Following the same steps as in Sec.~\ref{sec:cmft-sdw}, we find that
$T_c$ is determined by an equation of the same form as
Eq.~\eqref{eq:cmft-sdw3}, but with the replacement
\begin{equation}
  \label{eq:96}
  \tilde\gamma_{\rm sdw} \rightarrow \tilde \Gamma_{\rm sdw} =  {\tilde \gamma}_{\rm{sdw}} + A_1^2 J''/2 = A_1^2( J' \sin\delta + J''/2).
\end{equation}
Fig.~\ref{fig:sdwTc} shows that $T_{\rm{sdw}}$ is mildly enhanced by $J''$ at low magnetization $M$.

\subsubsection{transverse/cone order}
\label{sec:cone-phase}

\begin{figure}[h]
  \centering 
  \includegraphics[width=3.4in]{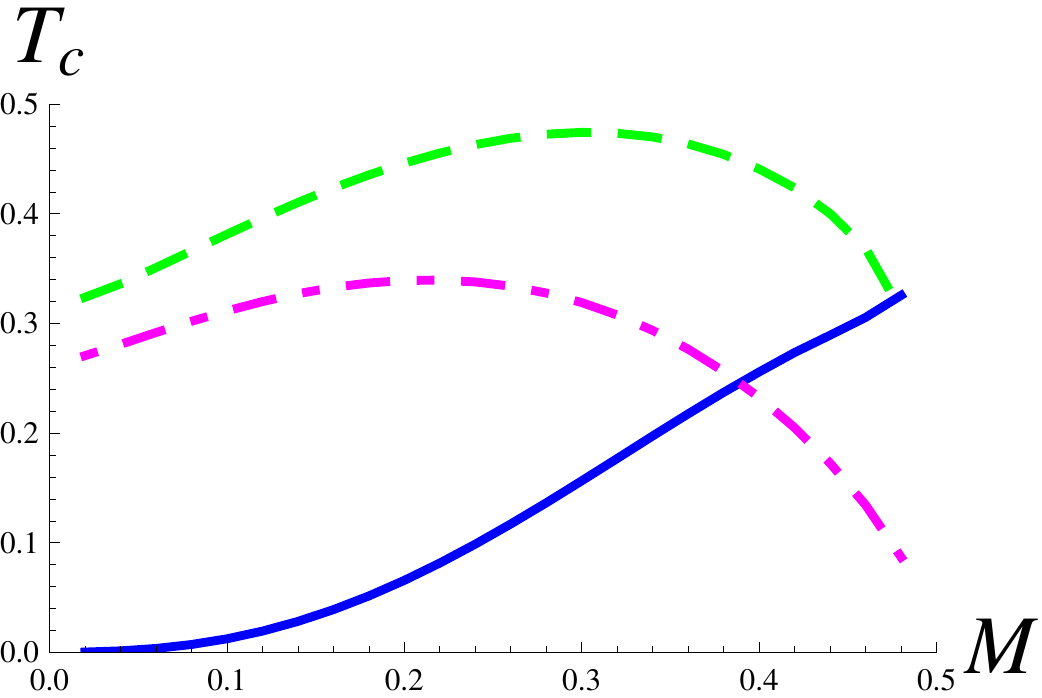}
  \caption{(Color online) Critical temperature of cone ordering as
    obtained in the ideal 2d model, Eq.~\eqref{eq:H1cone}, (solid (blue)
    line) and in the 3d model (dashed (green) line) as given by
    Eq.~\eqref{eq:cone-system2}.  The dot-dashed (magneta) curve
    represents the critical temperature for $T_{\rm{il}}$ of the
    interlayer-driven order, relevant for the AF and IC phases, which we
    obtain by setting $J' =0$ in Eq.~\eqref{eq:3dcone}.}
  \label{fig:coneTc}
\end{figure}

Interlayer exchange $J''$ strongly enhances the  transverse (to the field)
order (underlying the cone, AF, and IC states) for magnetic fields
along the $b$ and $c$ axes, for which the DM interaction $D$ is
ineffective.  In this case, we need to account
for the $\gamma''_\pm$ ($\tilde\gamma''_\pm$) term represented in
Eq.~\eqref{eq:5}, which is bosonized in Eq.~\eqref{eq:27}.  To bring the
latter into canonical form for CMFT, we change its sign by a simple
shift $\theta_{y,z} \to \theta_{y,z} + \pi z/\beta $, which does not
affect any of the manipulations in subsection~\ref{sec:cmft-cone}.
Transforming next to $\tilde{\theta}_{y,z}$ as in
Eq.~\eqref{eq:theta-shift}, we arrive at the modification of
Eq.~\eqref{eq:cmft-cone2} where ${\hat \gamma}_{\rm{cone}}$ is replaced
by $\hat\Gamma^{b-c}_{\rm{cone}}$,
\begin{equation}
{\hat \gamma}_{\rm{cone}} \to \hat\Gamma^{b-c}_{\rm{cone}} = A_3^2 (J' q_0 + J'').
\label{eq:3dcone}
\end{equation}
As in Sec.~\ref{sec:cmft-cone}, we obtain two equations for $q_0$ and
$T$ by maximizing the right-hand side of the modified
Eq.~\eqref{eq:cmft-cone2} with respect to $q_0$, and using the equation
itself.  The result reads, in terms of dimensionless pair $(r,s)$ introduced
in Eqs.~\eqref{eq:cone-system},
\begin{eqnarray}
4~ {\rm Im} \Psi(\frac{\Delta_\pm}{2} + i r) &=& \frac{2\pi \sinh[2\pi r]}{\cosh[2\pi r] - \cos[\pi \Delta_\pm]} + \frac{2 J' s}{2J' s r + J''},\nonumber\\
\frac{s^{2-2\Delta_\pm}}{2J' s r + J''} &=& \frac{A_3^2}{2 \pi v} \frac{\Gamma(1-\Delta_\pm)}{\Gamma(\Delta_\pm)}
\Big|\Gamma(\frac{\Delta_\pm}{2} + i r)\Big|^4 \nonumber\\
&&\times (\cosh[2\pi r] - \cos[\pi \Delta_\pm]) .
\label{eq:cone-system2}
\end{eqnarray}
The cone ordering temperature $T_{\rm cone}$ is plotted in
Fig.~\ref{fig:coneTc}.  We observe that the interlayer coupling enhances
$T_{\rm{cone}}$ dramatically, and even leads to a substantial $T_c$ when
$J'=0$.  The reason for this is simply that the
non-frustrated nature of the interlayer exchange leads to an appreciable
inter-chain coupling of transverse spin components even when $q_0\ll 1$,
as Eq.~\eqref{eq:3dcone} shows.  Indeed, in the AF and IC phases, we
also have transverse ordering, but $J'$ does not actually contribute to $T_c$,
and the plot with of $T_c(J'=0)$ is relevant in those cases.

\begin{figure}[h]
  \centering 
  \includegraphics[width=3.4in]{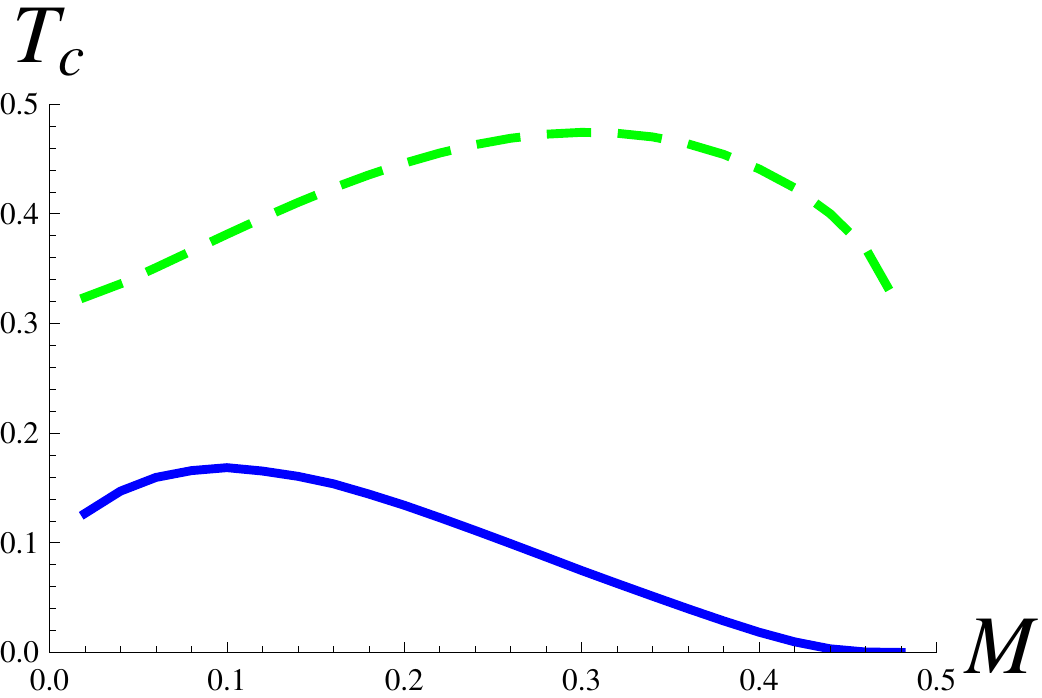}
  \caption{(Color online) $T_{\rm{sdw}}$ (solid (blue) line) and
    $T_{\rm{cone}}$ (dashed (green) line) for $J'' = 0.045 J$.  Compare
    this with ideal 2d situation in Fig,~\ref{fig:RG1} where the two
    orders compete strongly at intermediate values of $M$.}
  \label{fig:sdw-cone-Jpp}
\end{figure}

The effect of interlayer exchange $J''$ on the two orders can now be compared.
Notably, despite its smallness -- $J'' = 0.045 J$ -- interlayer coupling
completely eliminates SDW order in a system of weakly coupled layers,
$T_{\rm{sdw}} < T_{\rm{cone}}$ for all magnetizations from $0$ to $1/2$,
see Fig.~\ref{fig:sdw-cone-Jpp}.

\subsection{Field along $a$ axis}
\label{sec:cmft-dm}

\begin{figure}[h]
  \centering 
  \includegraphics[width=3.4in]{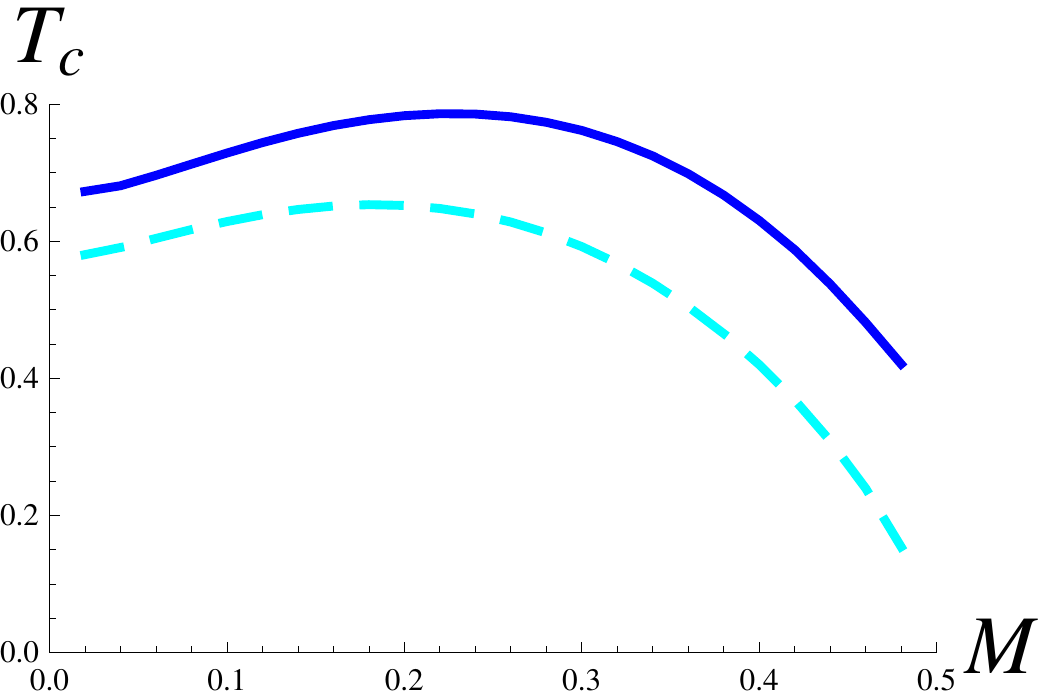}
  \caption{ (Color online) Ordering temperature with the field along the DM ($a$ axis) axis.
  Dashed (cyan) curve: $T_D(J'=0)$ due to DM interaction only ($J'=0$ in Eq.~\eqref{eq:cmft-dm}).
  Solid (blue) curve: $T_D$ obtained with both $D$ and $J'$ interactions present.} 
  \label{fig:dmTc}
\end{figure}

Here we describe how to calculate $T_c$ in CMFT in the presence of the
DM interaction $D$ for a field along the $a$ axis,
which is the arrangement considered in Sec.~\ref{sec:field-along-axis}
of the main text.  Here $J''$ is unimportant, as we will see.
Bosonization of the DM term in Eq.~\eqref{eq:3} gives
\begin{equation}
H'_{2,\rm{dm}} =  \sum_{y,z} \int\! dx\, (-1)^z 2 DA_3^2 \cos [\beta(\theta_{y,z} - \theta_{y+1,z})],
\label{eq:H2dm}
\end{equation}
while the cone term is given by Eq.~\eqref{eq:H1cone}, with
$\tilde{\gamma}_{\rm{cone}}=J' A_3^2 \beta/2$.  We observe
(c.f. Section~\ref{sec:incomm-order-state} where the corresponding $T=0$
state is discussed) that the two interactions can enhance each other if
the sign of the ordering vector is correlated with the sign of the DM
vector.  We therefore make the layer-dependent shift, which corresponds
to Eq.~\eqref{eq:43}, 
\begin{equation}
\theta_{y,z}(x) = (-1)^z q_0 x/\beta + \tilde{\theta}_{y,z}(x) .
\label{eq:theta-shift-z}
\end{equation}
This should be contrasted with Eq.~\eqref{eq:theta-shift}, which
describes the situation without any DM vector.  The transformation in
Eq.~\eqref{eq:theta-shift-z} makes the
competition between the staggered DM and interlayer interactions
(discussed in Sec.~\ref{sec:field-along-axis}) obvious, since the
argument of the interlayer cosine term, $\tilde\gamma''_\pm$ in Eq.~\eqref{eq:27}, acquires a
position-dependent phase
\begin{equation}
\theta_{y,z} - \theta_{y,z+1} = (-1)^z 2 q_0 x/\beta + \tilde{\theta}_{y,z} - \tilde{\theta}_{y,z+1}.
\end{equation}
The resulting oscillations eliminates the contribution of $J''$ to the
energy within CMFT.  Proceeding as described in
subsection~\ref{sec:cmft-cone}, we again obtain an equation for the
critical temperature in the same form as Eq.~\eqref{eq:cmft-cone3}, but
with $\hat\gamma_{\rm cone}$ replaced by $\hat\Gamma_{\rm cone}^{a}$:
\begin{equation}
  \hat\gamma_{\rm cone} \rightarrow \hat\Gamma^a_{\rm cone} =  A_3^2(J' q_0+2 D).
 \label{eq:cmft-dm}
\end{equation}
Note the great similarity of the above coupling with that in Eq.~\eqref{eq:3dcone}:
the two situations are related by exchanging $J'' \leftrightarrow 2 D$.
Hence the critical temperature, $T_D$, for transverse (cone) type
ordering follows from solving Eq.~\eqref{eq:cone-system2} with $J''$ replaced by $2 D$.  
The result is plotted in Fig.~\ref{fig:dmTc}, which compares
the case of DM interaction only $T_D(J'=0)$ with that of general $D\neq
0, J'\neq 0$ situation.  One observes that $J'$ leads to only a modest enhancement of
$T_D$ relative to the case with DM interaction present only.   Note that Fig.\ref{fig:RG2} in the main text
shows the solution with $D\neq 0, J'=0$.

\subsection{CMFT at $T=0$}
\label{sec:cmft-T=0}
Here we outline calculation leading to Eq.~\eqref{eq:28}. We start by
changing the sign of Eq.~\eqref{eq:27} via a shift: $\theta_{y,z} \to \theta_{y,z} +
(-1)^z \pi/(2\beta)$. Within CMFT, $H'_2$ is
replaced by a single-chain sine-Gordon Hamiltonian,
\begin{equation}
H'_{2,\rm{sG}} = - 2 \tilde\psi  J'' A_3^2 \int dx \cos\beta\theta(x) .
\label{eq:sg}
\end{equation}
Here $\tilde\psi = \langle \cos\beta\theta\rangle$ is to be determined self-consistently.
The brackets stand for averaging with the sine-Gordon action which, upon rescaling 
of the temporal coordinate $\tau=y/v$, reads
\begin{equation}
S_{\rm{sG}} = \int dx dy \Big(\frac{1}{2}(\partial_x \theta)^2 + \frac{1}{2}(\partial_y \theta)^2 - 2 \mu
\cos\beta\theta \Big) .
\label{eq:action-sg}
\end{equation}
Here $\mu = \tilde\psi J'' A_3^2/v$. The exact solution of
Eq.~\eqref{eq:action-sg} from Ref.~\onlinecite{LZ1997},
gives the ground state energy density
\begin{equation}
F(\mu) = -\frac{1}{4} M^2 \tan \left( \frac{\pi \xi}{2} \right),
\label{eq:sg2}
\end{equation}
which is expressed in terms of the parameter $\xi$ and mass gap $M$.
These are determined by
\begin{eqnarray}
\xi & = & \frac{\beta'^2}{1-\beta'^2}=\frac{\beta^2}{8\pi - \beta^2},\nonumber\\
\mu & = & \frac{\Gamma(\beta'^2)}{\pi \Gamma(1-\beta'^2)} 
\Big[ M \frac{\sqrt{\pi} \Gamma((1+\xi)/2)}{2 \Gamma(\xi/2)}\Big]^{2-2\beta'^2},
\label{eq:sg3}
\end{eqnarray}
with $\beta' = \beta/\sqrt{8\pi}$.  Using the obvious relation
\begin{equation}
\langle \cos\beta\theta\rangle = -\frac{1}{2} \frac{d F(\mu)}{d\mu} ,
\end{equation}
we obtain, after some algebra,
\begin{eqnarray}
\label{eq:sg4}
\tilde\psi &=& \sigma'(M)  \left( \frac{J'' A_3^2}{v}\right)^{\frac{\pi R^2}{2-2\pi R^2}},\\
\sigma'(M) &=&\frac{\tan[\pi \xi/2]}{2\pi (1-\beta'^2)} \left[\frac{\Gamma(\frac{\xi}{2})}{\Gamma(\frac{1+\xi}{2})}\right]^2
\Big[\frac{\pi \Gamma(1-\beta'^2)}{\Gamma(\beta'^2)}\Big]^{1/(1-\beta'^2)}. \nonumber
\end{eqnarray} 
The order parameter $\psi$ in Eq.~\eqref{eq:28} is related to the
self-consistently calculated $\tilde\psi$ 
very simply, $\psi = A_3 \tilde\psi$. Hence the prefactor in \eqref{eq:28} follows as
\begin{equation}
  \label{eq:97}
  \sigma(M) = A_3^{1/(1-\Delta_\pm)}  \sigma'(M).
\end{equation}

\begin{figure}[h]
  \centering 
  \includegraphics[width=3.4in]{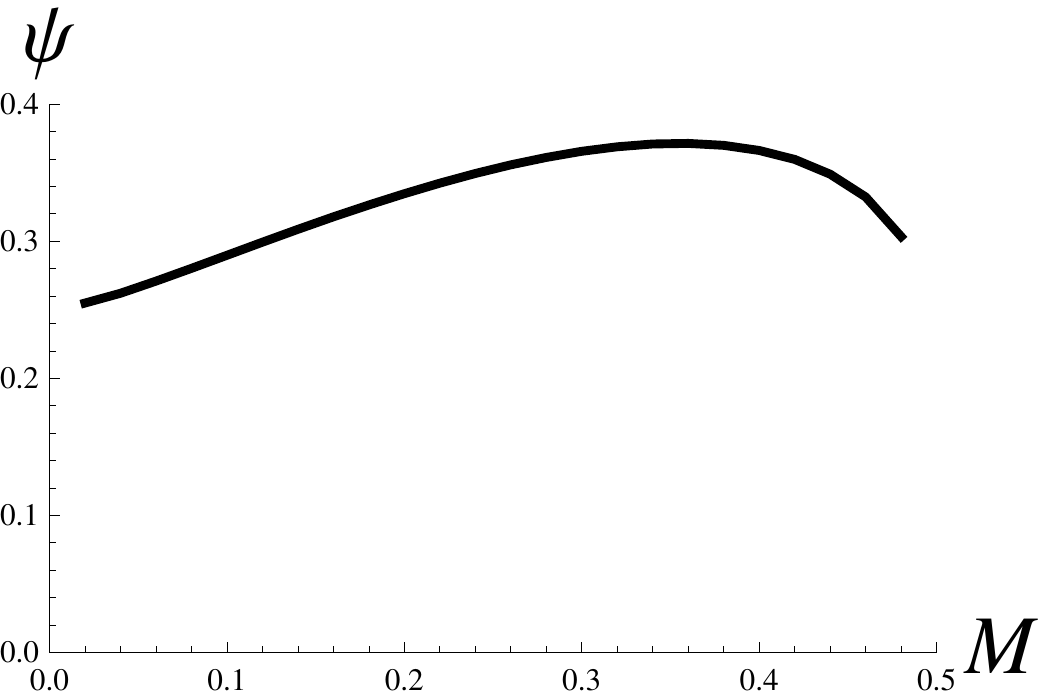}
  \caption{The $T=0$ order parameter $\psi$ versus $M$, calculated
    within CMFT.   } 
  \label{fig:orderparam}
\end{figure}

The resulting order parameter, $\psi(M)$, is plotted in
Fig.~\ref{fig:orderparam} as a function of magnetization.  Because the
exponent $\pi R^2/(2-2\pi R^2)$ in Eq.~\eqref{eq:sg4} is always small
(it varies from $1/2$ at $M=0$ to $1/6$ at $M=1/2$), one observes that
$\psi(M)$ is a slow-varying function of magnetization $M$.  The overall  non-monotonic shape of $\psi(M)$ is in
excellent agreement with the experimental data reported in Fig. 3b of
Ref.~\onlinecite{Coldea2002PRL}: the order parameter first rises with
increasing magnetization, reflecting the increasing relevance of
transverse spin correlations, and then falls down rapidly as $M\to 1/2$
as a result of diminishing density of magnons.   However, the CMFT
result shown in the figure does suffer one problem, discussed in the
next subsection, leading it to break down close to saturation (where it
incorrectly predicts that $\psi$ remains finite as $M\rightarrow 1/2$).  

\subsection{Limitations of CMFT}
\label{sec:cmft-limits}

The results of CMFT for the critical temperature and
order parameter, discussed in the previous subsections, exhibit unphysical behavior on approaching the saturation magnetization:
in these calculations, the $T_c$ for the cone state remains finite in
this limit, as does the zero temperature order parameter $\psi$.  These
features are both clearly incorrect, as both $T_c$ and $\psi$ must
decrease to zero as the spins become fully polarized.  We will
investigate the breakdown of CMFT in this subsection in more detail, and
determine the proper scaling theory for the vicinity of magnetization
saturation.  

First, let us observe the failure of  CMFT more directly.  Consider the solution of
Eqs.~\eqref{eq:cone-system} in the $\Delta_{\pm}\to 1/4$ limit in
more detail. First, note that the first equation there is
readily solved by $r=O(1)$ which is not particularly sensitive to the
value of $M$. This immediately tells us that $s \sim (A_3^2 J'/v)^2$ for
$\Delta_{\pm}=1/4$.   Near the saturation \cite{hikihara2004correlation} $A_3^2
\sim (\tfrac{1}{2} - M)^{1/2}$ while $v \sim (\tfrac{1}{2} - M) J$,
which implies $s \sim (J'/J)^2/(\tfrac{1}{2} - M)$.  Hence $T_c \sim v s
= (J')^2/J$ approaches a constant value while the ordering momentum $q_0 = 2
r s \sim (J'/J)^2/(\tfrac{1}{2} - M)$ diverges as $M\to
\tfrac{1}{2}$.   The latter divergence is a clear indication of the
failure of CMFT.  It can be traced the fact that
CMFT is by construction an expansion about the 1d chain limit.  The
natural parameter of this expansion is $J'/v$ which is supposed to be
small everywhere.  This assumption clearly breaks down near saturation,
where the spin velocity $v$ vanishes and the expansion is not justified
anymore.

Physically, near saturation, one has a dilute gas of spin flips, which
can be thought of as hard-core bosons or, in one dimension,
equivalently, as spinless fermions. Their density $n=\tfrac{1}{2} -M$
determines Fermi-momentum $k_F = \pi n$, which in turn determines the
(Fermi) velocity as $v = k_F/m$. Since the mass $m$ is of the order of
inverse chain exchange constant $J$, we obtain the scaling quoted above,
$v \sim (\tfrac{1}{2} - M) J$. The velocity vanishes because precisely
at saturation, the hard-core magnons possess a {\em quadratic}
dispersion relation, which is beyond the Luttinger liquid paradigm of
{\em linearly} dispersing collective excitations. (The scaling of $A_3
\sim n^{1/4}$ follows from the fact that the scaling dimension of the
spin-flip operator ${\cal S}^+_\pi$ is $\pi R^2 = 1/4$.)

To solve the problem correctly we need to start with a 2d description,
which is actually simple near the saturation.  We define the ``order
parameter'' field $\Psi_y(x) \sim S^+_{y;\pi}(x)$, which is just the
annihilation operator for a spin flip.  Because the flips are dilute, we
expand their kinetic energy near the bottom of their 1d band, and the
Hamiltonian of the layer is given by
\begin{eqnarray}
H_{\rm{sat}} &=& \sum_y \int\! dx\, \Psi_y^\dagger(x) (-\frac{\partial_x^2}{2 m} - \mu)\Psi^{\vphantom\dagger}_y(x) \nonumber\\
&&- J' (\Psi_y^\dagger(x) i \partial_x \Psi^{\vphantom\dagger}_{y+1}(x)
+ {\rm{h.c.}}) + ... \label{eq:98}
\end{eqnarray}
where the chemical potential $\mu \sim h_{\rm{sat}} - h$ describes the
deviation from the saturating magnetic field $h_{\rm{sat}}$ and dots
stand for the interaction terms (which are irrelevant at the 2d critical
point).   The scaling behavior for small $J'/J$ and small deviations
from saturation, can already be read off from
Eq.~\eqref{eq:98}, which should be considered for this purpose as a
(1+1)-dimensional quantum field theory.  The magnetization density, relative to saturation is
$\tfrac{1}{2}-M \sim \Psi^\dagger \Psi^{\vphantom\dagger}$ which scales like an
inverse length.  Furthermore, the $J'$ term above scales in the same
way, since it is missing one $x$ derivative.  Thus physical quantities should be scaling functions of
$\Xi=(J'/J)/(\tfrac{1}{2}-M)$, as claimed in Eq.~\eqref{eq:37}.  The
overall scaling behavior of $T_c$ is determined by the dynamical
critical exponent, which is $z=2$ due to the quadratically-dispersing
magnons.  Since $T_c$ therefore scales as the square of an inverse
length, the result in Eq.~\eqref{eq:38} follows.

We may understand the result more physically, and in particular the
behavior of the scaling function $\mathcal{F}(\Xi)$ by considering
Eq.~\eqref{eq:98} and its consequences in more detail.  First, the
behavior for small $\Xi$ corresponds just to the CMFT result in the
limit of $M\rightarrow 1/2$.  The limit of large $\Xi$ is more
interesting, and represents 2+1-dimensional physics.  Transforming to
momentum space, we find that the magnons have the following dispersion:
\begin{eqnarray}
\epsilon_k & = & \frac{k_x^2}{2 m} - \mu - J' k_x \cos k_y   \\
& = & \frac{(k_x- m J' \cos k_y)^2}{2m} - \frac{m(J')^2 }{2} \cos^2 k_y
-\mu .\nonumber
\end{eqnarray}
We expand around the minimum, writing $k_x=mJ'+q_x$, $k_y=q_y$, which
gives the continuum dispersion relation,
\begin{equation}
  \label{eq:99}
  \epsilon_k \approx -(\mu + \frac{m(J')^2 }{2} ) + \frac{q_x^2}{2m} +
  \frac{ m(J')^2 q_y^2}{2}.
\end{equation}
The continuum theory describes a free Bose gas with anisotropic
effective mass.  It of course forms a Bose condensate at low
temperature, which is described in the usual way by taking the continuum
limit $\Psi_y(x) \rightarrow e^{imJ' x} \Psi(x,y)$, with
\begin{equation}
 \Psi(x,y) = \sqrt{n_s}\, e^{i \theta(x,y)} .
\end{equation}
Here $\theta$ is the 2d superfluid phase, and $n_s \sim \tfrac{1}{2}-M$.  For the above anisotropic
Bose condensate, standard manipulations give the phase-only effective
Hamiltonian (superfluid kinetic energy), 
\begin{equation}
H_{\rm{sat}} = \frac{1}{2} \int\! dx\,dy\, \Big(\rho_x (\partial_x
\theta)^2 + \rho_y (\partial_y \theta)^2 \Big) .
\label{eq:xy}
\end{equation}

Eq. \eqref{eq:xy} describes an anisotropic 2d XY model with stiffnesses
$\rho_x = n_s/m \sim n_s J$ and $\rho_y \sim m (J')^2 n_s = n_s (J')^2/J$.  Its
critical temperature is determined by the geometric mean of the two
stiffnesses: $T_c = \pi \sqrt{\rho_x
  \rho_y}/2 \sim n_s J'$.   This argument describes the {\sl large} $\Xi$
limit of the scaling function $\mathcal{F}(\Xi)$.   Note also that this argument
implies that the ordering momentum $q_0 \sim T_{\rm{cone}}/v$ remains
small even near the saturation.  This we saw already in spin-wave theory
in  Appendix \ref{sec:swa}, where we observed $q_0 \sim J'/J$.

Away from saturation, so that $M - \tfrac{1}{2} \gg J'/J$, the CMFT
is applicable, and Eq.~\eqref{eq:cone-system} predicts
small $q_0$ again. For this reason we expect that $q_0$ is uniformly
small for all magnetizations, and lattice effects, of the
kind mentioned in the end of Section \ref{sec:cmft-cone}, remain
unimportant in all the regimes considered.   For this
reason we chose to keep only the leading terms of the small wavevector
expansion (that is, approximate lattice differences by spatial
derivatives) throughout the main text.

\subsection{Calculation of susceptibilities}
\label{sec:Fcos}

Here we present some technical details of the evaluation of
susceptibilites used in this appendix.  We define for convenience the
susceptibility in space and time,
\begin{eqnarray}
  \label{eq:92}
  \chi_{\Delta}(x,\tau,T) & = & \left\langle {\mathcal
      O}_{\Delta}(x,\tau) {\mathcal O}_{\Delta}(0,0) \right\rangle_{0;T},
\end{eqnarray}
evaluated at temperature $T$.  
It is straightforward to perform the average in Eq.~\eqref{eq:92} with
the free boson Hamiltonian Eq.~\eqref{eq:H-bos-0}, see for example,
Ref.~\onlinecite{giamarchi_book}:
\begin{eqnarray}
&& \chi_{\Delta}(x,\tau;T) =\frac{1}{2} \exp\{-4\pi\Delta \times I(y,\tau)\}, \\
&&I(y,\tau) = T \sum_{\omega_n} \int_{-\infty}^{\infty}  \frac{dq}{2\pi} e^{-\alpha |q|}
\frac{1-\cos[q y] \cos[\omega_n \tau]}{q^2 +\omega_n^2} \nonumber .
\end{eqnarray}
Here $y = x/v$, $\alpha=a_0/v$ is the short-distance cutoff needed to
regularize the integral, $\omega_n = 2\pi T n$ is the standard bosonic
Matsubara frequency, and $0 \leq \tau \leq 1/T$ is the Matsubara time.
The frequency summation is standard (we use GR 1.445.2 in
Ref.~\onlinecite{Gradshteyn_Ryzhik}) and leads to
\begin{eqnarray}
&& I(y,\tau) = \frac{1}{4\pi}\int_0^\infty \frac{dq}{q} \frac{e^{-\alpha q}}{1 - e^{-q/T}} \times 
\Big( 2 - e^{i q y - q\tau} \nonumber \\
&& - e^{- i q y - q\tau} + e^{-q/T}[ 2 - e^{i q y + q\tau}  - e^{- i q y + q\tau}]\Big).
\end{eqnarray}
Next we expand the denominator in series which is evaluated term by term with the help
of the identity 
\begin{equation}
\int_0^\infty \frac{dq}{q} \Big( e^{- A q} - e^{-B q}\Big) = \ln(\frac{B}{A}).
\end{equation}
In this way we arrive at
\begin{eqnarray}
&&4\pi I = \ln\frac{(\alpha + \tau)^2 + y^2}{\alpha^2} + 
\nonumber\\
&&+ \sum_{m=1}^\infty\Big(
\ln\frac{(m + \alpha T)^2 - T^2 (\tau - i y)^2}{(m + \alpha T)^2} \nonumber\\
&& + \ln\frac{(m + \alpha T)^2 - T^2 (\tau + i y)^2}{(m + \alpha T)^2}\Big).
\label{eq:I1sum}
\end{eqnarray}
For later use we note here that small $(\tau, y) \sim 0$ behavior is described
by the first term in the right-hand side of the above equation. It is easy to check
that small $(t,y)$ behavior, where $t = 1/T - \tau$, is described by a similar
$\ln[(\alpha + t)^2 + y^2]$ term (which is contained in $m=1$ contributions).

Focusing for the moment on the regime where $|t|\gg \alpha$
(i.e. including the small $\tau$ limit but not the limit near
$\tau=1/T$), we next observe that $m\neq 0$ contain no singular dependence on $\alpha$,
which allows us to set $\alpha = 0$ there. The well-known identity
$\ln[\sin(x)/x] = \sum_{k=1}^\infty \ln[1 - x^2/(\pi^2 k^2)]$ leads us to
\begin{equation}
4\pi I = \ln\Big[\frac{(\alpha + \tau)^2 + y^2}{\alpha^2} \frac{\sin[\pi T(\tau - i y)] \sin[\pi T (\tau + i y)]}
{\pi^2 T^2 (\tau^2 + y^2)}\Big].
\end{equation}
Hence we obtain, when  $|t|\gg \alpha$,
\begin{eqnarray}
\label{eq:93}
&&2 \chi_{\Delta}(x,\tau;T)  = \Big\{ \frac{\tau^2+y^2}{(\alpha + \tau)^2 + y^2}  \nonumber\\
&&\times\frac{\pi^2 T^2 \alpha^2}{\sin[\pi T(\tau - i y)] \sin[\pi T (\tau + i y)]}\Big\}^{\Delta} ,
\label{eq:Fcos-full}
\end{eqnarray}
which is imaginary time version of the expression obtained in
Ref.~\onlinecite{schulz86}.  
The appearance of the cut-off $\alpha$ in the temporal $\tau$ direction
is a generic 
feature of the bosonization technique (see Ref. \onlinecite{oal2005} for
more examples).  The first factor in Eq.~\eqref{eq:93} clearly carries
the information on the small $\tau$ limit.  To account for {\sl both}
$\tau\approx 0$ and $\tau \approx 1/T$ properly, we must include another
similar factor,
\begin{eqnarray}
  \label{eq:94}
  &&2 \chi_{\Delta}(x,\tau;T)  = \Big\{ \frac{\tau^2+y^2}{(\alpha + \tau)^2 + y^2} \frac{t^2+y^2}{(\alpha + t)^2 + y^2} \nonumber\\
&&\times \frac{\pi^2 T^2 \alpha^2}{\sin[\pi T(\tau - i y)] \sin[\pi T (\tau + i y)]}\Big\}^{\Delta} .
\end{eqnarray}
Eq.~\eqref{eq:94} then is correct for the full imaginary time interval.
To separate the short and long time behaviors of $\chi_{\Delta}(x,\tau;T)$
we approximate it as
\begin{equation}
\chi_{\Delta}(x,\tau;T) = \chi^>_{\Delta}(x,\tau;T)  + \chi^<_{\Delta}(x,\tau;T) .
\end{equation}
Here the first term describes long-distance behavior,
\begin{equation}
\chi^>_{\Delta}(x,\tau;T)  = \Big\{ \frac{\pi^2 T^2 \alpha^2}{\sin[\pi T(\tau - i y)] \sin[\pi T (\tau + i y)]}\Big\}^{\Delta} ,
\label{eq:Fcos-long}
\end{equation}
which gives the na\"ive limit of Eq.~\eqref{eq:94} when
$\alpha\rightarrow 0$, valid away from the endpoints.  The second,
\begin{eqnarray}
\label{eq:Fcos<}
&& \chi^<_{\Delta}(x,\tau;T)  =  
 \Big\{\frac{\alpha^2}{(\alpha + \tau)^2 + y^2}\Big\}^{\Delta}
 - \Big\{\frac{\alpha^2}{\tau^2 + y^2}\Big\}^{\Delta} \nonumber \\
&& + \Big\{\frac{\alpha^2}{(\alpha + t)^2 + y^2}\Big\}^{\Delta}
 - \Big\{\frac{\alpha^2}{t^2 + y^2}\Big\}^{\Delta} ,
\end{eqnarray}
accounts for an important difference in the short-distance behavior of
the full Eq.~\eqref{eq:94} and the approximate Eq.~\eqref{eq:Fcos-long}
expressions for $\chi_{\Delta}(x,\tau;T)$. Observe that the
short-distance behavior is not sensitive to temperature as it takes
place on the scale determined by $\alpha$ while the long-distance one
describes correlations on a much-longer thermal scale $1/T$.

For CMFT, we require certain limits of the Fourier transform of the susceptibility,
Eq.~\eqref{eq:91}, or
\begin{eqnarray}
  \label{eq:95}
&& \chi_{\Delta}(q,\omega_n;T) \\
&& =  \int_{-\infty}^{\infty} \!\!\! dx\,
  \int_0^{1/T} \!\!\! d\tau\, e^{i q x + i \omega_n \tau}  \chi_{\Delta}(x,\tau;T).\nonumber 
\end{eqnarray}

\subsubsection{SDW case}
\label{sec:12delta1}

For the SDW case, we need the static ($\omega_n=0$), zero momentum
($q=0$) limit of Eq.~\eqref{eq:95} in the range $1/2<\Delta<1$.  Here we
must take some care to keep the short and long time contributions
separate.  We split $\chi_{\Delta}(0,0;T)  =
\chi^>_{\Delta}(0,0;T) +\chi^<_{\Delta}(0,0;T) $, Fourier
transforming separately Eq.~\eqref{eq:Fcos-long} and Eq.~\eqref{eq:Fcos<}.
The first contribution is rather standard and given by
\begin{eqnarray}
\label{eq:A>}
&&\chi^>_{\Delta}(0,0;T)  = \\
&&\frac{\pi v \alpha^2}{2} \Big(2\pi T \alpha\Big)^{2\Delta -2} \frac{\Gamma(1-\Delta) \Gamma^2(\Delta/2)}
{\Gamma(\Delta) \Gamma^2(1-\Delta/2)}.\nonumber
\end{eqnarray}
Observe that Eq.~\eqref{eq:A>} diverges as $1/(1-\Delta)$ when
$\Delta\to 1$.  In the case of the SDW order this limit corresponds to
the behavior near saturation where $\Delta$ approaches $1$ at the full
magnetization, $M=1/2$ (see Table. \ref{tab:dims} and \eqref{eq: sat-field}).  The divergence is
not physical and stems from the incorrect short-distance behavior of
Eq.~\eqref{eq:Fcos-long}. It is compensated by $\chi^<_{\Delta}(0,0;T)$,
which yields two identical contributions from the two terms in
Eq.~\eqref{eq:Fcos<}.   Substituting $\tau = \alpha t, y = \alpha z$ and
using $\int_0^\infty dz/(1 + z^2)^\Delta = \sqrt{\pi} \Gamma(\Delta-1/2)/(2 \Gamma(\Delta))$
we arrive at
\begin{eqnarray}
&& \chi^<_{\Delta}(0,0;T)  =  \\
 && \frac{v \alpha^2 \sqrt{\pi} \Gamma(\Delta-1/2)}{(2-2\Delta) \Gamma(\Delta)}
\Big((1+L)^{2-2\Delta} - L^{2-2\Delta} -1\Big) ,\nonumber 
\end{eqnarray}
where $L = 1/(\alpha T) \gg 1$. Simple calculation shows that this expression
diverges logarithmically near $\Delta=1/2$, $\chi^<_{\Delta \approx 1/2}(0,0;T) \sim \ln(L e)$,
which however represents a small subleading correction to Eq.~\eqref{eq:A>},  which diverges linearly,
$\chi^>_{\Delta \approx 1/2}(0,0;T)  \sim L$, in this region. Thus $\chi^<_{\Delta \approx 1/2}(0,0;T) $ can be safely neglected.
Near $\Delta=1$ limit things are different: here $\chi^<_{\Delta \approx 1}(0,0;T) $ results in a large
$T$-independent contribution
\begin{equation}
\chi^<_{\Delta \approx 1}(0,0;T)  = - \frac{v \alpha^2 \sqrt{\pi} \Gamma(\Delta-1/2)}
{(2-2\Delta) \Gamma(\Delta)}
\approx - \frac{v \alpha^2 \pi}{2(1-\Delta)} .
\label{eq:DeltaA}
\end{equation}
Similar short-distance correction can be found in Refs.~\onlinecite{schulz83,schulz86}.
Collecting both contributions we finally obtain, for $1/2 < \Delta < 1$,
\begin{eqnarray}
\chi_{\Delta}(0,0;T)  &=& \frac{\pi v \alpha^2}{2} \Big\{\Big(2\pi T \alpha\Big)^{2\Delta -2} 
\frac{\Gamma(1-\Delta) \Gamma^2(\Delta/2)}
{\Gamma(\Delta) \Gamma^2(1-\Delta/2)} -\nonumber\\
&& - \frac{\Gamma(\Delta-1/2)}{\sqrt{\pi} (1-\Delta) \Gamma(\Delta)}\Big\}.
\label{eq:A}
\end{eqnarray}
At $\Delta=1$ this expression reduces to
\begin{equation}
\chi_{\Delta=1}(0,0;T)  = - \pi v \alpha^2 \ln[2\pi T \alpha] ,
\end{equation}
which is free of unphysical $(1-\Delta)^{-1}$ divergence.
The resulting $T_c(\Delta=1)$ is exponentially small 
in $v/{\tilde \gamma}_{\rm{sdw}}$ ratio.\cite{schulz83} 

\subsubsection{Cone order}
\label{sec:cone-order}

For the CMFT treatment of the cone state, one requires the
static susceptibility at non-zero wavevector, with $\Delta=\Delta_\pm =
\pi R^2$.  This is always less than or equal to  $1/2$, making the
short-time corrections in Eq.~\eqref{eq:Fcos<} negligible.
Therefore we may directly Fourier transform only the long time term,
Eq.~\eqref{eq:Fcos-long}.  The result is well-known (see,
e.g. Ref.~\onlinecite{bocquet2001}): 
\begin{eqnarray}
&&\chi_\Delta(q,0;0) = \frac{\pi \alpha^2}{2} \Big(2\pi T \alpha\Big)^{2\Delta -2} 
\frac{\Gamma(1-\Delta)}{\Gamma(\Delta)} \nonumber\\
&&\times \Big|\frac{\Gamma(\Delta/2 + i v q/(4\pi T))}{\Gamma(1-\Delta/2 + i v q/(4\pi T))}\Big|^2 .
\label{eq:chi-xx}
\end{eqnarray}

\section{Generation of biquadratic interaction}
\label{sec:biquadratic}

In this appendix, we detail the generation of the biquadratic
interaction, Eq.~\eqref{eq:14}. We use a standard Wilsonian RG, in which
one derives the low-energy theory by integrating out high-energy modes.

We begin by passing from the Hamiltonian formulation to the (Euclidean) Lagrangian
one, integrating out the conjugate field $\phi$ in the path integral.
Furthermore, we rescale the temporal direction, introducing $x_0 =
v\tau$, $x_1=x$, 
in order to render the free action rotationally invariant in the ${\bm
  x}=(x_0,x_1)$ plane.  The free action of the $\theta$ fields,
corresponding to Eq.~\eqref{eq:H-bos-0}, is
\begin{eqnarray}
S_0 &=& \frac{1}{2} \int d^2{\bm x}\, \left|{\boldsymbol\nabla}\theta_y\right|^2  \nonumber\\
&=& \frac{1}{2} \int_0^{\Lambda} \frac{d^2 {\bm k}}{(2\pi)^2} ~k^2
~\theta_y(k) \theta_y(-k). \label{eq:101}
\end{eqnarray}
Here we have introduced a momentum space cut-off $\Lambda$, which is
actually of $O(1)$ (the lattice spacing).
Furthermore, throughout this appendix, we have suppressed the $z$ index
of the chains to simplify the formulae.  Note that, due to the change
from $\tau$ to $x_0$, the interaction terms become perturbations with
dimensionless couplings, given by original ones divided by $v$.

The RG proceeds in the standard way, by progressively integrating out
modes within a shell of width $d\Lambda = \Lambda d\ell$ near the cut-off, thereby
reducing the latter to a ``running'' cut-off $\Lambda_\ell = \Lambda
e^{-\ell}$, which defines the logarithmic scaling variable $\ell \in \{0,\infty\}$.  
Following the convention used in the main text, we do not perform any
iterative rescaling of length and time scales, thereby allowing the
cutoff to ``run'' to increasingly smaller value as the RG proceeds.  

Formally, the integration of modes is accomplished, in each iteration, by writing
\begin{equation}
  \label{eq:100}
  \theta_y = \theta_y^<+ \theta_y^>,
\end{equation}
where the ``slow'' field $\theta_y^<$ contains non-zero Fourier components with
$k<\Lambda_{\ell+d\ell}= \Lambda_\ell e^{-d\ell}$, and the ``fast'' field $\theta_y^>$
contains the remaining ones with $\Lambda_{\ell+d\ell}<k<\Lambda_\ell$.
We integrate out the fast field at each iteration, perturbatively in the
interactions.  After this, we relabel $\theta_y^< \rightarrow \theta_y$,
which then defines the theory at the reduced cut-off
$\Lambda_{\ell+d\ell}$. 

At zeroth order in the interactions, the free action renormalizes
trivially, since the slow and fast modes are decoupled.  It remains in
the form of Eq.~(\ref{eq:101}), with $\Lambda$ replaced by
$\Lambda_\ell$.   To first order, the perturbations $H'_{1,2,3}$
renormalize very simply, according to their scaling dimensions.  We
illustrate this explicitly for the cone/twist interaction under
consideration here.  Ignoring the SDW term, we have the action
corresponding to $H'_1$ in Eq.~(\ref{eq:Hsg1}):
\begin{eqnarray}
  \label{eq:102}
  S'_1 & = & -\tilde\gamma_{\rm cone} \sum_y \int\! dx d\tau\, (\partial_x\theta_y
  + \partial_x \theta_{y+1})\cos \beta(\theta_y-\theta_{y+1}) \nonumber
  \\
  & = & \frac{i \tilde\gamma_{\rm cone}}{\beta v} \sum_y \int\!
  d^2{\bm x}\, \left( n_{y+1}^-\partial_x n_y^+ - n_{y+1}^+ \partial_x n_y^-\right).
\end{eqnarray}
Here we introduced the shorthand $n_y^\pm = e^{\pm i \beta \theta_y}$.
To first order in the RG, we use Eq.~(\ref{eq:100}) and average
Eq.~(\ref{eq:102}) over the fast fields using the free action.  Since
the fields are decoupled at each $y$, the two $n^\pm$ factors average
independently.  One has
\begin{eqnarray}
  \label{eq:103}
  \langle n^\pm_y\rangle_> & = & e^{\pm i \beta \theta^<_y} \left\langle
    e^{\pm i \beta \theta^>_y}\right\rangle_> \nonumber \\
  & = & e^{\pm i \beta \theta^<_y}  \exp[-\frac{\beta^2}{2}
  \int_{\Lambda_{\ell+d\ell}}^{\Lambda_\ell} \frac{d^2 {\bm
      k}}{(2\pi)^2} \frac{1}{k^2}] \nonumber \\
  & = & e^{\pm i \beta \theta^<_y}  e^{-\Delta_\pm d\ell},
\end{eqnarray}
where $\Delta_\pm = \beta^2/(4\pi) = \pi R^2$ is just the scaling
dimension of the $n_y^\pm$ field.  Letting $\theta^< \rightarrow
\theta$, we see that Eq.~(\ref{eq:103}) applied to Eq.~(\ref{eq:102})
simply multiplies $\tilde\gamma_{\rm cone}$ by the constant
$e^{-2\Delta_\pm d\ell}$.  Hence, we have $\tilde\gamma_{\rm
  cone}(\ell+d\ell) = (1-2\Delta_\pm d\ell) \tilde\gamma_{\rm
  cone}(\ell)$, or
\begin{equation}
  \label{eq:104}
  \partial_\ell \tilde\gamma_{\rm cone} = -2\Delta_\pm \tilde\gamma_{\rm cone}.
\end{equation}
This of course integrates to 
\begin{equation}
  \label{eq:106}
  \tilde\gamma_{\rm cone}(\ell) = \tilde\gamma_{\rm cone}(0) e^{-2\Delta_\pm \ell} =
  \tilde\gamma_{\rm cone}(0) \left(\frac{\Lambda_\ell}{\Lambda}\right)^{2\Delta_\pm}.
\end{equation}

The same treatment holds for the interlayer coupling $\tilde\gamma''_\pm$
(see Eq.~(\ref{eq:27}):
\begin{equation}
  \label{eq:105}
  \tilde\gamma''_\pm(\ell) = \tilde\gamma''_\pm \left(\frac{\Lambda_\ell}{\Lambda}\right)^{2\Delta_\pm}.
\end{equation}
As we have discussed, this is the most strongly relevant interaction for
fields in the $b$-$c$ plane, which we consider here.  The RG can be
considered perturbative provided the dimensionless rescaled coupling,
$\tilde\gamma''_\pm(\ell)/v$, remains small compared with the typical value of
the bare action at the corresponding scale, $\Lambda_\ell^2$.  This
fixes the value of the cut-off, $\Lambda''$, at which the coupled chains form correlated
$a$-$b$ planes:
\begin{equation}
\frac{\tilde\gamma''_\pm}{v} (\Lambda''/\Lambda)^{2\Delta_\pm} = (\Lambda'')^2 .
\end{equation}
Solving for $\Lambda''$, and using $\tilde\gamma''_\pm = A_3^2 J''$,
\begin{equation}
\Lambda'' \sim \Big(\frac{J'' A_3^2}{v}\Big)^{1/(2 -2\Delta_\pm)} ,
\end{equation}
where we have used that the bare cutoff $\Lambda$ is $O(1)$.  
Thus the corresponding spatial scale  which determines the renormalization
of all interactions is given by $\xi'' =1/\Lambda'' \sim (v/J'')^{1/(2 -2\Delta_\pm)}$,
in agreement with Eq.~\eqref{eq:7}.

Let us now, finally, generate the biquadratic term.  This occurs as a
{\sl second} order contribution of $\tilde\gamma_{\rm cone}$ to the
effective action.  
Expanding the action in powers of this term, we get in second order 
($Z = \int e^{-S_0}[1 + S_{(1)} + S_{(2)} +...]$)
\begin{eqnarray}
&&S_{(2)} =\nonumber \\
&& \frac{1}{2} ( \frac{\tilde\gamma_{\rm{cone}}}{\beta v})^2 \sum_y \int d^2{\bm x} d^2{\bm x'} 
\Big\{ \partial_{x'} (n_y^+({\bm x}) n_y^+({\bm x'}))\times \nonumber\\
&&\times \partial_x (n_{y+1}^-({\bm x}) n_{y+1}^-({\bm x'})) + {\rm h.c.}
\Big\}.
\end{eqnarray}
Terms which do not have the necessary $e^{i2\beta\theta_y}$ structure are omitted here.
Now we integrate out the fast fields in each chain. Then, in chain $y$ we obtain the combination
like this
\begin{eqnarray}
&& \partial_{x'} \left\{~e^{i\beta[\theta^<_y({\bm x}) + \theta^<_y({\bm x'})]} 
e^{-\beta^2 \int^> \frac{d^2{\bm k}}{(2\pi)^2}  \frac{1 + \cos[{\bm k}\cdot({\bm x}-{\bm x'})]}{k^2}} \right\}\nonumber\\
&&\approx e^{i2\beta\theta_y^<({\bm X})}
(-\beta^2) \int^> \frac{d^2{\bm k}}{(2\pi)^2} \frac{k_{x} \sin[{\bm k}\cdot {\bm \rho}]}{k^2} ,
\end{eqnarray}
where ${\bm X}= ({\bm x}+{\bm x}')/2$ and ${\bm \rho}={\bm x}-{\bm x}'$
are the center-of-mass and relative coordinates.  The most relevant term
that emerges has the spatial derivative acting on the c-function which
is produced by fast modes (the superscript on the integral indicates
that it is over the support of the fast modes only).  In the second line
above,  the
derivative has been carried out, bringing down the integral factor shown
from the derivative of the exponential.  After doing so, we have
approximated the exponential itself by $1$. This approximation is exact
to first order in $d\ell$, which is infinitesimally small.  Then
\begin{eqnarray}
&&S_{(2)} = -\frac{1}{2} ( \frac{\tilde\gamma_{\rm{cone}}}{\beta v})^2 \beta^4 \int d^2{\bm \rho} \int^> \frac{d^2{\bm k}_1}{(2\pi)^2} 
 \int^>\frac{d^2{\bm k}_2}{(2\pi)^2} \nonumber\\
&& \times\frac{k_{1x} \sin[{\bm k}_1\cdot{\bm \rho}]}{k_1^2}  \frac{k_{2x} \sin[{\bm k}_2\cdot{\bm \rho}]}{k_2^2} \nonumber\\
&& \times\int d^2{\bm X} ~2 \cos[2\beta(\theta_y^< - \theta_{y+1}^<)] .
\end{eqnarray}
The integral over the relative distance ${\bm \rho}$ produces difference of two delta-functions,
$\delta({\bm k}_1 + {\bm k}_2) - \delta({\bm k}_1 - {\bm k}_2)$, which, thanks to the $k_{1x} k_{2x}$
factor in the numerator, only doubles the final result. The whole of the momentum-shell integration
reduces to 
\begin{equation}
\int^> \frac{d^2{\bm k}_1}{(2\pi)^2} \frac{k_{1x}^2}{k_1^4} =
\frac{1}{4\pi} \int_{\Lambda_{\ell+d\ell}}^{\Lambda_\ell}  \frac{dk}{k}
=  \frac{d\ell}{4\pi} .
\end{equation}
The generated biquadratic correction to the action is $-S_{(2)}$.
Taking $\theta^< \rightarrow \theta$, we
see that we indeed generate a biquadratic interaction of the form
\begin{equation}
  \label{eq:107}
  H_{\rm bq} = \tilde\gamma_{\rm bq} \sum_y \int \! dx\, \cos[2\beta(\theta_y(x) - \theta_{y+1}(x))] ,
\end{equation}
with
\begin{equation}
d \tilde\gamma_{\rm bq} =  v\frac{\beta^2 d\ell \tilde\gamma_{\rm{cone}}^2}{4\pi v^2}   .
\label{eq:biq3}
\end{equation}
Here we have added a factor $v$ to the generated interaction, accounting
for the transformation back to imaginary time $\tau$ from $x_0$.  
The scaling dimension of this term is $2\times (2\beta)^2/(4\pi) = 8 \Delta_\pm$.
Hence, the RG flow equation for $\tilde\gamma_{\rm bq}$ is
\begin{equation}
\partial_\ell \tilde\gamma_{\rm{bq}} = - 8 \Delta_\pm \tilde\gamma_{\rm{bq}}
+ \frac{\beta^2 \tilde\gamma_{\rm{cone}}^2}{4\pi v} . 
\label{eq:biq4}
\end{equation}
Note that $\tilde\gamma_{\rm{cone}}$ here is itself a function of
running RG scale $\ell$, as specified in Eq.~(\ref{eq:106}).
Solving Eq.~(\ref{eq:biq4})  is easy and leads to 
\begin{equation}
\tilde\gamma_{\rm{bq}}(\ell) = \frac{\beta^2 \tilde\gamma_{\rm{cone}}(0)^2}{16 \pi \Delta v} \Big(e^{-4\Delta \ell} - e^{-8\Delta \ell}\Big),
\label{eq:biq5}
\end{equation}
which shows that $\ell \gg 1$ behavior is controlled by the driving term $\tilde\gamma_{\rm{cone}}^2/v$ in the right-hand-side
of Eq.~\eqref{eq:biq4}. 

As discussed above, the chains enter the strongly coupled limit at $\xi'' =
\exp[\ell''] \sim (v/J'')^{1/(2 -2\Delta_\pm)}$, where Eq.~\eqref{eq:biq5}
must be stopped.  At this point, the phases $\theta_y$ may be regarded
as no longer fluctuating, and hence reduce to the classical phases
$\vartheta_y$ of the main text.  Thus $\tilde\gamma_{\rm bq}(\ell'')$
corresponds directly to $g_{\rm bq}$ defined in Eq.~(\ref{eq:14}).
Combining therefore Eq.~(\ref{eq:biq5}) with Eq.~(\ref{eq:103}), which tells us
that spontaneous moment of the $a$-$b$ planes $|\psi| \sim
(\xi'')^{-\Delta}$, we arrive at the estimate Eq.~\eqref{eq:15},
$g_{\rm{bq}} \sim (J')^2 |\psi|^4/v$.  

\section{Negligible DM couplings}
\label{sec:negl-dm-coupl}

In Sec.~\ref{sec:dm-ology}, it was stated that three of the five allowed
DM couplings, $D_a$, $D'_b$ and $D'_c$, can be safely neglected.  In
this appendix, we explain why this is the case.

\subsection{$D_a$ }
\label{sec:d_a}

First consider the $D_a$ term.  As with all the DM couplings, this is
only effective for fields parallel to its D-vector, in this case the $a$
axis.  For such fields, like the $D_c$ term studied in
Sec.~\ref{sec:field-along-c}, it can be ``gauged away'' for a single
chain, by an $x-dependent$ spin rotation about the $z$ axis of spin.
Unlike the $D_c$ coupling, however, the $D_a$ interaction is constant
within each triangular plane.  Therefore this rotation has negligible
effects upon the other in-plane couplings, most importantly $D'_a=D$,
which we have argued dominates the physics in this field orientation,
but also $J'$, which plays a subsidiary but still important role.  This
gauge rotation {\sl does} affect the $J''$ interaction, however, since
the $D_a$ term is staggered along the $z$ axis, see
Eq.~\eqref{DM_chain}.  We have seen already in
Sec.~\ref{sec:interl-corr} that $J''$ itself is already (without $D_a$) ineffective in
establishing interlayer correlations, and its only effects arise through
generating the $J''_2$ interaction between second neighbor layers,
Eq.~\eqref{eq:66}.  The $J''_2$ interaction is, happily, also unaffected
by the gauge rotation, as the second neighbor layers involved rotate
identically.  Thus even if some $D_a$ is present, the analysis of
Sec.~\ref{sec:field-along-axis} remains unchanged.

\subsection{$D'_b$}
\label{sec:d_b}

Next consider the $D'_b$ interaction.  Unlike the $D=D'_a$ interaction,
this coupling has the same sign on both diagonal bonds between chains
(${\bm D}_{y,z}^+ = {\bf D}^-_{y,z}$ in Eq.~\eqref{DM_diagonal}).  This
means that, like the $J'$ coupling, this interaction is highly
frustrated.  As a consequence, the leading order contributions arising
from this term involve a gradient, analogous to the twist/cone term in,
e.g. Eq.~\eqref{eq:H1p}.  Thus the effects of this term are generally
strongly suppressed, both by this gradient (and associated increased
scaling dimension) {\sl and} by its small magnitude, which is of at most
a few percent.  In other words, it carries the same scaling dimension as
the twist/cone coupling, but is probably at least a factor of 10 smaller
in magnitude.  Thus it is always negligible.

\subsection{$D'_c$}
\label{sec:d_c}

Finally, we turn to the $D'_c$ term.  This interaction is similar in
some ways to the $D=D'_a$ interaction, which dominates for fields along
$a$.  Both are unfrustrated, as they have opposite signs on the two
diagonals, and both are staggered along the $a$ ($z$) direction.
However, $D'_c$ is also staggered along $c$ ($y$), while $D$ was
constant within the triangular planes.  

The fate of $D'_c$ is less clear than that of the prior two terms under
consideration.  It is neither trivially gauged away nor obviously
negligible.  However, it is easy to establish that it does compete with
many of the key interactions that have already been identified as
driving forces in the system.  As such, provided $D'_c$ is not too
large, it loses this competition and has minimal effects.

First, we see that $D'_c$ has the same scaling dimension and hence
relevance as the $J''$ term.  Moreover, like the $D$ term, it competes
with the $J''$ interaction because of the staggering along $z$.  Hence,
if $D'_c$ is not comparable to $J''$, it will lose this competition.
Indeed, if one assumes the form, Eq.~\eqref{eq:8}, which satisfies the
$\gamma''_\pm$ coupling ($\propto J''$), the $D'_c$ term identically
vanishes.  

Second, the $D'_c$ term also competes with the $D_c$ term, since the
latter favors opposite rotations on neighboring chains, which the $D'_c$
term attempts to couple.  Transforming to the rotating frame favored by
$D_c$, Eq.~\eqref{eq:22}, will make the $D'_c$ term oscillate, and hence
average out over long distances.

Thus to have any significant effect, the $D'_c$ term would need to be
large enough to overcome at least two competing interactions.  At least
for small $D'_c$, we conclude that the phase diagrams established in the
main text are unchanged.  Evidently, this is the case in \ccc.

\section{Spin-wave analysis in a high field}
\label{sec:swa}
In this appendix, we study the effect of the DM interaction and the
inter-layer interaction $J''$ on the high field magnons, and
particularly on the ordering wavevector infinitesimally below the
saturation field. In a strong magnetic field, the ground state is a
fully polarized ferromagnetic state and one can easily solve the
single-magnon problem exactly. By comparing measurements of the high
field magnons with such calculations, the microscopic parameters of the
standard Hamiltonian Eq.~(\ref{standard_hamiltonian}) were
determined.\cite{Coldea2002PRL,veillette2005incomm,
  veillette2006commensurate} However, in the standard model, all the
possible DM vectors are not included. We present here a complete
analysis based on Eq.~(\ref{eq:11}), which was derived in
Appendix \ref{sec:dmv}.

Let us first show that the components of the DM vector perpendicular to
an applied field can be negligible in the spin-wave analysis.  To show
this, we decompose ${\bm S}_{i}$ into $\langle {\bm S} \rangle+\delta
{\bm S}_{i}$, where $\langle {\bm S} \rangle$ is the ordered moment
parallel to the applied field ${\bm h}$.  In the linear spinwave theory
we neglect $\delta {\bm S}_{i}$ parallel to ${\bm h}$, which means
$\delta {\bm S}_{i} \times \delta \bm S_{j}$ is always parallel to ${\bm
  h}$ and does not couple to the component of the DM vector
perpendicular to ${\bm h}$. One can also show that the DM interaction
does not produce single magnon terms which is proportional to $\delta
{\bm S}_{i}$ using the symmetry argument in Appendix~\ref{sec:dmv}. In
what follows, we only retain $D_{\zeta}={\bm D}\cdot {\hat \zeta}$ where
$\zeta \equiv {\bm h}/h$.  We now take the direction of the field
(${\hat \zeta}$) as a quantization axis of spins.  Introducing
$S^\nu_i={\bm S}_i \cdot {\hat \nu}$ and $S^\pm \equiv S^\xi_i \pm i
S^\eta_i$ such that ${\hat \nu}={\hat \xi}, {\hat \eta}, {\hat \zeta}$
form an orthonormal basis, the local Hamiltonian for the bond $ij$ is
written as
\begin{eqnarray}
H_{ij}&=&J_{ij} {\bm S}_i \cdot {\bm S}_j +D_{ij,\zeta} (S^\xi_i S^\eta_j-S^\eta_i S^\xi_j) \nonumber \\
&=& \frac{\tilde J_{ij}}{2}(e^{i\phi_{ij}}S^+_i S^-_j+{\rm h.c.})+J_{ij}S^\zeta_i S^\zeta_j,
\end{eqnarray}
where ${\tilde J}_{ij}=\sqrt{J^2_{ij}+D^2_{ij,\zeta}}$ and $\tan \phi_{ij}=D_{ij,\zeta}/J_{ij}$.  
In the following, we focus on ${\hat \zeta}={\hat a}$, ${\hat b}$, and ${\hat c}$ cases and introduce 
\begin{eqnarray}
{\tilde J}_\zeta = \sqrt{J^2+(D_\zeta)^2},~~~\tan \phi_\zeta=D_\zeta/J, \\
{\tilde J}'_\zeta= \sqrt{(J')^2+(D'_\zeta)^2}~~~\tan \phi'_\zeta=D'_\zeta/J'
\end{eqnarray}
for on-chain and diagonal bonds, respectively.
We now apply the Holstein-Primakoff transformation:
\begin{equation}
S^\zeta_i=S-n_i,~S^+_i=(2S-n_i)^{\frac{1}{2}} b_i,
~S^-_i=b^\dagger_i (2S-n_i)^{\frac{1}{2}},
\end{equation}
with $n_i=b^\dagger_i b_i$ and $S=1/2$, and obtain
\begin{equation}
H_{ij}\sim {\tilde J}_{ij} S(e^{i\phi_{ij}}b^\dagger_i b_j+{\rm h.c.})-J_{ij}S(n_i+n_j)+JS^2.
\end{equation}
Denoting by $b_{\alpha,{\bm k}}$ the Fourier transform of the boson at the position ${\bm R}+{\bm \delta}_{\alpha}$, 
the spinwave Hamiltonian is written as
\begin{equation}
H_{\rm SW}=\sum_{\bm k}\Psi^\dagger_{\bm k} [{\cal H}({\bm k})+h-(2J+4J'+2J'')S]\Psi_{\bm k},
\label{eq: spinwave_ham}
\end{equation}
where ${\Psi}_{\bm k}=(b_{1,{\bm k}}, b_{2,{\bm k}}, b_{3,{\bm k}},
b_{4,{\bm k}})^{\rm T}$ and the $4\times 4$ matrix ${\cal H}({\bm k})$ depend on the field direction. 

\subsection{Field along $a$ axis}
Let us first consider the case of field along the $a$ axis.  In this
case, ${\cal H}({\bm k})$ in Eq.~(\ref{eq: spinwave_ham}) is given by
\begin{equation}
{\cal H}({\bm k})=\frac{1}{2} \left(\begin{array}{cc}
{\sf A}_{a,{\bm k}} (\phi_a,\phi'_a) & {\sf B}_{\bm k} \\
{\sf B}^\dagger_{\bm k} & {\sf A}_{a,{\bm k}} (-\phi_a, -\phi'_a)
\end{array}\right).
\end{equation}
Here, the matrices ${\sf A}_{a,{\bm k}}(\phi_a,\phi'_a)$ and ${\sf B}_{\bm k}$ are
\begin{equation}
{\sf A}_{a,{\bm k}} (\phi_a,\phi'_a) = \left(\begin{array}{cc}
2{\tilde J_a} \cos(k_b-\phi_a) & {\tilde J'_a} f_a (\phi'_a; {\bm k}) \\
{\tilde J'_a} f_a (-\phi'_a; -{\bm k}) & 2{\tilde J_a} \cos(k_b-\phi_a)
\end{array}\right),\nonumber
\end{equation}
and ${\sf B}_{\bm k} = J''(1+e^{ik_a})I$, where $I$ is the $2 \times 2$
identity matrix and $f_a(\phi'_a; {\bm k})=(e^{-i
  \phi'_a}e^{ik_b}+e^{i\phi'_a})(1+e^{ik_c})$.  We also note that
$k_\mu$ is defined by ${\bm k} \cdot \hat \mu$ for $\mu=a,b,$ and $c$.
We now try to find the location of the minimum of the spectrum of
1-magnon excitations, which is given in the form of ${\bm k}^*=(0,
2\pi(1/2+\epsilon), 0)$.  Using the relations such as $\cos
\phi_a=J/{\tilde J}_a$, the eigenvalues of ${\cal H}(0,k_b,0)$ are
explicitly obtained as
\begin{eqnarray}
\omega_{1,\pm}= J \cos k_b + 2J' \cos (k_b/2) ~~~~~~~~~~~~~~~~~~~~ \nonumber \\
\pm \sqrt{(J'')^2 + (D_a \sin k_b + 2D'_a \sin(k_b/2))^2}, \\
\omega_{2,\pm}= J \cos k_b - 2J' \cos (k_b/2) ~~~~~~~~~~~~~~~~~~~~ \nonumber \\
\pm \sqrt{(J'')^2 + (D_a \sin k_b - 2D'_a \sin(k_b/2))^2}.
\label {eq: omega2_pm}
\end{eqnarray} 
Putting $D_a=0$, one finds
the results consistent with Ref.~\onlinecite{Coldea2002PRL}.  Among the
four solutions, $\omega_{1,\pm}$ and $\omega_{2,\pm}$, $\omega_{1,-}$
has the lowest energy around $k_b=2\pi (1/2+\epsilon_0)$, where
$\epsilon_0=J'/(2 \pi J)$ which is the incommensuration in the absence
of $D_a$ and $D'_a$.  It is convenient to introduce a variable
$X=\cos(k_b/2)$.  Then we rewrite $\omega_{1,-}$ as
\begin{eqnarray}
\frac{\omega_{1,-}}{J}=2X^2-1+2 \frac{J'}{J}X ~~~~~~~~~~~~~~~~~~~~~~~~~~\nonumber \\
-\sqrt{\frac{(J'')^2}{J^2}+4(1-X^2)\left(\frac{D_a}{J}X+\frac{D'_a}{J} \right)^2}
\end{eqnarray}
Assuming $X$ is small, we expand the above equation and have the following approximate expression for the incommensuration:
\begin{eqnarray}
\label{eq:111}
&&\sin(\pi \epsilon)=\frac{J'}{2J} \Big( 1-\frac{2D_a D'_a}{J' \sqrt{(J'')^2+4(D'_a)^2}} \\
&& + \frac{(D_a)^2-(D'_a)^2}{J \sqrt{(J'')^2+4(D'_a)^2}}
-\frac{4(D_a D'_a)^2}{J [(J'')^2+4(D'_a)^2]^{3/2}} \Big).\nonumber 
\end{eqnarray}
Whether the incommensuraton is enhanced or not depends on the signs of
$D_a$ and $D'_a$ and the subtle balance of them.  We note that, if $D_a$
is present, and not too small, the incommensuration can be substantially
modified from the DM-free value, since the second term in the brackets
in the first line of Eq.~(\ref{eq:111}) is small only in the ratio
$D_a/J'$.  Since $D_a$ was neglected in the experimental fits in
Ref.\onlinecite{Coldea2002PRL}, this might lead to small errors in
the magnetic parameters, at perhaps a level of ten percent of their
estimated values, i.e. an uncertainty in $J'$ of $\pm 0.1 J'_{\rm
  estimated}$, and similarly for $D=D'_a$.  Errors of the order of
$10\%$ of the largest interaction, $J$, are clearly ruled out by the
fits.\cite{private-coldea}

\subsection{Field along $b$ axis}
Next we consider the case of field along the $b$ axis. 
In this case, we have 
\begin{equation}
{\cal H}({\bm k})=\frac{1}{2} \left(\begin{array}{cc}
{\sf A}_{b,{\bm k}} (\phi'_b) & {\sf B}_{\bm k} \\
{\sf B}^\dagger_{\bm k} & {\sf A}_{b,{\bm k}} (-\phi'_b)
\end{array}\right),
\end{equation}
where
\begin{equation}
{\sf A}_{b,{\bm k}}(\phi'_b)=\left(\begin{array}{cc}
2J\cos k_b & {\tilde J}'_b f_b(\phi'_b; {\bm k}) \\
{\tilde J}'_b f_b(-\phi'_b; -{\bm k}) & 2J \cos k_b
\end{array}\right) 
\end{equation}
with $f_b (\phi'_b; {\bm k})=e^{i\phi'_b} (1+e^{ik_b})(1+e^{ik_c})$.   
We minimize the excitation energy to find the ordering wavevector of the form ${\bm k}^*=(0, 2\pi(1/2+\epsilon),0)$. 
The lowest eigenvalue of ${\cal H}(0,k_b,0)$ is 
\begin{eqnarray}
\omega =J\cos k_b ~~~~~~~~~~~~~~~~~~~~~~~~~~~~~~~~~~~~~~~~~~~~~~~&&
\nonumber \\
- \sqrt{(J'')^2-4 J' J'' \cos (k_b/2)+4({\tilde J}'_b)^2 \cos^2 (k_b/2)}.&&
\end{eqnarray}
Similarly to the previous subsection, we rewrite the above as
\begin{equation}
\frac{\omega}{J}=2X^2-1-\sqrt{\frac{(J'')^2}{J^2}-4\frac{J' J''}{J^2}X+4\frac{({\tilde J}'_b)^2}{J^2}X^2},
\end{equation}
where $({\tilde J}'_b)^2=(J')^2+(D'_b)^2$.  One can find the
incommensuration $\sin (\pi \epsilon)=-X$ from the minimum of the above
equation and observe that DM interaction, $D'_b$, always enhances the
incommensuration from its $D'_b=0$ value $J'/(2J)$.

\subsection{Field along $c$ axis}
Finally, we consider the case of field along the $c$ axis.
In this case, ${\cal H}({\bm k})$ in Eq. (\ref{eq: spinwave_ham}) is given by
\begin{equation}
{\cal H}({\bm k}) = \frac{1}{2}\left(\begin{array}{cc}
{\sf A}_{c,{\bm k}} (\phi_c,\phi'_c) & {\sf B}_{\bm k} \\
{\sf B}^\dagger_{\bm k} & {\sf A}_{c,{\bm k}} (\phi_c,-\phi'_c)
\end{array}\right),
\end{equation}
where 
\begin{equation}
{\sf A}_{c,{\bm k}}(\phi_c,\phi'_c)=\left(\begin{array}{cc}
2{\tilde J}_c \cos (k_b-\phi_c) & {\tilde J}'_c f_c(\phi'_c; {\bm k}) \\
{\tilde J}'_c f_c(-\phi'_c; -{\bm k}) & 2{\tilde J}_c \cos (k_b+\phi_c)
\end{array}\right) 
\end{equation}
with $f_c(\phi'_c; \bm k) = e^{i\phi'_c} (1+e^{ik_b+ik_c}) + e^{-i\phi'_c} (e^{ik_b}+e^{ik_c})$. 
The minimum of the spectrum of 1-magnon excitation is also of the form ${\bm k}^*=(0,2\pi(1/2+\epsilon),0)$.  
The lowest eigenvalue of ${\cal H}(0,k_b,0)$ is explicitly obtained as
\begin{equation}
\omega_1=-J''+J \cos k_b -\sqrt{(D_c)^2 \sin^2 k_b +4 (J')^2 \cos^2 (k_b/2)}.
\label{eq for X}
\end{equation}
Here we have used the relations such as $\cos \phi_c = J/{\tilde J}_c$. 
The remarkable point here is that the minimum and hence the incommensuration $\epsilon$ is independent of $D'_c$ and $J''$. So once we know $\epsilon$ and $J, J'$, it uniquely determine the strength of $D_c$. 
Let us now assume that $\epsilon$ is of the order $J'/J$, which is true if $D_c=0$, and obtain approximate eigenenergy as
\begin{equation}
\frac{\omega_1}{J}=2 \sin^2 (\pi \epsilon)-2\frac{\sqrt{(J')^2+(D_c)^2}}{J}\sin(\pi \epsilon) -\frac{J''}{J}-1,
\end{equation}
where 
we have neglected a term proportional to $\sin^4 (\pi \epsilon)$. 
From the above equation, we can obtain the incommensuration $\epsilon$ as a function of $J$, $J'$, and $D_c$ as 
\begin{equation}
\sin (\pi \epsilon) = \frac{\sqrt{(J')^2+(D_c)^2}}{2J} 
\label{eq:90}
\end{equation}
From this relation, we see that the $D_c$ on the $J$ bonds enhances the
incommensuration $\epsilon$. This is in contrast to the measured incommensurability,\cite{veillette2005incomm} which is reduced compared to the expected one from the ideal standard model. 


\bibliography{sensitivity}

\begin{thebibliography}{83}
\bibitem[{pri()}]{private-coldea}
\bibinfo{howpublished}{We would like to thank R. Coldea for pointing this
  experimental finding to us.}
\end{thebibliography}

\end{document}